\documentclass[ 
superscriptaddress,
showpacs,
showkeys,
preprintnumbers,
 amsmath,amssymb,nofootinbib,
 aps,
prd, twocolumn,  
floatfix,
usenames,
dvipsnames
]{revtex4-1}

\usepackage{epsfig}
\usepackage{graphicx}
\usepackage{dcolumn}
\usepackage{bm}
\usepackage[colorlinks=true,linkcolor=blue,citecolor=blue]{hyperref}
\usepackage{natbib}
\usepackage{comment}
\usepackage{xcolor}
\usepackage{enumerate}
\usepackage{makecell, boldline}
\usepackage{gensymb}
\usepackage[normalem]{ulem}

\definecolor{forestgreen}{rgb}{0.1,0.49,0.07}
\DeclareMathOperator*{\maxi}{max}

\newcommand{\yr}{\, \text{yr}}
\newcommand{\degsq}{\, \rm deg^2}

\newcommand{\mchirp}{\mathcal{M}}

\newcommand{\msun}{{\, \rm M}_\odot}
\newcommand{\mtot}{ {\, \rm M_{\rm tot}}  }
\newcommand{\tc}{ {\, \rm t_{\rm c}}  }
\newcommand{\uncer}{uncertainties }  
\newcommand{\fisher}{\texttt{fisher} }
\newcommand{\ptmcmc}{\texttt{ptemcee} }
\newcommand{\mcmc}{\texttt{PTMCMCSampler} }

\newcommand\mnras{Monthly Notices of the Royal Astronomical Society}
\newcommand\apjl{The Astrophysical Journal Letters}
\newcommand\apjs{The Astrophysical Journal Supplement Series}
\newcommand\pasa{Publications of the Astronomical Society of Australia}
\begin{document}

\preprint{APS/123-QED}

\title{Observing the inspiral of coalescing massive black hole binaries with LISA\\ in the era of Multi-Messenger Astrophysics }
\author{Alberto Mangiagli}
 \email[E-mail: ]{a.mangiagli@campus.unimib.it}
 \thanks{\\ Data can be found at \url{https://github.com/amangiagli/Fits-for-parameter-estimation-of-MBHBs-in-LISA}}
\affiliation{%
Department of Physics G. Occhialini, University of Milano - Bicocca, Piazza della Scienza 3, 20126 Milano, Italy
}%
\affiliation{%
National Institute of Nuclear Physics INFN, Milano - Bicocca, Piazza della Scienza 3, 20126 Milano, Italy}%

\author{Antoine Klein}
\affiliation{School of Physics and Astronomy \& Institute for Gravitational Wave Astronomy, University of Birmingham, Birmingham, B15 2TT, UK}

\author{Matteo Bonetti}
\affiliation{%
Department of Physics G. Occhialini, University of Milano - Bicocca, Piazza della Scienza 3, 20126 Milano, Italy
}%
\affiliation{%
National Institute of Nuclear Physics INFN, Milano - Bicocca, Piazza della Scienza 3, 20126 Milano, Italy}%

\author{Michael L. Katz}
\affiliation{Department of Physics and Astronomy, Northwestern University, Evanston, IL 60201, U.S.A}
\affiliation{Center for Interdisciplinary Exploration and Research in Astrophysics, Evanston, IL 60201, U.S.A}

\author{Alberto Sesana}
\affiliation{%
Department of Physics G. Occhialini, University of Milano - Bicocca, Piazza della Scienza 3, 20126 Milano, Italy
}%
\affiliation{%
National Institute of Nuclear Physics INFN, Milano - Bicocca, Piazza della Scienza 3, 20126 Milano, Italy}%

\author{Marta Volonteri}
\affiliation{Institut d'Astrophysique de Paris, Sorbonne Universit\'e, CNRS, UMR 7095, 98 bis bd Arago, 75014 Paris, France}

\author{Monica Colpi}
\affiliation{%
Department of Physics G. Occhialini, University of Milano - Bicocca, Piazza della Scienza 3, 20126 Milano, Italy
}%
\affiliation{%
National Institute of Nuclear Physics INFN, Milano - Bicocca, Piazza della Scienza 3, 20126 Milano, Italy}%

\author{Sylvain Marsat}
\affiliation{%
Universit\'{e} de Paris, CNRS, Astroparticule et Cosmologie, F-75006 Paris, France}

\author{Stanislav Babak}
\affiliation{%
Universit\'{e} de Paris, CNRS, Astroparticule et Cosmologie, F-75006 Paris, France}

\affiliation{%
Moscow Institute of Physics and Technology, Dolgoprudny, Moscow region, Russia}

\date{\today}%

\date{\today}

\begin{abstract}
Massive black hole binaries (MBHBs) of $10^5 \, \rm M_\odot - 3 \times 10^7 \, \rm M_\odot $ merging in low redshift galaxies ($z\le4$) are sufficiently loud to be detected weeks before coalescence with the Laser Interferometer Space Antenna (LISA). This allows us to perform the parameter estimation {\it on the fly}, i.e. as a function of the time to coalescence during the inspiral phase, relevant for early warning of the planned LISA protected periods and for searches of electromagnetic signals.
In this work, we study the evolution of the sky position, luminosity distance, chirp mass and mass ratio uncertainties as function of time left before merger.
Overall, light systems with total intrinsic mass $\rm M_{\rm tot} = 3 \times 10^5 \, \rm M_\odot$ are characterized by smaller uncertainties than heavy ones ($\rm M_{\rm tot} = 10^7 \, \rm M_\odot$) during the inspiral. Luminosity distance, chirp mass and mass ratio are well constrained at the end of the inspiral. Concerning sky position, at $z=1$, MBHBs with $\rm M_{\rm tot} = 3 \times 10^5 \, \rm M_\odot$ can be localized with a median precision of $\simeq 10^2 \, \rm deg^2 (\simeq 1 \, \rm deg^2)$ at 1 month (1  hour) from merger, while the sky position of heavy MBHBs can be determined to $10 \, \rm deg^2$  only 1 hour before merger. However the uncertainty around the median values broadens with time, ranging in between 0.04 -- 20 $\rm deg^2$ (0.3 -- 3 $\times 10^3 \, \rm deg^2$) for light (heavy) systems at 1 hour before merger. At merger the sky localization improves down to $\simeq 10^{-1} \, \rm deg^2$ for all masses.
For the benefit of the observer community, we provide the full set of data from our simulations and simple and ready-to-use analytical fits to describe the time evolution of uncertainties in the aforementioned parameters, valid for systems with total mass between $10^5$--$10^7 \, \rm M_\odot$ and redshift $0.3$--$3$.

\end{abstract}



\maketitle

\section{\label{sec:intro}Introduction}

Massive black hole binaries (MBHBs) in the range $10^5-10^7 \msun$ are key targets of the space mission LISA, the Laser Interferometer Space Antenna  \cite{2017arXiv170200786A}. Coalescing MBHBs will be amongst the loudest sources of gravitational waves (GWs) in the LISA band \cite{PhysRevD.93.024003}. Their detection will enable to test for the first time and in a unique way the true nature of the massive dark objects nested at the centers of galaxies, whether they are  massive black holes (MBHs) indeed or more exotic compact objects of yet unknown nature \cite{PhysRevD.73.064030, Hughes:2004vw}. In the standard $\Lambda\text{CDM}$ scenario, MBHs of $10^{4-5}\msun$ were expected to form as early as redshift $z\sim 10-15$ when the Universe was less than a few million years old \cite{Umeda16,Valiante16,Johnson16}. Later, they grew through episodes of accretion and mergers, the last occurring in the aftermath of galaxy collisions  \cite{10.1093/mnras/183.3.341, 10.1046/j.1365-8711.2000.03077.x, Volonteri_2003, 10.1111/j.1365-2966.2007.11734.x, 10.1111/j.1365-2966.2007.12162.x,Colpi2014}.
MBHBs appears to be the inevitable outcome of galaxy assembly and 
LISA, with its sensitivity, will enable a survey of the entire Universe, providing the first census of MBHBs to  uncover their origin and growth along the cosmic history \cite{Haehnelt:1994wt, Jaffe:2002rt, Wyithe:2002ep, Enoki:2004ew, Sesana_2004, Sesana:2004gf, 10.1111/j.1365-2966.2005.08987.x, 10.1111/j.1365-2966.2012.21057.x,PhysRevD.93.024003,LISA-2019}. 
 
Given the rich variety of galaxies involved in merger events, chances are that MBHBs do not evolve in vacuum, but in gas rich environments, possibly surrounded by cold circumbinary gas disks \cite{ROSA2020101525}. In these circumstances, accretion of material onto the holes could produce copious amounts of electromagnetic (EM) radiation that can be triggered during the inspiral, merger and post-merger phase \cite{Armitage_2002, Milosavljevic:2004cg, Dotti:2006zn, Kocsis_2006}. Although it is well established that the MBHB torques open a cavity in the circumbinary disk, it has been recently pointed out that gas leaks through the disk edge feeding minidisks around the binary all the way to the final coalescence \cite{10.1093/mnras/sty423}. Full GRMHD simulations suggest that the circumbinary disk, the streams feeding the minidisks and the minidisks themselves contribute to variable EM emission at all wavelengths \cite{d_Ascoli_2018}. High energy emission coming from minidisks can be modulated by the binary orbital phase \cite{Bowen_2018} to an extent that might be possibly observed by future X-ray probes \cite{Dal_Canton_2019,2020NatAs...4...26M}.

The observation of an EM counterpart to a LISA event would be of paramount importance. MBHBs detected with GWs are standard sirens, since the luminosity distance to the source is one of the parameters entering the gravitational waveform and can therefore be directly measured \cite{Schutz1986}. The GW signal, however, does not allow to measure the source redshift, for which identification of an EM counterpart is needed.  
Therefore, the contemporaneous observation of the GW signal and the associated EM counterpart, will allow us to test the expansion of the Universe through the distance-redshift relation \cite{Schutz1986, 1475-7516-2016-04-002, Petiteau_2011, Holz_2005}, and the propagation properties of gravitational waves over cosmic distances \cite{PhysRevD.71.084025}.
In addition, joint GW and EM observations would provide unique insights on the physics of accretion in  violently changing space-time across the merger. The possibility of carrying out those tests depends critically on LISA's capability to localize the source  accurately and for telescopes to be able to point where the source is located. 

Several authors have studied LISA's capability to localize the source and to measure its parameters \cite{PhysRevD.57.7089, PhysRevD.69.082005, 10.1046/j.1365-8711.2002.05247.x, PhysRevD.71.084025, PhysRevD.70.042001,  PhysRevD.93.024003,PhysRevD.76.104016,Lang_2008, Kocsis_2008, PhysRevD.77.024030,PhysRevD.84.064003,  2020arXiv200110011B, Marsat:2020rtl}. Most of these works focused on estimating the binary parameters 
from the full signal, i.e. from the inspiral to the ringdown, while only few of them explored LISA capability to constrain the source parameters as a function of the time to coalescence \cite{Lang_2008, Kocsis_2008, PhysRevD.77.024030, PhysRevD.84.064003}. 

Lang and Hughes \cite{Lang_2008} estimated the uncertainties in the sky position and luminosity distance for a set of MBHBs as function of the time to coalescence exploring the relative impact of spin-precession in the last days before merger.
They found that spin-precession reduces the error in the major and minor axis of the error ellipse by a factor $1.5 - 9$ at the end of the inspiral. A similar improvement was found for the luminosity distance. Notably, from their simulations, LISA should be able to localize MBHB with total intrinsic mass $ \lesssim 4 \times 10^6 \msun$ at $z = 1$ with a precision of $\simeq 10 \degsq$ already one month before merger.
They also found that LISA would localize with better precision sources lying outside the Galactic plane.
Kocsis \emph{et al.} \cite{Kocsis_2008} performed a similar analysis, adopting the `harmonic mode decomposition' method to compute the 3D localization volume as a function of time to coalescence. They also included a detailed discussion of the possible EM counterparts one could expect from MBHBs and their implications for wide- and narrow-field instruments.
Trias and Sintes \cite{PhysRevD.77.024030} adopted a full post-Newtonian waveform, and found an overall improvement in the parameter estimation for massive MBHB ($\mtot > 5 \times 10^6 \msun$). In particular the inclusion of amplitude corrections leads to earlier warning and improved sky-position accuracy as a function of time to coalescence for massive unequal MBHBs.
McWilliams \emph{et al.} \cite{PhysRevD.84.064003} explored LISA ability to constrain the sky-position and luminosity distance of non-spinning equal mass systems using the full inspiral-merger-ringdown waveform and including the orbital motion of the detector plus the three-channel LISA response. They found a final median sky-position error of $\simeq 3 \, \rm arcmin$ for a system with total intrinsic mass of $2 \times 10^6 \msun$ at $z=1$.

All these studies were performed with the {\it classic} LISA design, with a five-million kilometer constellation and a low frequency sensitivity extending down to $10^{-5}$ Hz. Following the re-design of the LISA mission, the armlength of the interferometer is reduced to 2.5 million kilometers, featuring a steeper low frequency sensitivity extending possibly down to $2\times 10^{-5}$Hz, based on the in-flight LISA Pathfinder performance \cite{armano2019lisa}. Loss in the low frequency sensitivity, as compared to the {\it classic} LISA, leads to a shorter (detectable) duration of the GW signal and, therefore,
to some degradation of our abilities to localize the sources. It also implies that more weight is given to the merger and post-merger part of the signal. In light of these changes, it is therefore necessary to reassess earlier findings to establish the performance of the current design in producing {\it on the fly} estimates of source parameter errors.

In this paper we explore the power of a time dependent parameter estimation analysis by considering the current LISA sensitivity curve. We estimate the relative uncertainties in the measure of the sky position, luminosity distance, chirp mass and mass ratio as a function of the time to coalescence. To this end, we use the frequency-domain waveform based on the shifted uniform asymptotics (SUA) method developed in \cite{PhysRevD.90.124029}. The waveform describes the inspiral portion of the signal, including spin precession and higher harmonics. The contribution of merger and ringdown to the parameter estimation precision is folded in by rescaling the errors according to the full signal-to-noise ratio ($S/N$) of the event provided by the PhenomC waveform \cite{PhysRevD.82.064016}, following the procedure of \cite{PhysRevD.93.024003}. 
The GW signals are integrated starting from $10^{-5}$ Hz, and we also explore the impact of a putative degradation of LISA performance below $10^{-4}$ Hz.

Eventually, Markov-Chain Monte Carlo (MCMC) simulations will be required for an accurate estimate of source parameters. For this scope, the LISA Data Challenge (LDC) project is being developed and it constitutes now the largest and most updated library available to obtain the information needed to tackle these problems. However, the parameter space of merging MBHBs is vast, and its full exploration with  MCMC techniques is time consuming.
Several studies \citep[e.g.][]{Cornish_2007, PhysRevD.91.104001} demonstrated that parameter estimation via Fisher information matrix evaluation reproduces reasonably well MCMC results for high $S/N$ sources, which are the  targets of this study. Although Fisher matrix-based estimates allow to drastically cut down computational cost, there is still the need to perform large set of simulations and, up to now, there is no public available code for this. Berti \emph{et al.} \cite{PhysRevD.71.084025} provided analytical formulas for the mass and distance errors for non-spinning MBHB with total mass $\mtot = 2 \times 10^6 \msun$ but for the old LISA design and as function of redshift, not as function of time to coalescence. 

The scope of this paper is to provide a vast library of LISA parameter estimates as a function of time to merger for MBHBs. All results are obtained by consistently evaluating the Fisher matrix of the signal as the source evolves, and are tested, for a subset of selected sources, against full MCMC calculations.
To the benefit of the community, we provide analytical formulas to describe how parameter estimates improve {\it on the fly}, i.e. while the MBHB is approaching the merger. We perform an extensive set of simulations exploring the parameter space of MBHBs in the mass range $10^5\msun< \mtot <3\times10^7\msun$ at $z<4$ and fit the results with polynomial expressions. The resulting formulas are cast in terms of three variables: the total (intrinsic) mass of the system, the redshift of the source and the (observed) time to coalescence. 

With this work, we also release the full set of data on which these formulas are based. The data can be found at \url{https://github.com/amangiagli/Fits-for-parameter-estimation-of-MBHBs-in-LISA}.

The paper is organized as follows. In Section \ref{sec:LISA_noise} we describe
the LISA sensitivity curve and the low-frequency degradation adopted for this study. In Section \ref{sec:Theory} we describe our theoretical framework and introduce GW analysis concepts, as the signal-to-noise ratio ($S/N$) and the Fisher matrix formalism, our expression for the error \uncer and how we rescale the sky position and luminosity distance errors at merger. We also provide a brief introduction to the MCMC formalism. Our main results are presented in Section \ref{sec:res}. In Section \ref{sec:time_progress_pe} we adopt the analytical formulas to estimate LISA ability to constrain source parameters in advance. In Section \ref{sec:multimessenger_view} we discuss possible synergies with EM facilities and in Section \ref{sec:conclusion} we conclude with some final remarks and draw our conclusions.
In Appendix \ref{sec:app_to_discuss_random_params} we study the effect of binary parameters in shaping the \uncer distribution.
In Appendix \ref{sec:comparison_old_curve} we compare our current results with the ones in \cite{Lang_2008}.
In Appendix \ref{sec:binary_angular_momentum}
we focus on LISA ability to estimate the binary angular momentum.
Finally in Appendix \ref{sec:app_for_tables} we report the table with the coefficient values for the analytical formulas.

\section{\label{sec:LISA_noise} LISA observatory}
For the LISA sensitivity curve, we adopt a six-link laser configuration labeled as ``Payload Description Document Allocation'' in the ``LISA strain curves'' document \footnote{See also \url{ https://atrium.in2p3.fr/nuxeo/nxpath/default/Atrium/sections/Public/LISA@view_documents?tabIds=\%3A}} and described also in \cite{Cornish:2018dyw}.
The total noise power spectral density in a single LISA data channel is modeled as 

\begin{equation}
\begin{split}
\label{eq:strain_noise}
S_n(f) = \frac{1}{L^2}\biggl( & P_{\mathrm{OMS}}(f) +  \frac{4P_{\mathrm{acc}}(f)}{(2\pi f)^4}\biggr) \times \\
		& \times \biggl(1 + \frac{6}{10}\biggl( \frac{f}{f_{*}}\biggr)^2\biggr) + S_c(f) \, \text{Hz}^{-1}
\end{split}
\end{equation}
with $L = 2.5 \, \text{Gm}$, $f_{*} = 19.09 \, \text{mHz}$ and
\begin{equation}
P_{\mathrm{OMS}}(f) = (1.5 \times 10^{-11})^2 \left( 1+ \left( \frac{2 \, \text{mHz}}{f} \right)^4\right) \text{m}^2 \text{Hz}^{-1},
\end{equation}
\begin{equation}
\begin{split}
    P_{\mathrm{acc}} = (3 \times 10^{-15})^2  & \left( 1 + \left( \frac{0.4\, \text{mHz}}{f}\right)^2\right)  \times \\   
      &  \times \left( 1 + \left( \frac{f}{8 \,  \text{mHz}}\right)^4\right)         \text{m}^2 \text{s}^{-4} \text{Hz}^{-1}
\end{split}
\end{equation}
In addition to the instrument noise, unresolved galactic binaries are expected to form a confusion background noise below $ \lesssim 1 \, \rm mHz$. While the LISA mission progresses, more galactic binaries will be detected and this noise source will reduce. We model the background noise contribution as
\begin{equation}
    S_c(f) = A f^{-7/3} e^{-K f^{\alpha}} [1 + \tanh(-\gamma(f-f_k))] \rm Hz^{-1}
\end{equation}
where $A = (3/10) \, 3.26651613 \cdot 10^{-44}$ and $\alpha = 1.18300266$. Parameters $K, \, \gamma, \, f_k$ change as the mission progresses and their values are reported in Tab. \ref{tab:coeff_bckgr}
\cite{PhysRevD.95.103012}.

\begin{table}
\caption{\label{tab:coeff_bckgr} Coefficients describing the expected stochastic galactic background as a function of mission duration.}
\begin{ruledtabular}
\begin{tabular}{c | m{2cm} | m{2cm} | m{2cm} } 
$T_{obs}$   & \centering $K$  & \centering $\gamma$  & $f_k \rm [mHz]$ \\
\Xhline{0.1pt}
1 day  & 941.315118  & 103.239773 & 11.5120924 \\
3 months & 1368.87568 & 1033.51646 & 4.01884128 \\
6 months & 1687.29474 & 1622.04855 & 3.47302482 \\ 
1 year & 1763.27234 & 1686.31844 & 2.77606177 \\
2 years & 2326.78814 & 2068.21665 & 2.41178384 \\
4 years & 3014.30978 & 2957.74596 & 2.09278117 \\
10 years & 3749.70124 & 3151.99454 & 1.57362626 \\
\end{tabular}
\end{ruledtabular}
\end{table}

For simplicity, our fiducial sensitivity assumes the long-wavelength approximation and, therefore, it does not include the high frequency oscillations. Nevertheless, most of the sources in the parameter space explored in this study emit GWs with $f < 0.05 \rm \, Hz$ so we expect it to be a valid approximation. 

LISA sensitivity is set by the mission requirement in the range $10^{-4}-1 \rm \, Hz$ \cite{lisa_requirement_docs}: besides the extrapolation from Eq.~\ref{eq:strain_noise}, we also consider a less favorable version of the curve in the low-frequency region ($f < 2 \times 10^{-4} \rm \, Hz$).
The new curve is obtained adding an additional source of noise in the acceleration term, i.e
\begin{equation}
    P_{\mathrm{acc, \,degr}} = P_{\mathrm{acc}}  \times  \left( 1 + \left( \frac{0.1 \,  \text{mHz}}{f}\right)^2\right) 
\end{equation}
and substituting this new acceleration noise to the one in eq. \ref{eq:strain_noise} \cite{lisa_noise_performance}.

LISA follows a heliocentric Earth-trailing orbit with an opening angle of $20\degree$. The constellation is made of three identical spacecrafts, set in an equilateral triangular configuration inclined at $60\degree$ respective to the ecliptic orbit, that rotate around their guiding center. The combination of the orbital motion of the overall configuration and the rotation of the single spacecraft imprint unique features in the observed signals, such as amplitude and phase modulation that help breaking degeneracies. The detector response to the incoming gravitational wave is modeled as in \cite{PhysRevD.57.7089, PhysRevD.93.024003}.

\begin{figure}
    \includegraphics[width=0.5\textwidth]{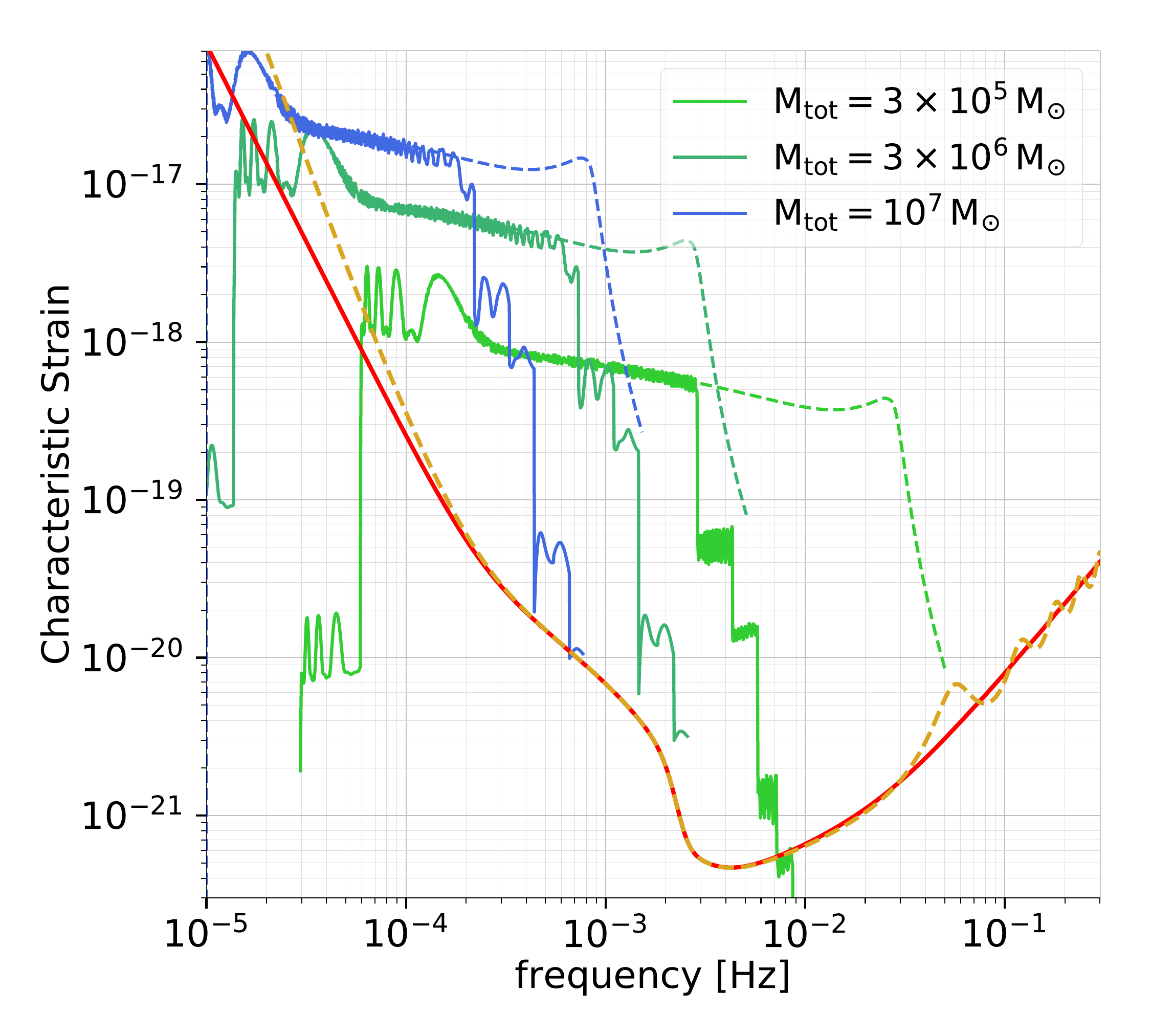} 
    \caption{Example of GW signals in the LISA band for three MBHBs with source-frame total mass as labeled at redshift $z = 1$. The solid red curve corresponds to the instrument sensitivity, while dashed golden line corresponds to the sensitivity with an additional noise contribution at low frequencies. Blue and green curves represent the typical tracks for  non-spinning MBHBs: dashed lines correspond to PhenomC waveform, while solid lines to the only-inspiral waveform. For the plot, we assume $q = 0.5$, $\iota = \psi = \pi/4$, $\theta_N = \cos(\pi/4)$, $\phi_N = \pi/4$, $t_c = 4 \yr$. For this figure, the stochastic background from the galactic binaries has been set to the level after 4 years of mission.}
    \label{fig:wf_and_noise}
\end{figure}

\section{\label{sec:Theory} Theory}
The signal from a precessing MBHB in quasi-circular orbit is described by 15 parameters: the two components masses in the source frame, $m_1$ and $m_2$, the colatitude $\cos\theta_N$ and longitude $\phi_N$  of the binary sky position in ecliptic coordinates, the luminosity distance $d_L$, the two time-varying angles describing the orientation of the binary orbital angular momentum, $\theta_L$ and $\phi_L$, the time to coalescence $t_c$, the initial phase $\phi_0$, the magnitude of the two spins $\chi_1$ and $\chi_2$ and 6 angles describing their orientation relative to the binary plane.
We further define the mass ratio $q = m_2/m_1$ with $m_1 > m_2$ and the chirp mass $\mathcal M = (m_1 m_2 )^{3/5} / \text{M}_{\text{tot}}^{1/5}$, where $ \text{M}_{\text{tot}} = m_1+ m_2$ is the total mass of the system. Moreover, instead of the binary angular momentum orientation, we can define the inclination $\iota$ of the orbital angular momentum with respect to the line of sight and the polarization $\psi$.

In this work we adopt the following  waveforms to describe the GW signal:
\begin{enumerate}[(I)]
\item A general precessing inspiral-only waveform based on the shifted uniform asymptotics (SUA) method (described in \cite{PhysRevD.90.124029}). The waveform is computed in the frequency domain for the inspiral part only. This model contains higher order harmonics and allows arbitrary orientation of spins (including orbital precession due to spin-orbit coupling). This waveform is used in the computation of the parameter uncertainties based on the Fisher information matrix, and to obtain the main results of this paper;
\item PhenomC \cite{PhysRevD.82.064016}, a spin-aligned inspiral-merger-ringdown waveform. This model includes only the dominant harmonic and ignores the orbital precession. It is used only to evaluate the signal-to-noise ratio for the full signal and to rescale the uncertainties on sky position and luminosity distance at merger;
\item PhenomHM waveform \cite{PhysRevLett.120.161102} which also describes inspiral, merger and ringdown parts of the GW signal. Similar to PhenomC this model ignores orbital precession but takes into account higher order harmonics.  It is used in assessing parameters uncertainties within Bayesian (Markov-chain Monte-Carlo) approach.
\end{enumerate}

The first model is the most complete and gives the best estimates for the inspiral part of the GW signal. All three models are reliable in estimating uncertainties in the intrinsic parameters of binary systems (masses, magnitude of spins).
Extrinsic parameters (such as distance, orbital inclination, sky position) are quite degenerate in PhenomC model,
higher order harmonics and orbital precession break this degeneracy (at least partially), this is the reason behind
using SUA-based inspiral-only and inspiral-merger-ringdown PhenomHM models in the parameter estimation.

In Fig.~\ref{fig:wf_and_noise} we plot the non-sky-averaged sensitivity curves adopted in this work with typical GW signals.

For a given system, we compute the signal-to-noise ratio as 

\begin{equation}
    (S/N)^2 = 4 \int_{f_{\rm min}}^{f_{\rm max}} \frac{|\tilde{h}(f)|^2 }{S_n(f)} df
\end{equation}
where $\tilde{h}(f)$ is the Fourier transform of the GW time-domain strain and $S_n(f)$ is the detector sensitivity as defined in Eq. \ref{eq:strain_noise}. In this study, the minimum frequency $f_{\rm min}$ is computed from the coalescence time of the system with a lower limit of $f_{\rm min} = 10^{-5} \rm \, Hz$ ($f_{\rm min} = 2 \times 10^{-5} \rm \, Hz$) for the nominal (degraded) curve, while the maximum frequency $f_{\rm max}$ corresponds to the system reaching 1 hour from coalescence. To compute the parameter errors, we adopt the Fisher matrix formalism \cite{PhysRevD.77.042001}. In particular we define the Fisher matrix as
\begin{equation}
    \Gamma_{ab} \equiv \left( \frac{\partial h}{\partial \theta^a} \bigg| \frac{\partial h}{\partial \theta^b}    \right)
\end{equation}
where $\partial h/\partial \theta^a$ is the partial derivative of the GW waveform respect to the parameter $\theta^a$ evaluated at the injected value and $(\cdot|\cdot)$ is the standard inner product between two complex quantities, defined as
\begin{equation}
    (a|b) \equiv 2 \int_{0}^{\infty} \frac{\tilde{a}^{*}(f)\tilde{b}(f) + \tilde{a}(f)\tilde{b}^{*}(f)}{S_n(f)} df \,.
\end{equation}
Since LISA can be seen as two independent detectors, we construct the total Fisher matrix as the sum of the Fisher matrix for each detector as
\begin{equation}
    \Gamma_{ab} = \Gamma^{\text{I}}_{ab} + \Gamma^{\text{II}}_{ab} \, .
\end{equation}
The correlation matrix is the inverse of the Fisher matrix, $ \Sigma = \Gamma^{-1}$. To avoid possible problems when computing the inverse of the Fisher matrix numerically, we implement an additional check. Following \cite{PhysRevD.83.044036}, we accept the inverse of the Fisher matrix only if 
\begin{equation}
    \maxi_{i,j}|I_{\rm num}^{ij} - \delta^{ij}   |< \epsilon_{\rm min}
\end{equation}
where $I_{\rm num}^{ij}$ is the ``numerical identity matrix'' obtained multiplying the Fisher matrix by its inverse and $\delta^{ij}$ is the standard Kronecker delta function. We set $\epsilon_{\rm min} = 10^{-3}$.
We also check that this additional condition on the Fisher matrix does not introduce any bias in the distribution of the initial parameters.
The expected statistical error on  a single parameter $ \Delta \theta^a$ is computed from the corresponding diagonal element of the correlation matrix as
\begin{equation}
    \Delta \theta^a \equiv \sqrt{ \langle (\delta \theta^a)^2 \rangle  } = \sqrt{\Sigma^{aa}}. 
\end{equation}
We adopt SUA as waveform to carry on the inspiral parameter estimation at different (observed) times to merger.
In particular we stop the waveform at 1 month, 1 week, 3 days, 1 day, 10 hours, 5 hours and 1 hour before coalescence. We compute the 2D sky position error ellipse $\Delta \Omega$ following \cite{PhysRevD.74.122001} (see  Eq. 4.13) so that the probability for the source to lie outside this region is $e^{-1}$ where $e$ is the source position ellipse eccentricity. 
The uncertainties on chirp mass and mass ratio are simply propagated as:
\begin{equation}
\begin{split}
   \biggl(  \frac{\Delta \mchirp }{\mchirp} \biggr)^2 & = \biggl( \frac{m_1}{\mchirp} \biggr)^2  \biggl( \frac{\partial \mchirp}{\partial m_1}\biggr)^2 \Sigma^{\ln m_1, \ln m_1} \\
   &  + \biggl( \frac{m_2}{\mchirp} \biggr)^2  \biggl( \frac{\partial \mchirp}{\partial m_2}\biggr)^2 \Sigma^{\ln m_2, \ln m_2} \\
   & + 2 \biggl( \frac{m_1 m_2}{\mchirp^2} \biggr) \biggl( \frac{\partial \mchirp}{\partial m_1}\biggr) 
   \biggl( \frac{\partial \mchirp}{\partial m_2}\biggr) \Sigma^{\ln m_1, \ln m_2} \\
\end{split}
\end{equation}
and similarly for the mass ratio, replacing $\mchirp$ with $q$ (here $\ln$ refers to the natural logarithm).

To model the effect of merger and ringdown in the parameter estimation (labeled as `merger' in our figures), we rescale the sky position and luminosity distance uncertainties at 1 hour before merger as 
 \cite{PhysRevD.93.024003}
\begin{align}
\label{eq:area_rescale}
    \Delta \Omega_{\mathrm{merger}}  &= \Delta \Omega_{\rm 1 h} \times \left[ \frac{(S/N)_{\rm SUA}} {(S/N)_{\mathrm{PhC}}} \right]^2 \\
    \label{eq:dl_rescale}
    \Delta d_{L, \,{\mathrm{merger}}}  &= \Delta d_{L, \,{\rm 1 h}} \times \left[ \frac{(S/N)_{\rm SUA}} {(S/N)_{\mathrm{PhC}}} \right]
\end{align}
where $S/N_{\rm SUA}$ is the $S/N$ accumulated with the waveform SUA at the end of the inspiral, and $S/N_{\mathrm{PhC}}$ is the one computed with the PhenomC waveform.

To asses the validity of our approach, we compare the sky position \uncer computed with our approach against MCMC simulations for a small subset of 20 samples.
For a system with true parameters $\overline{\theta}_0$ producing a data strain in the detector $d = h(\overline{\theta}_0) + n$ where $h$ is our waveform and $n$ is the noise realization, the likelihood can be computed as 
\begin{equation}
    p(d|\overline{\theta}) \propto e^{-\frac{1}{2} ( h(\overline{\theta}) - d |  h(\overline{\theta}) - d)},
\end{equation}
where we have neglected constant normalization factor. 

To compute the posterior distribution $p(\overline{\theta}|d)$ for the parameter $\overline{\theta}$, we adopt Bayes theorem
\begin{equation}
p(\overline{\theta}|d) = \frac{\mathcal{L}(d|\overline{\theta}) \pi(\overline{\theta})}{p(d)}
\end{equation}
where $\mathcal{L}(d|\overline{\theta})$ is the likelihood of data given $\overline{\theta}$, $\pi(\overline{\theta})$ is the prior distribution for the binary parameters and $p(d) = \int d \overline{\theta} \mathcal{L}(d|\overline{\theta}) \pi(\overline{\theta})$ is a normalization constant. To sample the  posterior distributions $p(\overline{\theta}|d)$ we adopt two MCMC algorithms: \texttt{PTMCMCSampler} \cite{justin_ellis_2017_1037579}
without parallel tempering  and \texttt{ptemcee} \cite{10.1093/mnras/stv2422}
with parallel tempering. Using \texttt{PTMCMCSampler} without parallel tempering allows sampling  the posterior around the true values in a very efficient way, while in the second method we explore the global posterior (including secondary modes) using GPU \cite{Katz:2020hku}.
For the first algorithm we adopt uniform distribution in chirp mass, while for the second one uniform distribution in total mass. The other parameters are sampled the same way for both treatments: uniform distribution in mass-ratio, source volume, binary orientation, polarization, phase and spin magnitude from -1 to 1.
For all 20 systems, we use the zero noise realization and adopted the PhenomHM waveform \cite{PhysRevLett.120.161102}, that takes into account higher modes, merger-ringdown phases but ignores precession. 
Those systems were randomly drawn from our parameter space with the only requirement to have sky localization \uncer $\Delta \Omega < 10 \degsq$ at 5 hours before merger. For simplicity, we remove the contribution of the galactic background from Eq. \ref{eq:strain_noise}.

A primary goal of our work is to provide analytical fits to describe how parameter uncertainties decrease as the signal is accumulated. Therefore we choose to report the median errors, without focusing on the fraction of systems for which a given parameter can be measured up to a certain precision \cite{PhysRevD.93.024003}. For the same reason,  we do not consider in this work semianalytical models (\cite{Barausse2012} and later expansions). 

The error on the luminosity distance for sources at $z > 0.25$ is expected to be dominated by weak lensing, due to the matter distribution between us and the source \cite{Wang_2002, 10.1111/j.1365-2966.2010.16317.x}. We do not include the effect of weak lensing  in our analysis: therefore the reported uncertainties on the luminosity distance and redshift refer to the pure GW measurements.

Signals from cosmological sources at redshift $z$ are affected by the expansion of the Universe. This implies that what is actually measured in the detector is a redshifted chirp mass, i.e. $\mathcal{M}_{z} = \mathcal{M}\,  (1+z)$. We will however refer to masses in the source frame since relative uncertainties are basically redshift independent.
Finally, we assume a fiducial $\Lambda$CDM cosmology with $h = 0.678$, $\Omega_m = 0.308$ and $\Omega_{\Lambda} = 0.692$.

\section{\label{sec:res}Results}
We run different sets of simulations to properly explore the parameter space. We keep fixed the total (source frame) mass of the system for a set of redshifts. 
 We consider:
\begin{enumerate}[(I)]    
    \item $\rm \mtot = 10^5, \,  3\times 10^5, \,5\times 10^5,\,  7.5 \times 10^5, \, 10^6, \, 3\times 10^6, \, 5\times 10^6, \, 7.5\times 10^6, \, 10^7, \, 3\times 10^7 \, \msun$;
    \item $ z = 0.1, \, 0.3, \,  0.5, \, 1, \, 2, \,3, \, 4$;
\end{enumerate}  

The mass ratio is randomized in $[0.1,\, 1]$, while the time to coalescence is drawn in $[0, \, 4] \yr$.   Spin magnitudes are flat distributed  in $[0, \, 1]$. Sky position, angular momentum angles and spin orientations are uniformly distributed over a sphere. Since we want to explore how parameter estimation improves as function of time to coalescence we did not take into account the possibility that LISA stops taking data while a signal is chirping in band.
Unless otherwise noted, for each combination of total mass and redshift, we perform $\rm N = 10^4$ random realizations.

\subsection{General trends in parameter determination precision}

\begin{figure*}
    \includegraphics[width=\textwidth]{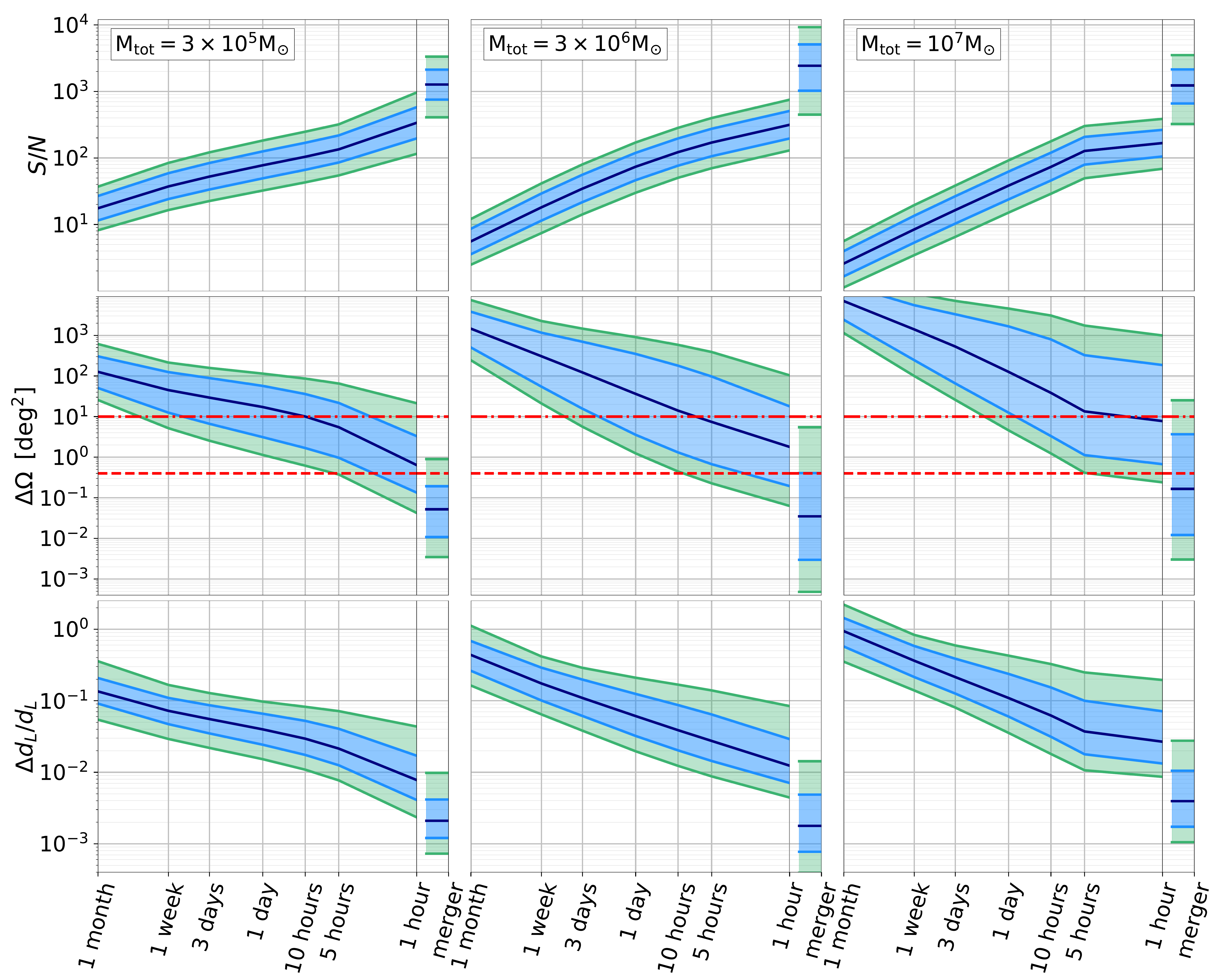} 
    \caption{Accumulated $S/N$, sky-position and luminosity distance uncertainties as function of time to coalescence for light (left column), intermediate (central column) and heavy (right column) systems. All sources are located at $z=1$. Blue line corresponds to the median of the distribution, while light blue and green areas correspond to the 68 and 95 percentiles. For each case, we also plot the $S/N$, the sky position and luminosity distance uncertainties when the full signal is considered (`merger'), inferred according to 
    the scaling in Eq. \ref{eq:area_rescale}-\ref{eq:dl_rescale}. In the mid panels, the dashed and dotted-dashed horizontal red lines correspond to the field of view of \emph{Athena} and LSST of $0.4 \degsq$ and $10 \degsq$ respectively. For all cases, while the $S/N$ monotonically increases, the median of the distributions decreases, leading to a progressively 
    more accurate parameter estimation. However with time the uncertainties around the median value broaden, implying different levels of parameter estimation accuracy for sources with the same mass and redshift. This is especially true for the sky localisation.}
    \label{fig:triple-sky-loc}
\end{figure*}
\begin{figure*}
    \includegraphics[width=\textwidth]{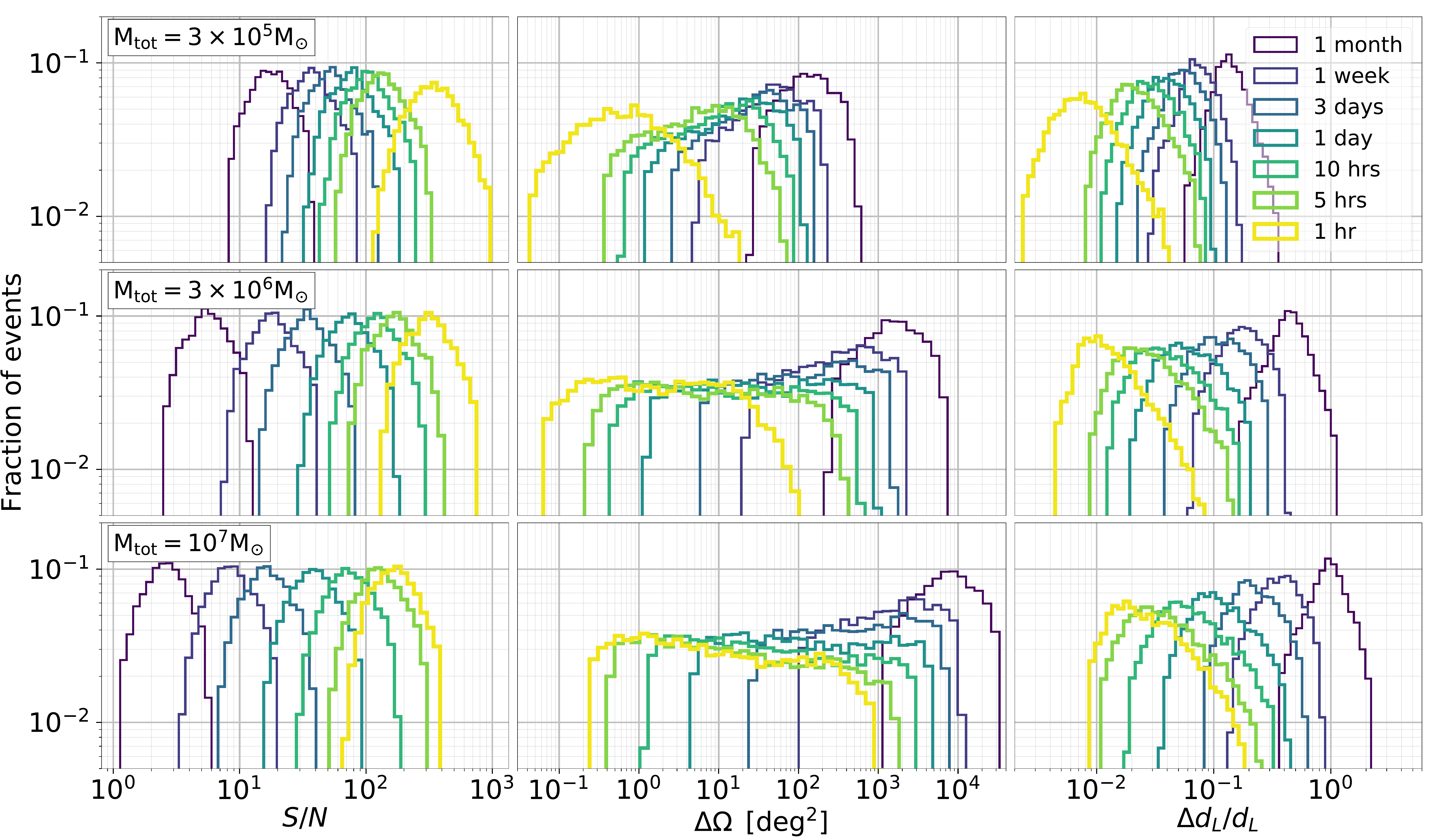} 
    \caption{$S/N$, $\Delta \Omega$ and $\Delta d_L/d_L$ distributions of the 95 percentile corresponding to different times to coalescence as indicated in the legend. The total source-frame mass and redshift are as in Fig.~\ref{fig:triple-sky-loc}. Upper panels: light systems. Central panels: intermediate systems. Lower panels: heavy systems.}
    \label{fig:triple-sky-loc-distro}
\end{figure*}
\begin{figure*}
    \includegraphics[width=\textwidth]{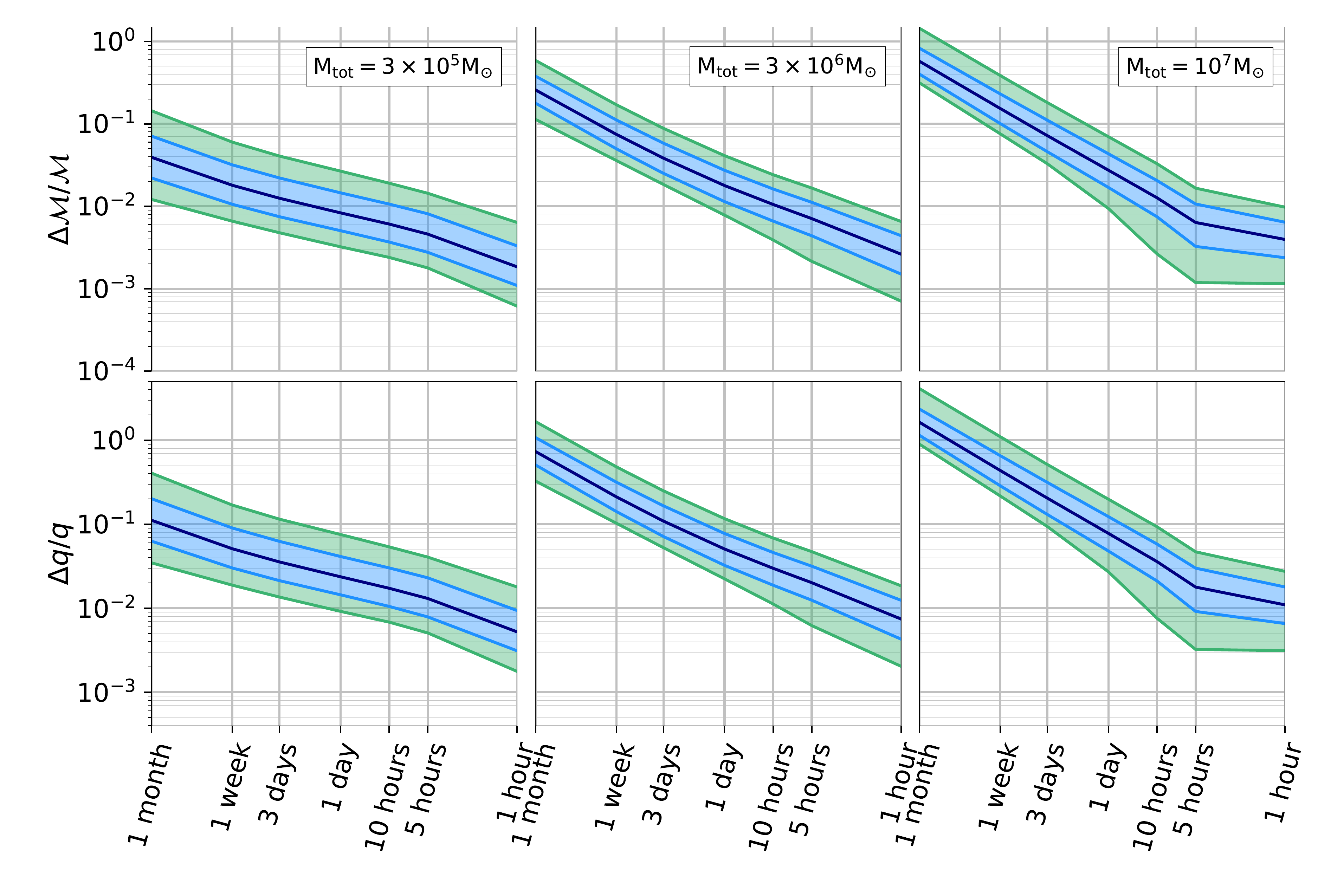} 
    \caption{Chirp mass and mass ratio relative uncertainties as a function of time to coalescence for the same systems considered in Fig.~\ref{fig:triple-sky-loc}.}
    \label{fig:all-params}
\end{figure*}
\begin{figure*}
    \includegraphics[width=\textwidth]{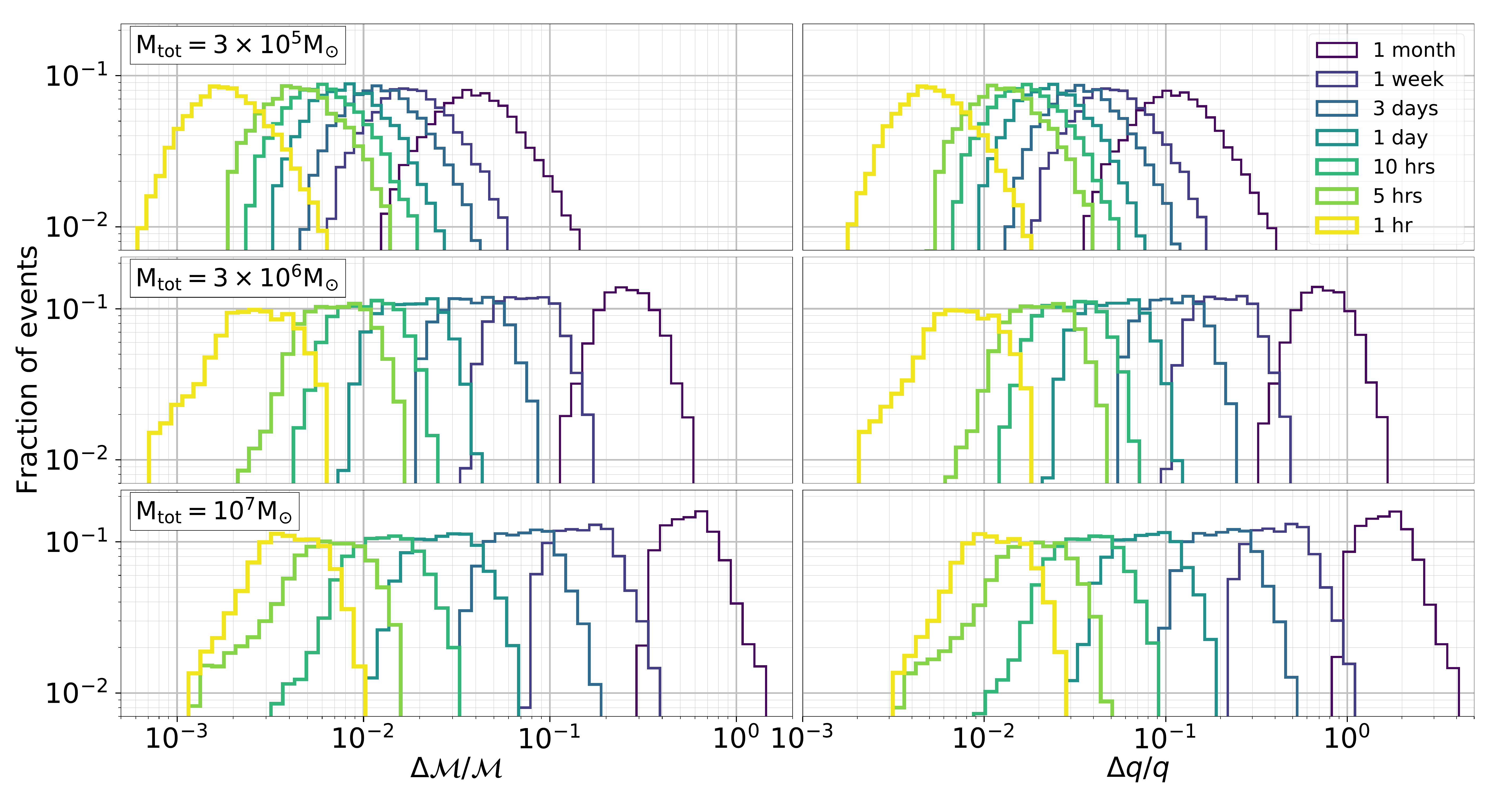} 
    \caption{Chirp mass and mass ratio uncertainty distributions of the 95 percentile corresponding to different times to coalescence as indicated in the legend. Total source-frame mass and redshift as for the systems in Fig.~\ref{fig:all-params}.}
    \label{fig:all-params-distro}
\end{figure*}

We start by selecting three representative systems to describe in detail the general behavior of the evolution of parameter estimation precision as a function of time to coalescence. We pick three systems with $\rm \mtot = 3 \times 10^5 \msun $ (`light' hereinafter), $\, 3 \times 10^6 \msun$  (`intermediate' hereinafter), $ 10^7 \, \msun$ (`heavy' hereinafter) at $z=1$. 

\subsubsection{Sky location and luminosity distance}
In Fig.~\ref{fig:triple-sky-loc} we show the time evolution of the $S/N$, the sky-position and luminosity distance estimate for the three systems. Light binaries live longer in the LISA band and accumulate a median $S/N \simeq 20$ already 1 month before coalescence,
 compared to $S/N$ of $\simeq 5$ and $\simeq 3$ of intermediate and heavy systems respectively. 
As systems approach merger, the $S/N$ increases. Including the full signal (`merger'), light systems have similar $S/N$ to heavy ones. 
However, due to the form of LISA sensitivity curve, the $S/N$ contribution from merger and ringdown for intermediate systems leads to a final value of $S/N \simeq 2 \times 10^3$. 

Turning to sky localization, lighter systems are typically better localized than heavier ones, especially at earlier time. 
Light binaries are localized with a median accuracy of $\simeq 100 \degsq$ already 1 month before coalescence, due to the modulations imprinted by the detector orbital motion and the higher $S/N$. As the binary approaches coalescence, the signal accumulates and the uncertainties in the sky localization reduce to $\simeq 10 \degsq$ at 10 hours before merger.  A similar improvement is also present at shorter timescales, i.e from 10 hours to 1 hour, where the uncertainty drops to $\simeq 1 \degsq$, since in this phase the accumulated $S/N$ increases rapidly and spin precession effects come into play. By rescaling the area at the end of the inspiral according to Eq.~\ref{eq:area_rescale}, we find final sky localization uncertainties of $ \lesssim 0.1 \degsq$. 
Intermediate systems are localized less precisely with a recovered area of $\simeq 10^3 \degsq$ 1 month prior to merger. This is due to the lower $S/N$ values. The angular resolution at 1 hour from merger is comparable to that of light systems with a median value of $\simeq 2 \degsq$. Eventually, the $S/N$ is dominated by the merger and ringdown part of the signal, allowing a further improvement to the binary's location and bringing the median value down to $\simeq 0.04 \degsq$.
Similar considerations apply to heavy MBHBs. The sky position is essentially unconstrained 1 month before coalescence, mostly due to the very low $S/N$. By the end of the inspiral, the source can be localized with a median uncertainty of $\simeq 10 \degsq$, further reduced to $\lesssim 0.2 \degsq$ when the full signal is accounted for.

The luminosity distance for light systems can be determined at $10 \%$ and $0.8\%$ level at 1 month and 1 hour from merger, respectively. Similarly to sky localization, the distance determination early in the inspiral is severely degraded when moving to more massive binaries.  Nonetheless at the end of the inspiral, the distance for the intermediate and heavy systems can be measured with $\simeq 1\%$ and $\simeq 3\%$ median precision, respectively. Rescaled uncertainties including merger and ringdown are around $0.2-0.4\%$ for all systems. 
We remind that the reported errors in luminosity distance do not include the weak lensing error and, therefore, are to be considered as optimistic.

We note that for intermediate (heavy) mass systems the $S/N$ is still below 10 at 1 month (1 week) from coalescence, therefore the Fisher matrix formalism should be applied with caution. However, at these early times and for these systems, the sky localization is so poor that it is hardly of any use when in search for a potential EM counterpart.

So far, we have discussed median values for the uncertainties on the parameters, but the full distribution is also of importance to interpret our results. Since we fix only the total mass and redshift, we expect that part of the uncertainties is inherited by the spread of the additional parameters affecting the binary signal. We defer to Appendix~\ref{sec:app_to_discuss_random_params} for an in-depth exploration of how each binary parameter shapes the 68 and 95 percentile distribution, while we here only summarize the overall trends and the main findings. As in previous studies, we find that the sky position estimate depends strongly on the true source position in the sky. Even if orbital modulations help to reduce the source position errors at earlier time, it becomes unimportant  at the end of the inspiral. However, most of the $S/N$ is accumulated close to merger and, therefore, the final $S/N$ will be small if the binary is located in a low sensitivity region. 
 
While the $S/N$ distributions are similar at all times and for all masses, higher mass systems display broader sky location distributions. Light (heavy) systems distributions extend over 2.5 ($\simeq 4$) order of magnitude at 1 hour before merger. Moreover far from coalescence the source position distributions show similar widths of $\simeq 1$ order of magnitude, almost independently from the total mass of the system. This is due to the fact that, typically, at that time the system has low $S/N$ so the actual true location of the source has a mild impact. 
Heavy systems show also larger uncertainties in the luminosity distance close to merger than light and intermediate ones.

In Fig.~\ref{fig:triple-sky-loc-distro} we show the 95 percentile distributions at each time for the three cases: light, intermediate and heavy binaries. The $S/N$ distributions look symmetrically distributed around the median at all times and for all systems. 
Also the sky position uncertainty distributions look similar at 1 month for all the three cases. When the binary approaches coalescence the median sky-position uncertainty decreases but the distribution widens. Light system distributions remain similar over all the inspiral with no major shape changes, while heavy mass system distributions flatten and skew towards lower uncertainties. 
For the intermediate and heavy systems the position uncertainty distributions are uniform at 1 day and 10 hours from coalescence. However these distributions are skewed to lower values at 5 hours and a 1 hour before merger with stronger effects for heavy mass systems.
We find that systems with lower values of the sky location uncertainties are those with low mass ratio and high spin magnitude of the primary BH due to the inclusion of higher harmonics and spin precession effects.
The wide spread for heavy systems is due to fast accumulation of most of their $S/N$ very close to the merger where LISA can be seen as static. The error in extrinsic parameters is particularly sensitive to the source position and orientation leading to a large spread.
Luminosity distance distributions are similar for all the three cases. While they are symmetric far from coalescence, higher mass systems distributions are skewed towards lower luminosity distance uncertainties at late times.

\subsubsection{Chirp mass and mass ratio}

In Fig.~\ref{fig:all-params} we show the chirp mass and mass ratio relative uncertainties for the three cases as function of time to coalescence. For light systems, the chirp mass is determined to few percent accuracy already 1 month from coalesce and it can be constrained to $\simeq 0.2\%$ at the end of the inspiral. This is due to the fact that chirp mass is inferred by phasing the signal during the inspiral, and lighter systems spend more time and wave cycles in band. Moving to higher mass systems, the median uncertainties shift to higher values without, however, a significant loss in LISA ability to constrain the chirp mass, especially at late times. The chirp mass is determined at $\simeq 30 \%$ ($\simeq 60\%$) precision at 1 month for intermediate and heavy systems and to less than $1\%$ at 1 hour from coalescence for both cases.

The mass ratio is constrained to $10 \%$ precision already 1 month from coalescence and $\simeq 0.5 \%$ 1 hour from merger for light systems. The mass ratio is basically undetermined 1 month from coalescence for intermediate and heavy systems, and only at the end of the inspiral the uncertainties reduce to $\simeq 1\%$ level for both cases.

Fig.~\ref{fig:all-params-distro} shows the $95\%$ percentile distributions for the chirp mass and mass ratio uncertainties. Overall, these distributions are narrower around the median compared to those describing the sky position. The chirp mass distributions for light binaries is symmetric around the median values at all times. However intermediate and heavy systems chirp mass \uncer distributions are  uniform inside the $95\%$ interval at early times and skewed close to merger. We check again that the systems contributing to smaller uncertainties are the ones with small mass ratio ($ q < 0.4$).
Similar considerations apply also to the mass ratio distributions, which are symmetric for the light systems, and skewed for intermediate and heavy ones.

\subsubsection{MCMC results}
\begin{figure}
    \centering
    \includegraphics[width=0.5\textwidth]{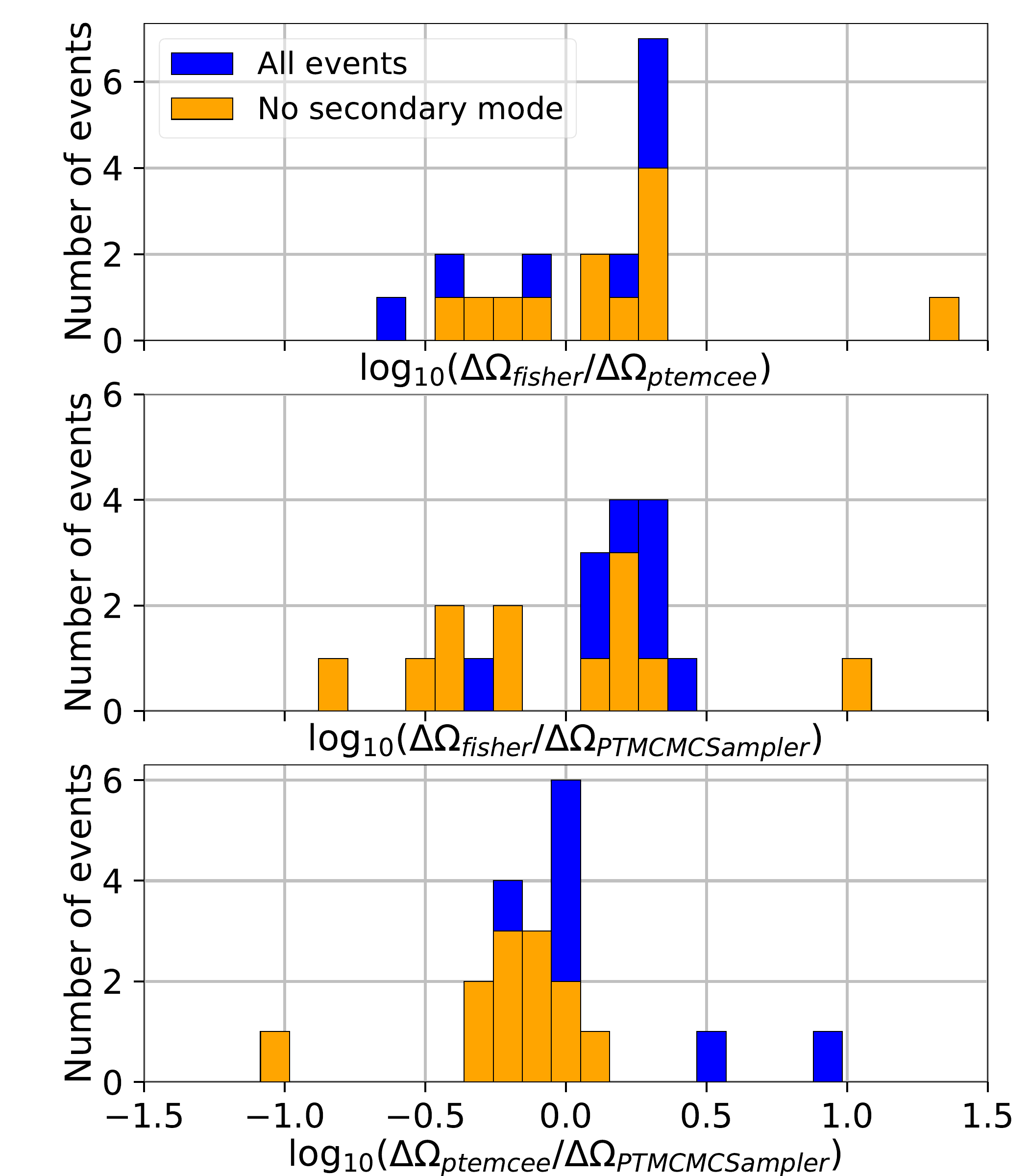}
    \caption{Sky localization log uncertainty ratio distribution obtained comparing different methods.
    Blue histograms refers to the whole set of 20 points\footnote{For one system, the sky position posterior distribution from \texttt{ptemcee} has not enough samples to make proper estimation, therefore blue histograms in the upper and lower panels have 19 entries.}, while orange histograms refer only to cases without secondary mode in the sky-position.
    Upper panel: \texttt{fisher} and \texttt{ptemcee}. Middle panel: \texttt{fisher} and \texttt{PTMCMCSampler}. Lower panel: \texttt{ptemcee} and \texttt{PTMCMCSampler}.}
    \label{fig:mcmc_comparison}
\end{figure}

In Fig.~\ref{fig:mcmc_comparison} we report the distributions for the log of the ratios between the sky position \uncer computed following our primary approach (\texttt{fisher}) and the results for the two MCMC simulations (\texttt{ptemcee}  and \texttt{PTMCMCSampler}). 
The distribution of the ratio between \fisher and \ptmcmc display two sub-populations with a mean value of $\simeq 0.11$. We find that the systems for which the \fisher approach produces better sky-position \uncer than \ptmcmc show small mass ratio, i.e. $q < 0.3$.
Similar considerations hold also for the comparison between \fisher and \mcmc areas, with a mean value of $0.05$.
We also compare the two MCMC estimates to check if the differences between the two implementations of the same technique are compatible with the one coming from \texttt{fisher}. In this case the \ptmcmc produces typically smaller areas than the \mcmc approach with a mean value of $\simeq -0.06$, close to the value coming from the comparison between \fisher and \texttt{PTMCMCSampler}. Use of global parameter exploration with \ptmcmc results in finding the secondary mode for some systems. 

The secondary modes correspond to the antipodal or reflected points, depending on the actual system considered, of the real binary position in the sky. Including the whole signal with the high frequency response of the detector and higher harmonics break degeneracies during the parameter estimation process \cite{Marsat:2020rtl} and disfavor these secondary modes
. In most of these cases,
the secondary mode has lower statistical significance.
Removing from the analysis the points with secondary mode in the sky position has no strong impact in the ratio distributions.

For each of the aforementioned 20 cases we perform an additional check comparing median sky position \uncer from two independent Fisher matrix codes, the one adopted for this study and another one from Marsat \emph{et al.} \cite{Marsat:2020rtl}. We keep fixed the total mass of the system, mass ratio, redshift and spin magnitude while we randomize over sky position, polarization, inclination, time to coalescence and initial phase.
For each case we perform $ \rm N = 10^3$ realizations. In Tab. \ref{tab:table_fisher_comparison} we report the mean value of the ratio between the median sky position \uncer obtained from the code adopted in this study and the alternative one, $\mathcal{R}^{\Delta \Omega}$, at different times from merger and when the full signal is included (`merger').
From 1 month and up to 3 days to merger, the code used for this study recover median sky position \uncer $\simeq 3$ times larger, while the opposite happen for times close to merger. Overall the agreement between the two results is quite good, especially at 1 day from merger.

\begin{table}[b]
\caption{\label{tab:table_fisher_comparison} Mean value of the ratio between the sky position \uncer computed with the same Fisher adopted in this study and the one from Marsat \emph{et al.} \cite{Marsat:2020rtl} at different times from merger and when the full signal is included (`merger').}
\begin{ruledtabular}
\begin{tabular}{cc}
\hspace{0.5cm} \textrm{Time}& 
\textrm{$\mathcal{R}^{\Delta \Omega}$}  \vspace{0.1cm} \hspace{0.5cm} \\
\colrule
1 month & 3.15 \hspace{0.8cm} \\
1 week & 2.05 \hspace{0.8cm} \\
3 days & 1.54 \hspace{0.8cm} \\
1 day & 0.84 \hspace{0.8cm} \\
10 hours & 0.47 \hspace{0.8cm} \\
5 hours & 0.30 \hspace{0.8cm} \\
1 hour & 0.87 \hspace{0.8cm} \\
merger & 0.35 \hspace{0.8cm} \\
\end{tabular}
\end{ruledtabular}
\end{table}

\subsection{\label{sec:analytical_pe}Analytical fits to parameter estimation uncertainties}

\begin{figure*}
    \centering
    \includegraphics[width=\textwidth]{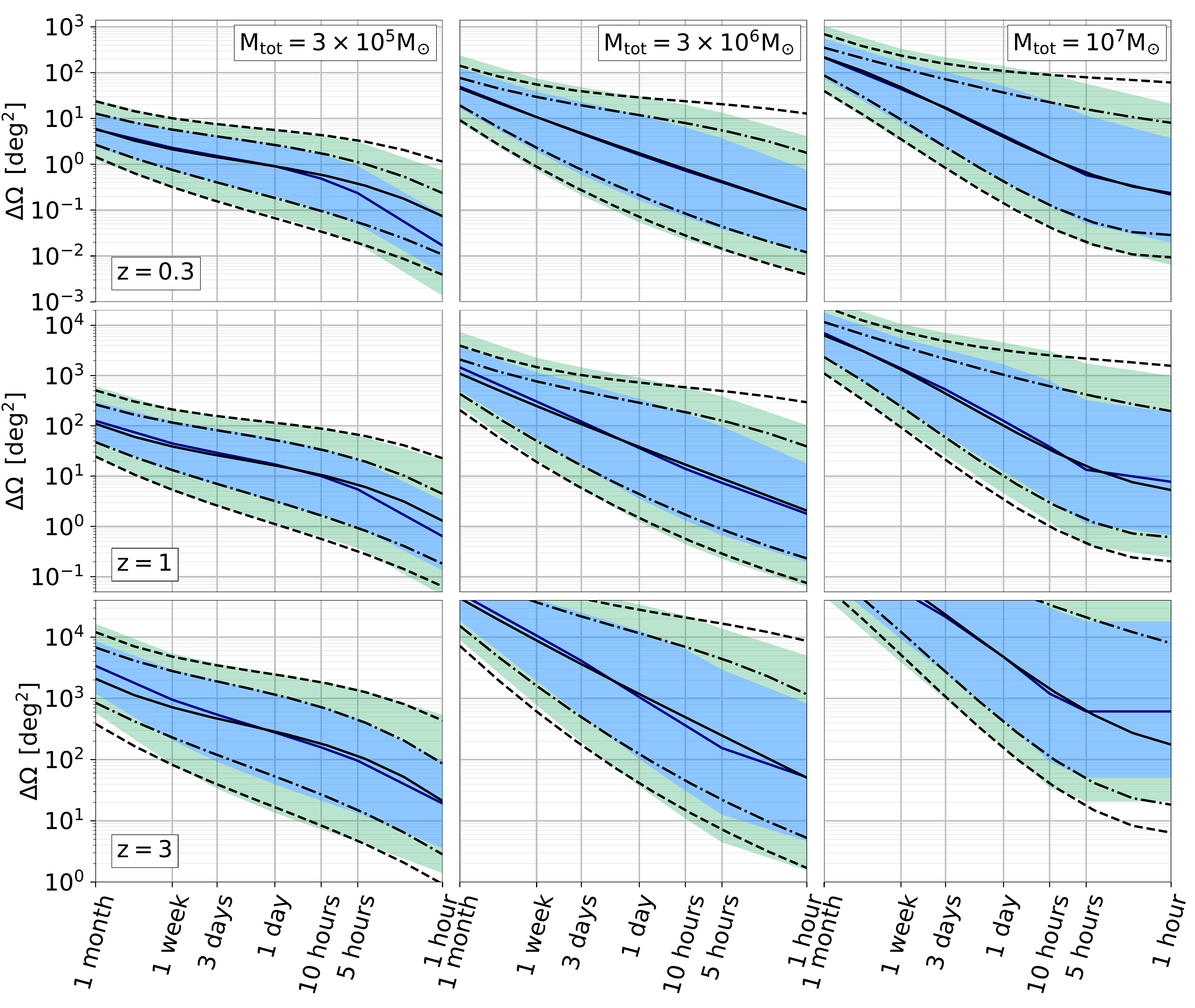}
    \caption{Time evolution of sky position uncertainties obtained with the Fisher Matrix approach compared to that recovered employing fits from Eq. \ref{eq:fit_expression} - \ref{eq:coeff_inspiral}. Blue lines correspond to the median of distribution while blue and green areas to the 68 and 95 percentiles. Solid, dashed-dotted, and dashed lines correspond to the fit outcome for the median, $68\%$ and $95\%$ regions, respectively. 
    Each column refers to a different source-frame MBHB total mass as labeled. Upper panels: MBHB at $z = 0.3$; middle panels: $z = 1$; lower panels: $z = 3$.}
    \label{fig:test_fit}
\end{figure*}
\begin{figure*}
    \centering
    \includegraphics[width=\textwidth]{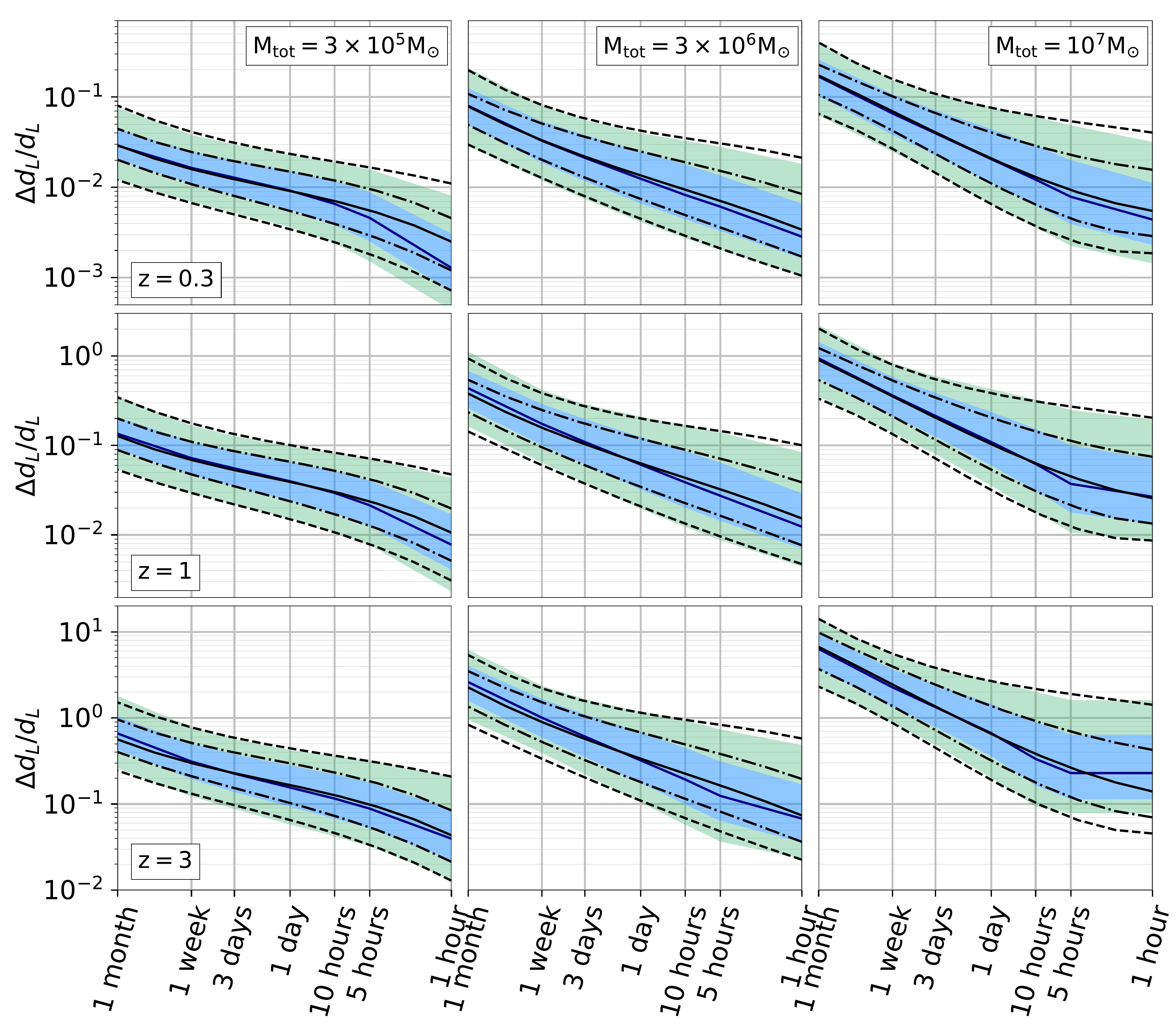}
    \caption{Same as Fig.~\ref{fig:test_fit}, but for the luminosity distance relative uncertainties.}
    \label{fig:test_fit_dl}
\end{figure*}
\begin{figure*}
    \centering
    \includegraphics[width=\textwidth]{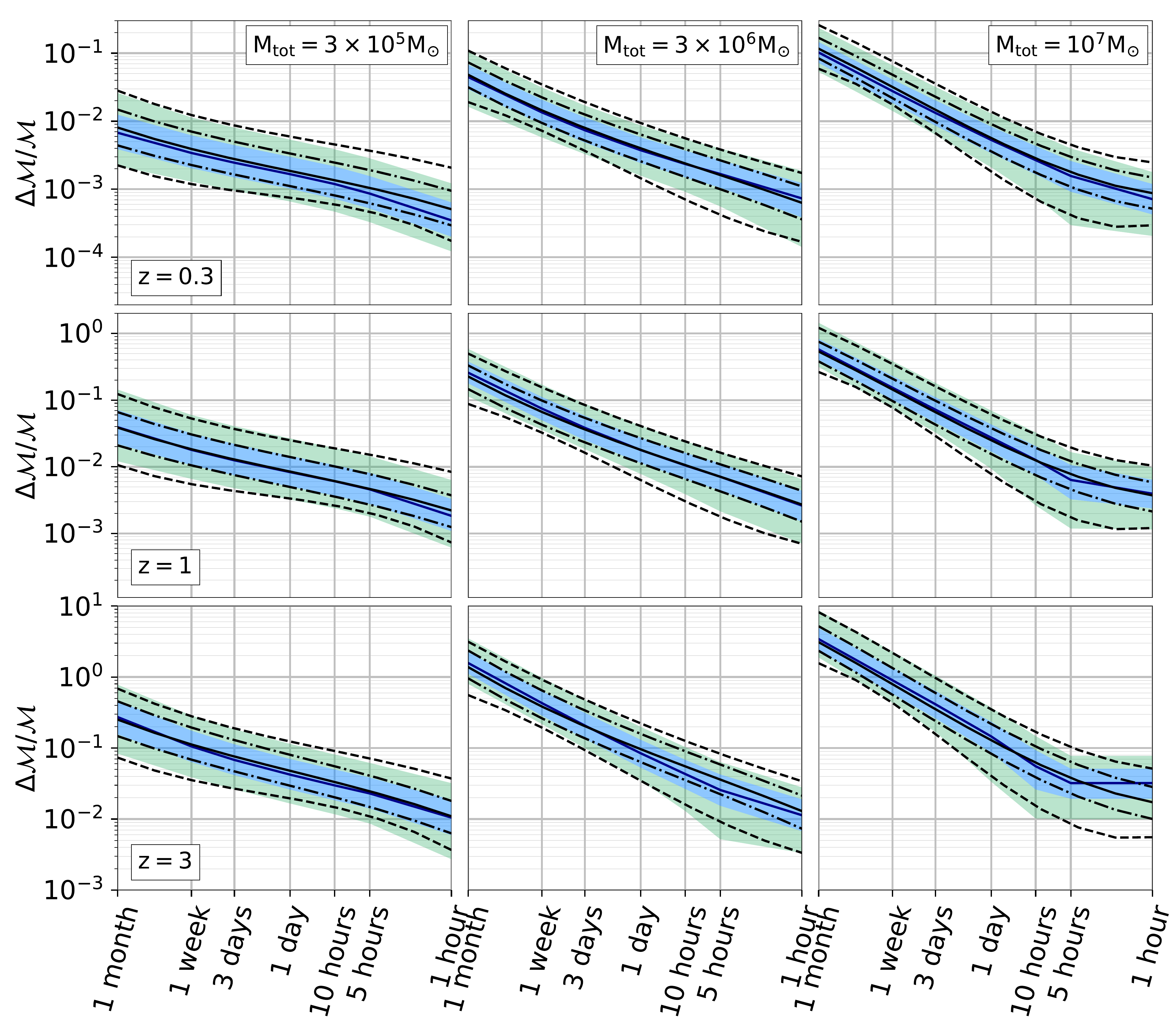}
    \caption{Same as Fig.~\ref{fig:test_fit}, but for the chirp mass relative  uncertainties.}
    \label{fig:test_fit_mchirp}
\end{figure*}
\begin{figure*}
    \centering
    \includegraphics[width=\textwidth]{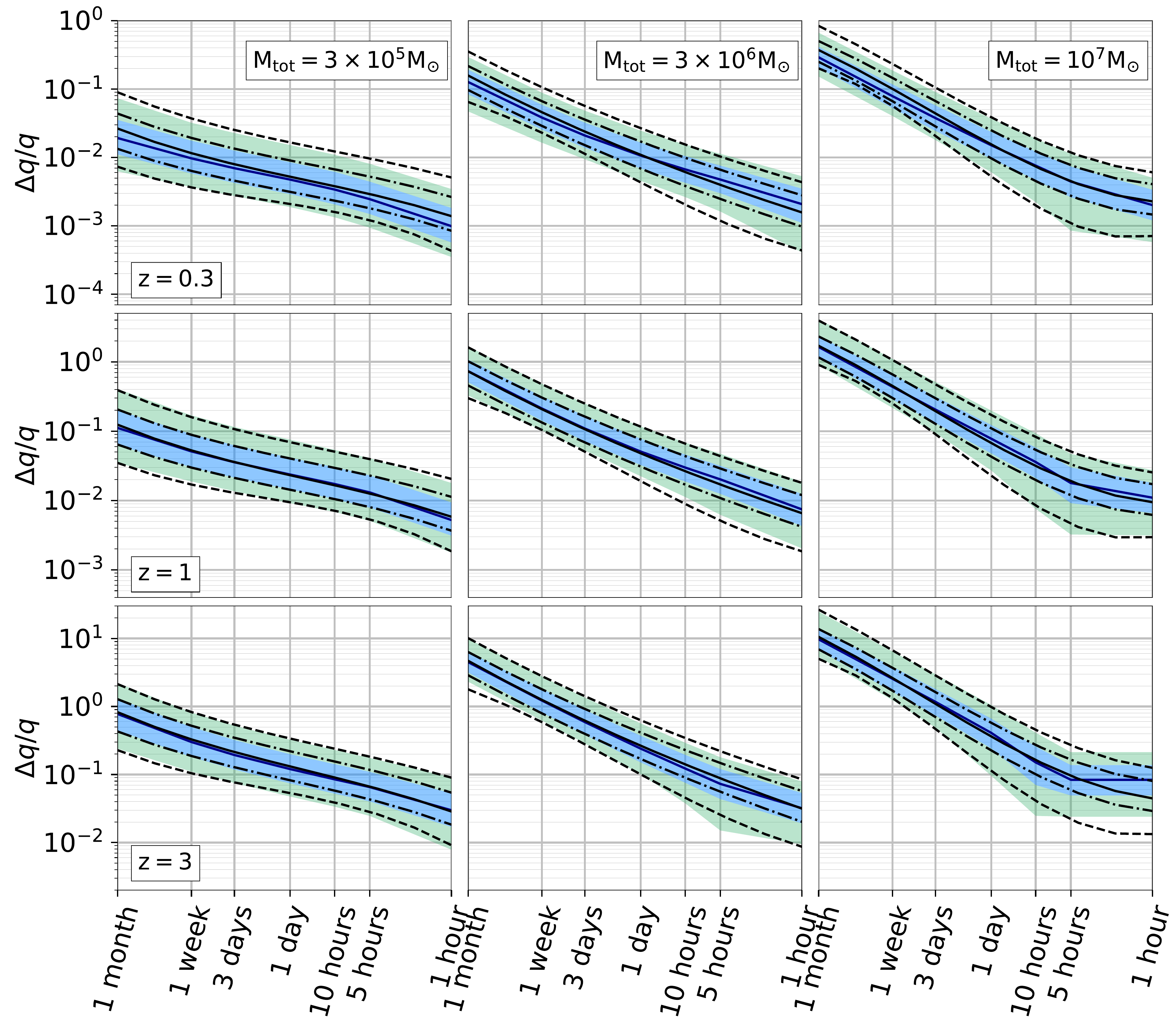}
    \caption{Same as Fig.~\ref{fig:test_fit}, but for the mass ratio relative uncertainties.}
    \label{fig:test_fit_mass-ratio}
\end{figure*}
\begin{figure*}
    \centering
    \includegraphics[width=\textwidth]{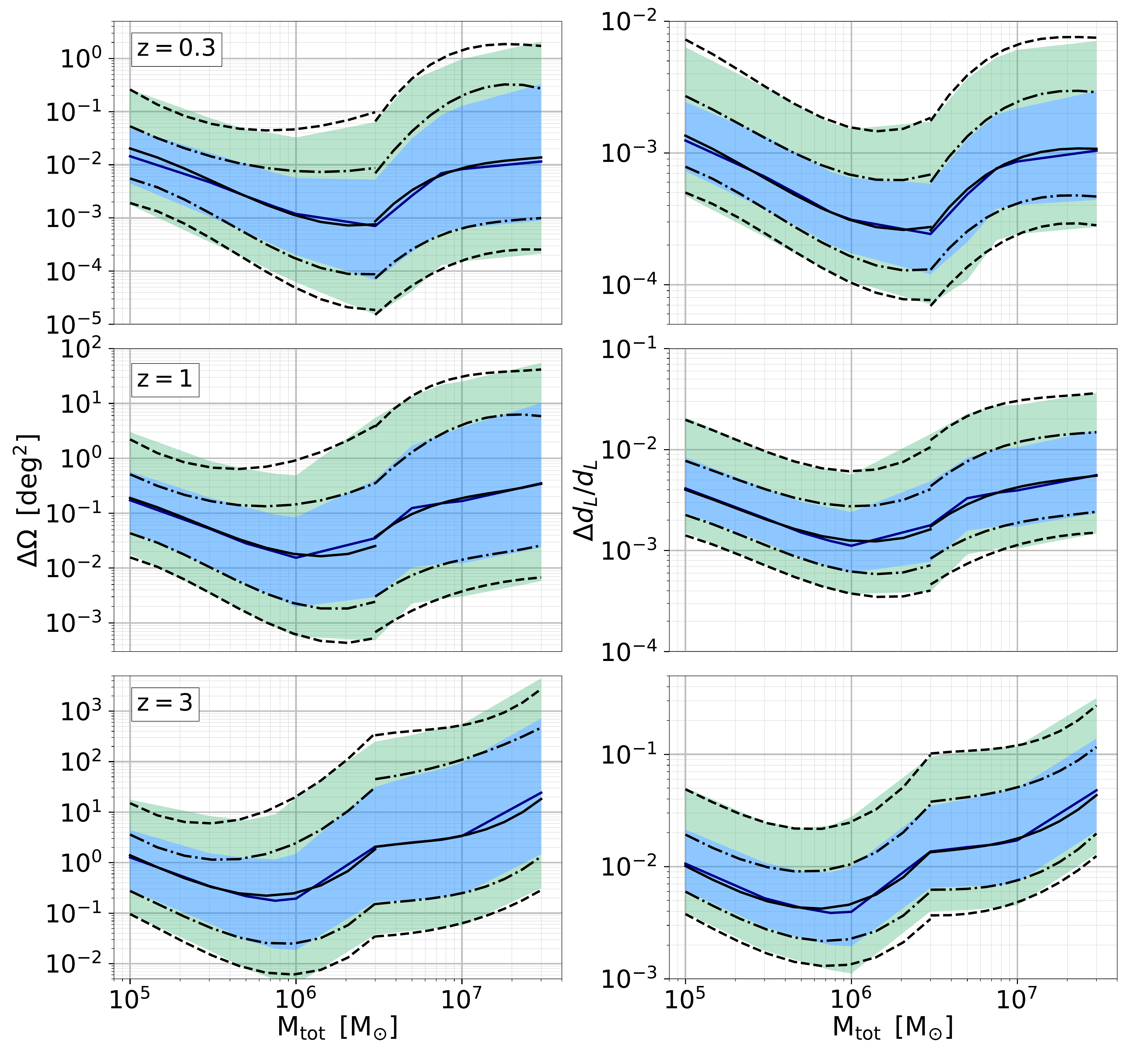}
    \caption{Sky localization (left column) and luminosity distance (right column) uncertainties at merger obtained with the Fisher Matrix approach compared to fits from Eq. \ref{eq:fit_expression_merger} - \ref{eq:fit_expression_merger_coeff} as function of the total mass of the binary. Color code and line style as in Fig.~\ref{fig:test_fit}. Upper panels: $z = 0.3$.  Middle panels: $z = 1$. Lower panels: $z = 3$.}
    \label{fig:test_fit_merger}
\end{figure*}

Having discussed the trends of the parameter estimation and precision for our selected sources, we now turn to the main aim of this study: provide {\it ready-to-use fitting formulas} that 
allow to infer on the fly the properties of an  emerging LISA source (sky localization, distance determination, chirp mass and mass ratio in particular).
For sky position and luminosity distance we provide additional formulas when the full signal is considered to inform astronomers  of a potential counterpart for targeted and instructed searches.

Since we propose to provide multidimensional formulas to model error uncertainties as the signal accumulates in band, we need to find a reasonable balance between the number of input parameters required and the accuracy of the proposed formulas. In other words we desire a formula that takes as input only a few key parameters and matches reasonably well  simulation results. 
The (observed) time left before binary coalescence has to be one of the input parameters. We also need the total (intrinsic) mass of the binary and the redshift of the source, since they will both affect the relative strain of the signal in the detector. Although the amplitude of the GW signal scales with the system chirp mass \cite{Maggiore:1900zz}, we choose to adopt the total mass of the system as input parameter since it is the quantity normally used within the astrophysical community.

In principle, the parameter estimation depends on all the other parameters described at the beginning of Section~\ref{sec:Theory}, i.e mass ratio, sky position, time to coalescence, spins, inclination, polarization and initial phase. Including all of them would dramatically increase the complexity of the fitting functions.

Averaging over some of the aforementioned parameters is also justified by what is expected in a real situation. MBHBs will be isotropically distributed in the sky, with uniform randomly distributed inclination and polarization. The situation is less clear for spin magnitude and directions. MBHs gain their mass mostly from accretion, which affects their spin in different ways depending on the coherency of the accretion flow \citep{Bardeen:1975zz,King2008,Berti2008,Perego2009,Dotti13}. Mutual spin orientations in MBHBs are further determined by their close environment; 'dry' mergers (where the binary evolution is primarily driven by stars) generally result in random spin orientations \cite{2007ApJ...661L.147B}, whereas 'wet' mergers (where the evolution is driven by gas) promote spin magnitude growth and spin-orbit alignment \citep{Dubois2014,Sesana2014,Fiacconi2018}.

In Appendix~\ref{sec:app_to_discuss_random_params} we explore in more detail how these parameters affect the sky position, luminosity distance, chirp mass and mass ratio uncertainties. However, we briefly note here that none of them produces clear trends, with the exception of the true sky position of the source (we refer to Appendix~\ref{sec:app_to_discuss_random_params} for further discussions) and therefore we chose to average over all of them and keep only three input parameters for our formulas, i.e. the total (intrinsic) mass of the system, its redshift and the (observed) time to coalescence.

Given total mass $\mtot$ at redshift $z$, the evolution of parameter uncertainty $\Delta{X}$  from 1 month to 1 hour before merger is described as  $\log_{10} \Delta{X}= \mathcal{F}(\log_{10} \tc, \log_{10} \mtot, z)$, where $\mathcal{F}$ is a third degree polynomial expression of $\log_{10} \tc$ and $\log_{10} \mtot$ with a supplementary dependence on redshift. In practice we use the form

\begin{equation}
\begin{split}
    \log_{10} {\Delta X} &=  c_1 + c_2 \, y  +c_3 \, y^2  + c_4 \, y^3  \\
    & + c_5 \, x  +c_6 \, x \, y  +c_7(z, \,  y) \, x\, y^2 \\
    & +c_8 \,x \, y^3  +c_9 \, x^2  +c_{10} \, x^2\, y \\
    & +c_{11} \,x^2 \, y^2  +c_{12} \,  x^2 \, y^3  +c_{13} \, x^3 \\
    & +c_{14} \, x^3\, y  +c_{15} \, x^3\, y^2 \\
    & +c_{16} \, x^3\, y^3 + z_c \, \left( \frac{z-0.5}{0.25 \, z + 0.25} \right ),
    \label{eq:fit_expression}
\end{split}
\end{equation}
where $x =\log_{10}( \tc / \rm sec)$, $y = \log_{10} (\mtot \rm / M_{\odot} ) $ and $z$ is the source redshift. $c_1, \dots, \, c_{16}$ and $z_c$ are numerical coefficients whereas $c_7(z, \, y)$ is a function of mass and redshift only, dependent on more four coefficients
\begin{equation}
\label{eq:coeff_inspiral}
\begin{split}
    c_7(z, \, y) & = d_1 + d_2 \, y \\
    & d_3 \, z + d_4 \, z \, y.
\end{split}
\end{equation}
Here we express $c_7$ as function of redshift and total mass however this is not true in general. For each case, we choose which parameter should be a combination of the previous input parameters by looking at the one that could match better the simulations.
$\mathcal{F}$ is therefore a function of 21 numerical coefficients. 

For the four key parameters discussed in the previous section -- $\Delta X= \Delta \Omega /{\rm \footnotesize sr}$, $ \rm \Delta d_L/d_L$, $\Delta \mathcal{M}/\mathcal{M}$, $\Delta q/q$ -- we adopt Eq. \eqref{eq:fit_expression} to fit the median, $68\%$ and $95\%$ confidence interval of the uncertainties in the LISA measurements. 
We report the value of the coefficients for the median, $68\%$ and $95\%$ confidence regions for the uncertainties on the sky localization and luminosity distance in Tab.~\ref{tab:coeff_tab_skyloc_dl}. The value for the chirp mass and mass ratio fits are reported in Tab.~\ref{tab:coeff_tab_mchirp_q}. In these table the coefficients computed according to Eq. \ref{eq:coeff_inspiral} are labeled with `$[z, \,  \log_{10} \mtot]$'.

For the degraded sensitivity curve we include two additional coefficients in Eq. \ref{eq:coeff_inspiral}, i.e. 
\begin{equation}
\label{eq:coeff_inspiral_degr}
\begin{split}
    c_{7, \rm{degr}}(z, \, y) & = d_1 + d_2 \, y \\
    & d_3 \, z + d_4 \, z \, y  \\
    & d_5 \, z^2   + d_6 \, z^2 \, y
\end{split}
\end{equation}
and replace the term $(z-0.5)$ in Eq. \eqref{eq:fit_expression} with $(z-1)$ to better fit the simulation results. The additional coefficients are reported in the aforementioned GitHub repository. 

Even if we perform simulations of systems with $\mtot \in [10^5, \, 3 \times 10^7] \, \msun$ and $z \in [0.1,4]$, our formulas can be applied only on a slightly smaller subset: they are valid for systems with $\mtot \in [10^5, \, 10^7] \, \msun$ and $z \in [0.3,3]$ because we choose to focus on providing formulas that can be used for the expected majority of LISA sources.

In the following, we provide a visual comparison between the analytical fits and the results of the Monte Carlo simulations for binaries with
$\mtot =3\times10^5,\, 3\times10^6,\,10^7\msun$ (i.e. the `light', `intermediate' and `heavy' systems considered in the previous subsection) placed at $z = 0.3$, $z = 1$ and $z = 3$.

In Fig.~\ref{fig:test_fit} we show results for the sky localization. Overall there is a good agreement between our fits and the outcomes of our simulations. However the fits overestimate the sky position uncertainty for light systems at $z = 0.3$ 1 hour from merger by a factor 3-4. This is because the precision in the source position for low-redshift light systems has a steep improvement close to merger, where the accumulated $S/N$ leads to a shrinkage of the localization area of an order of magnitude on a short timescale (from 5 hours to 1 hour before merger). 
We lose some precision when fitting the 69 and 95 percentile upper limit for intermediate and heavy systems, especially at $ z = 0.3$.
If the system is massive and at high redshift the orbital timescale at the ISCO is comparable to 1 hour and, since we truncate our waveform  when the MBHB separation reaches $6\mtot$ as we are not able to explore shorter time intervals. For this reason the sky position uncertainties are flat from 5 hours to 1 hours for massive and distant sources.

In Fig.~\ref{fig:test_fit_dl} we perform the same comparison for the luminosity distance. Also for this parameter,  our fits reproduce well the simulation outcomes with the exception of light systems at $z = 0.3$ and heavy systems at $z = 3$ both at 1 hour from coalescence. However here the results for the median from the fit differ from the simulation values only by a factor of 2.

Finally, in Fig.~\ref{fig:test_fit_mchirp} and Fig.~\ref{fig:test_fit_mass-ratio} we compare our formulas for the chirp mass and mass ratio to the simulation results. Again the fits match quite well the Fisher outcomes with the same caveats of the sky position and luminosity distance. The larger differences between our fits and the simulations are usually a factor $2$.

We also provide fits to the sky position and luminosity distance \uncer for the full signal, computed from Eq.~\ref{eq:area_rescale}-\ref{eq:dl_rescale}. Since for this case we do not have the time dependence, we reduce the number of coefficient necessary for the fit.
As a consequence we are able to extend the validity of these formulas to the full parameter space explored, i.e. the formulas describing the sky position and luminosity distance \uncer when the full signal is considered are valid for $\mtot \in [10^5, \, 3 \times 10^7] \, \msun$ and $z \in [0.1,4]$.

In this case, however, the noise due to Galactic binaries becomes important. Since the merger-ringdown $S/N$ is accumulated over a relatively narrow range of frequencies, the impact of the galactic foreground is highly mass-dependent and to obtain an acceptable fit
we have to split the mass range in two sub-intervals, namely $[10^5, \, 3 \times 10^6 ] \, \msun$ and $[3 \times 10^6, \, 3 \times 10^7] \, \msun$. The formulas are the same for the two ranges but, clearly, the coefficients are different.
As for the inspiral, one of the coefficients is given by a nested function of redshift. Using again $y = \log_{10} (\mtot \rm / M_{\odot} ) $ and the redshift $z$, uncertainties can be expressed as 

\begin{equation}
\begin{split}
    \log_{10} \Delta X &=  \frac{z}{m_1 (z)} + \\
    & +m_2 \, z  +m_3 \, y  + m_4 \, z^2  \\
    & + m_5 \, y^2  +m_6 \, z^3  +m_7 \,  y^3 \\
    & +m_8 \,z \,y  +m_9 \, z^2 \, y  +m_{10} \, z \, y^2 
    \label{eq:fit_expression_merger}
\end{split}
\end{equation}
where 
\begin{equation}
\begin{split}
    m_1(z) & = n_1 + n_2 \, z \\
    & n_3 \, z^2 + n_4 \, z^3  + n_5 \, z^4 ,
    \label{eq:fit_expression_merger_coeff}
\end{split}
\end{equation}

\noindent
and $\Delta X= \Delta \Omega /{\rm \footnotesize sr}$, $\Delta d_L/d_L$.
Similar to what done for the fits for the inspiral part of the signal, we provide expressions also for the $68\%$ and $95\%$ intervals. 
We report the coefficient value for the sky position and luminosity distance for $\mtot \in [10^5, \, 3 \times 10^6 ] \, \msun$ in Tab. \ref{tab:coeff_tab_skyloc_dl_merger_below} and for $\mtot \in [3 \times 10^6, \, 10^7 ] \, \msun$ in Tab. \ref{tab:coeff_tab_skyloc_dl_merger_above}.

For the degraded sensitivity curve we keep the same expression without adding any further coefficients. Also these coefficients can be found on the GitHub repository.

In Fig.~\ref{fig:test_fit_merger} we compare the above fits to the full inspiral-merger-ringdown sky location and luminosity distance errors obtained for our set of simulations, at three different redshifts.
At $\mtot = 3 \times 10^6$ there is a small gap between the fits due to the fact that we focus on fitting the overall behavior in both sub-intervals rather than requiring the continuity of the equation at their point of contact.
Overall our formulas follow closely the distributions of sky location and luminosity distance \uncer.

\section{\label{sec:time_progress_pe}Time progression of parameter estimation}

\begin{figure}
    \centering
    \includegraphics[width=0.5\textwidth]{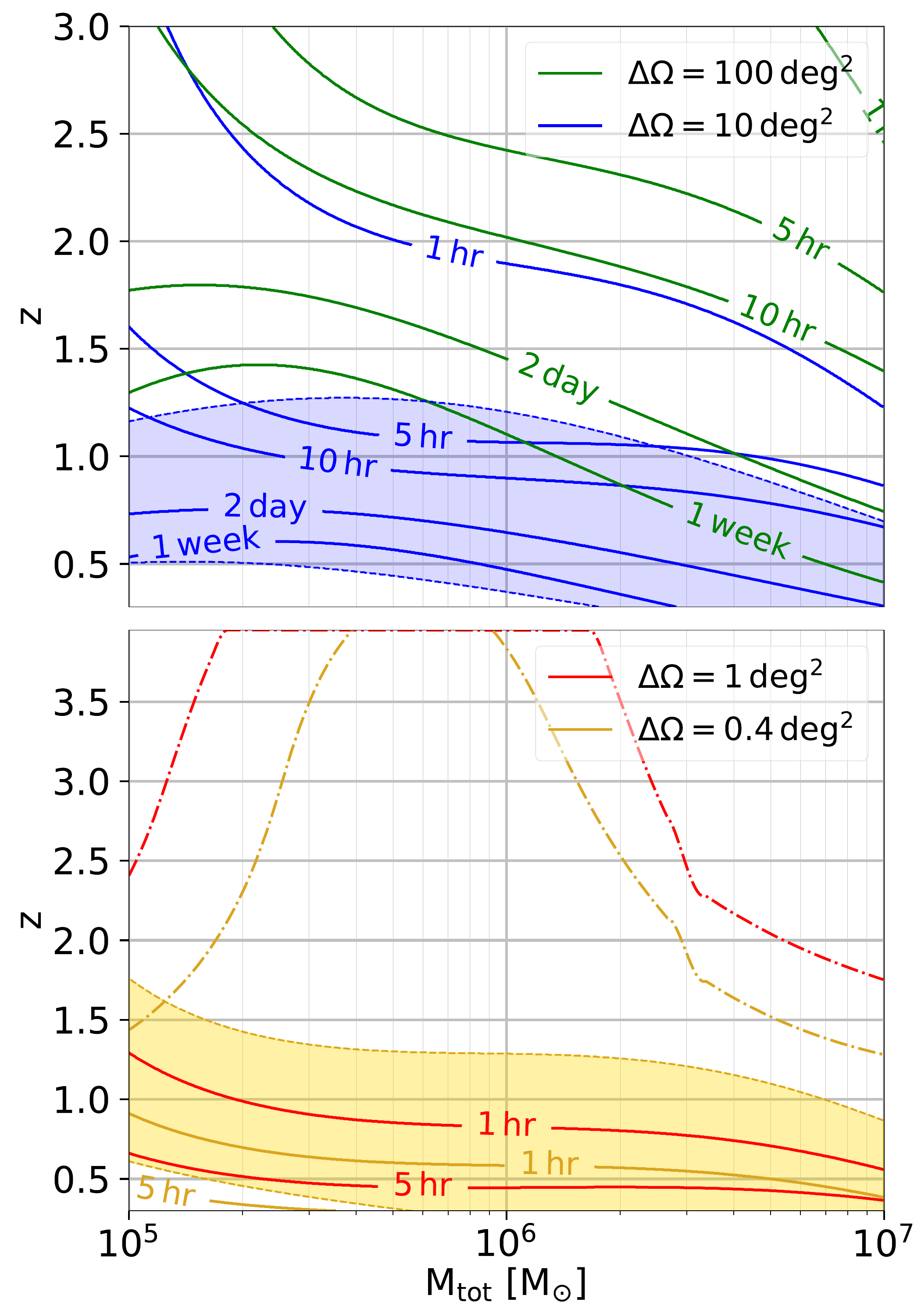}
    \caption{Contours of constant `remaining' time in the ($\mtot,z$) plane, for selected values of the median sky localization reached during the binary inspiral, as indicated in the legend. The shaded blue (yellow) area in the top (bottom) panel corresponds to the 68 percentile of the distribution around the median value of $\Delta \Omega=10 \degsq$ 2 days before merger ($\Delta \Omega=0.4 \degsq$ 1 hour before merger ). In the bottom panel the dash-dotted red (yellow) line gives the line of constant $\Delta \Omega$ equal to $1 \degsq$ ($0.4\degsq$) when we account for the full signal, i.e. including the 'merger'.}
    \label{fig:sky_loc_of_time}
\end{figure}
\begin{figure}
    \centering
    \includegraphics[width=0.5\textwidth]{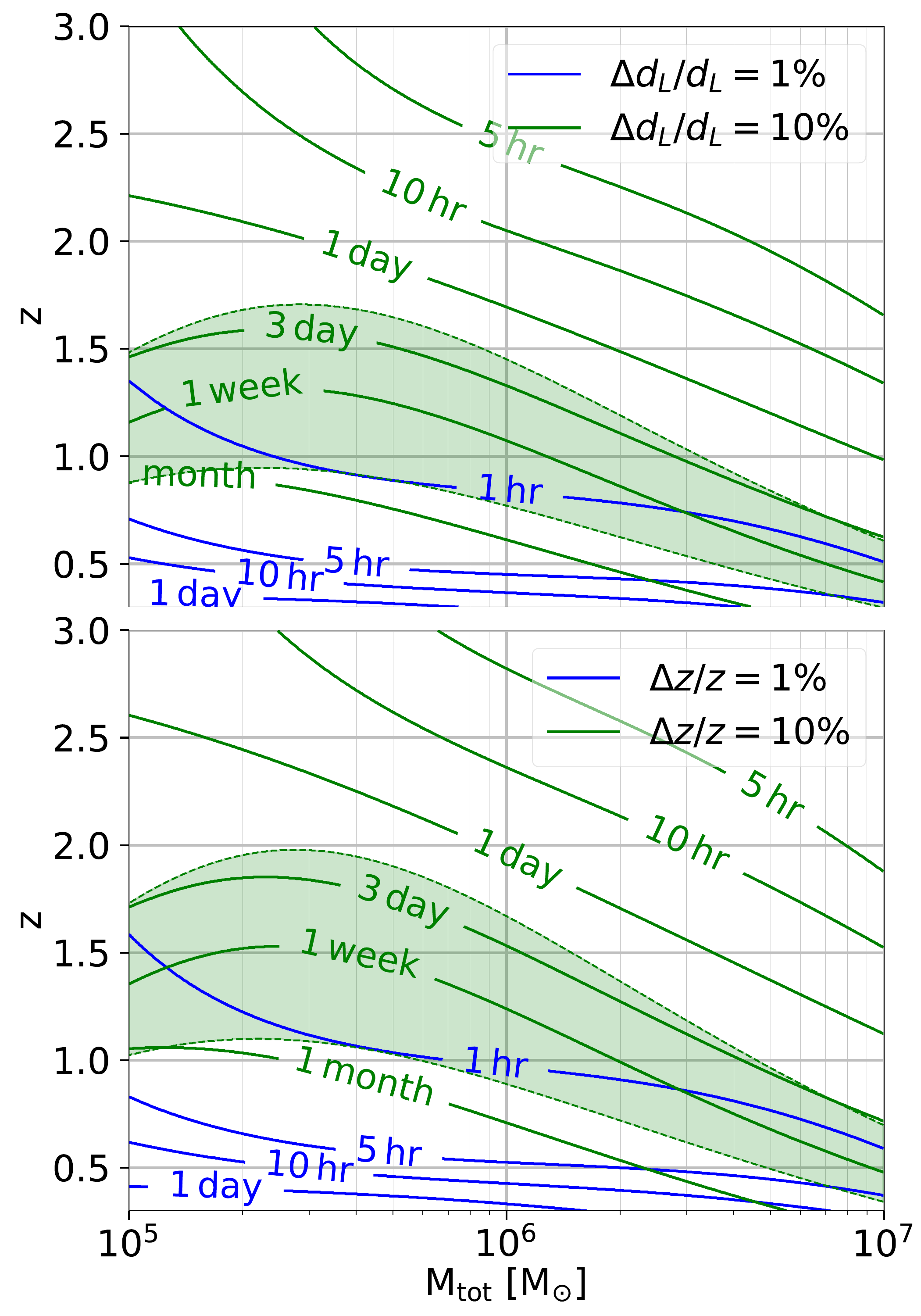}
    \caption{Contours of constant `remaining' time in the ($\mtot,z$) plane for two selected values of the precision on the luminosity distance (upper panel) and redshift (lower panel) estimates reached during the binary inspiral, as indicated in the legend. The green area corresponds to the 68 percentile of the relative uncertainty distribution on $d_L$ (upper panel) and $\Delta z/z$ (lower panel), calculated 1 week before coalescence, assuming $\Delta d_L/d_L=10\%$ and $\Delta z /z=10\% $, respectively.}
    \label{fig:dl_of_time}
\end{figure}

\begin{figure}
    \centering
    \includegraphics[width=0.5\textwidth]{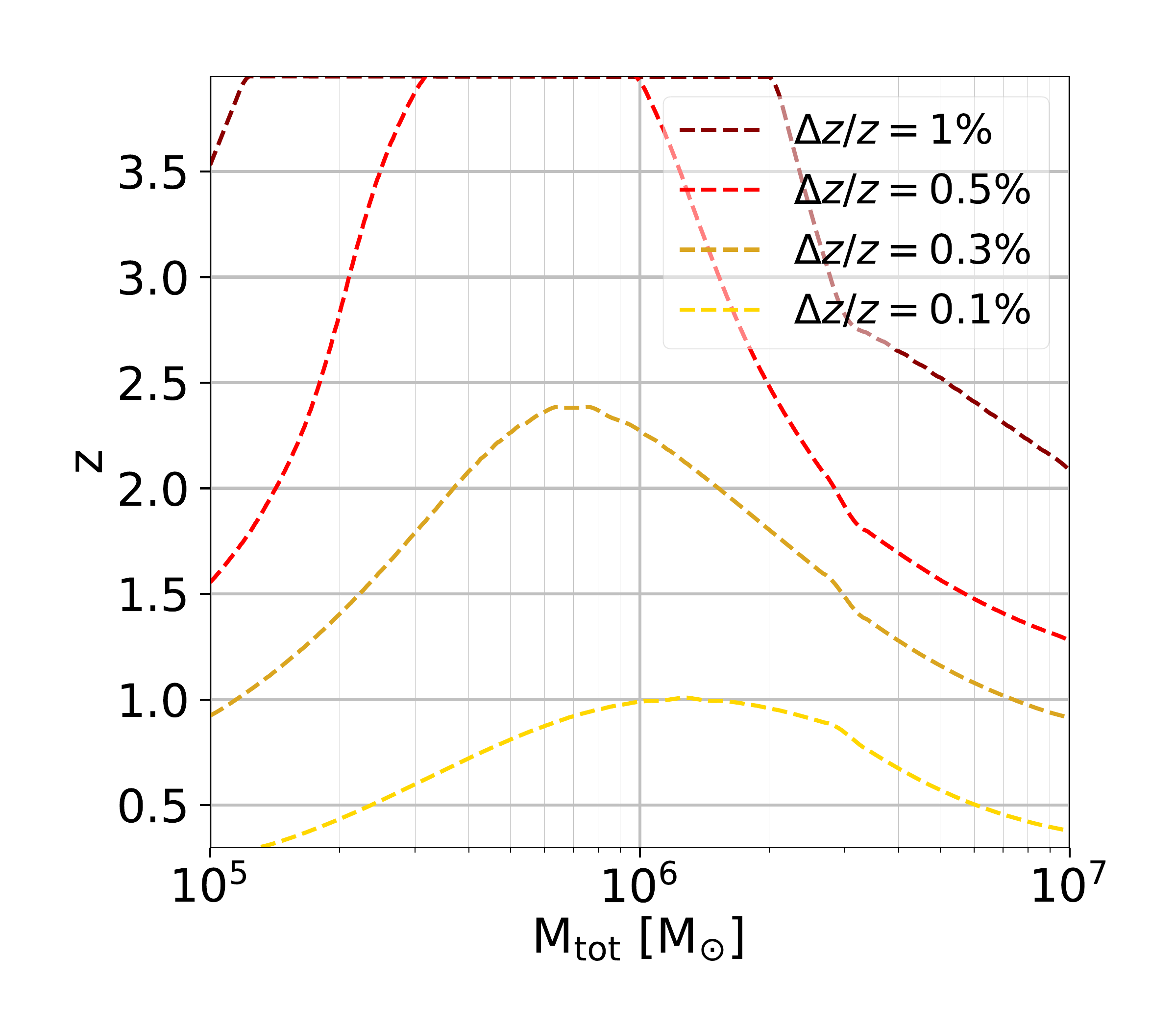}
    \caption{Lines of constant $\Delta z /z$ as labeled in the plot in the ($\mtot,z$) plane  when we account for the full signal, i.e. including the 'merger'.}
    \label{fig:dz_of_time_merger}
\end{figure}
\begin{figure*}
    \centering
    \begin{tabular}{cc}
    \includegraphics[width=0.5\textwidth]{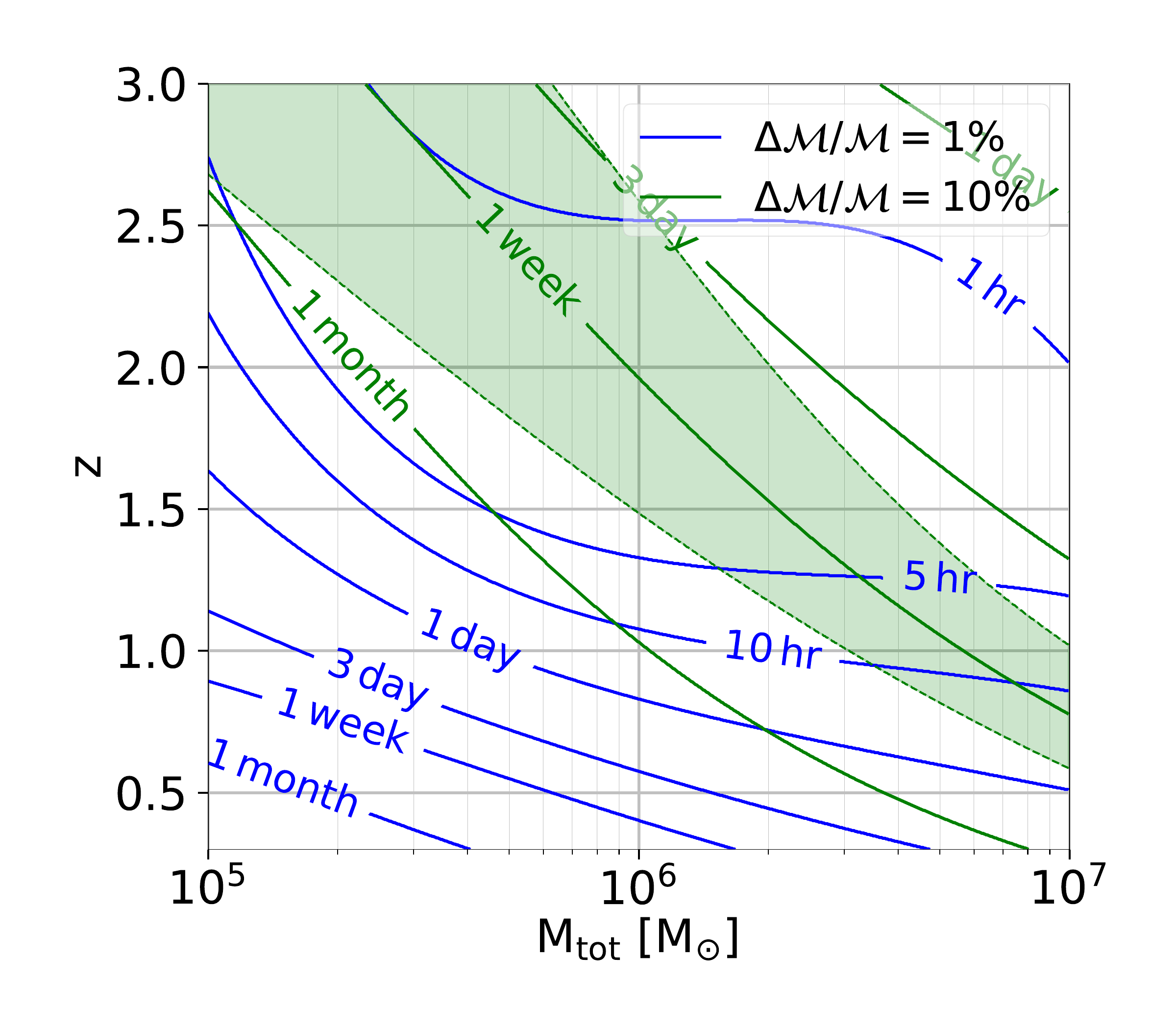}&
    \includegraphics[width=0.5\textwidth]{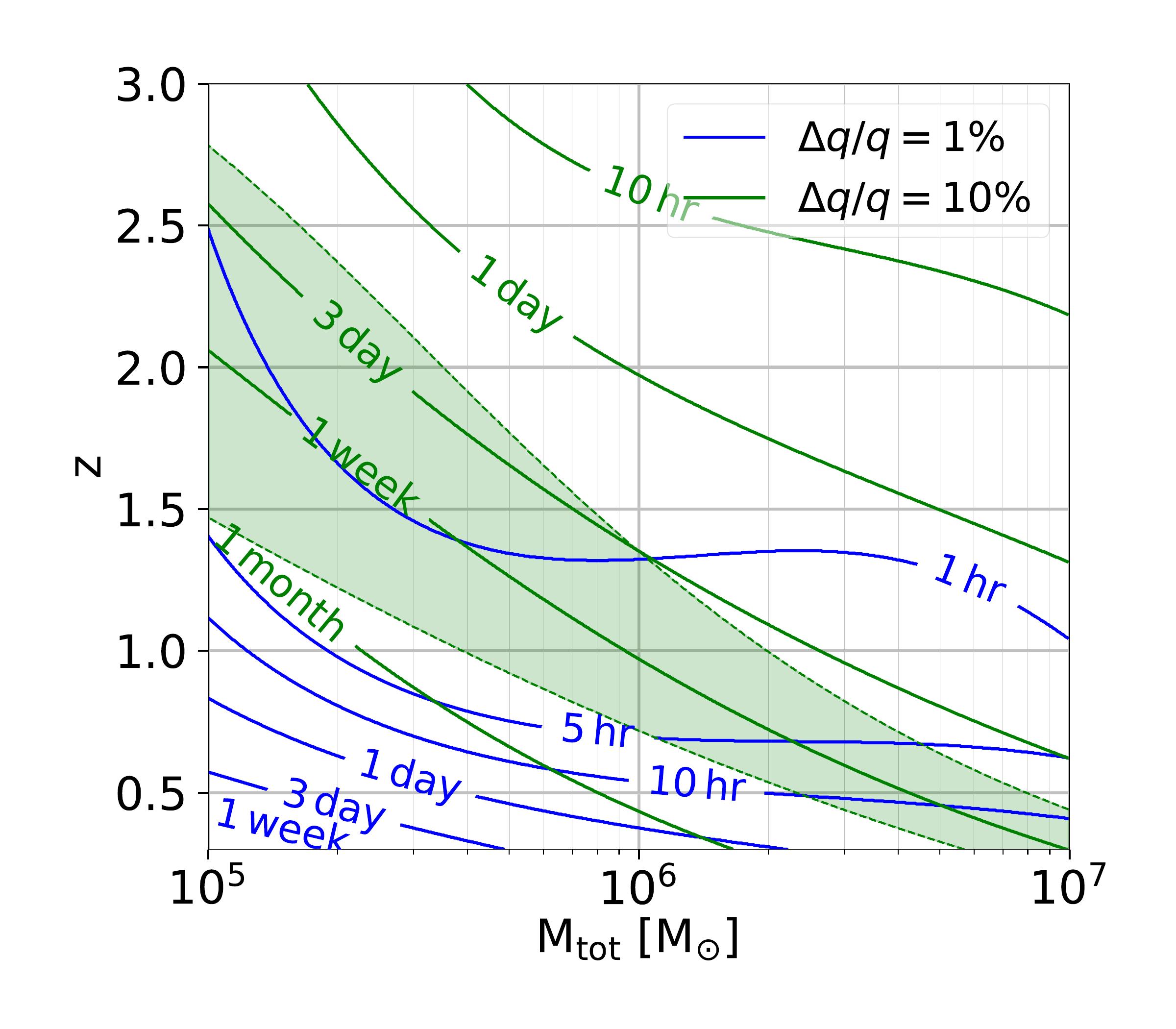}\\
    \end{tabular}    
    \caption{Contours of constant `remaining' time in the ($\mtot,z$) plane for selected values of the median relative error for chirp mass (left panel) and mass ratio (right panel) as labeled. Green and blue lines refer to a precision of $10\%$ and 1\%, respectively. The green area in each panel corresponds to the 68 percentile of the uncertainty distribution on ${\cal M}$ and $q$ evaluated 1 week before coalescence assuming a precision of $10\%$.}
    \label{fig:mchirp_of_time}
\end{figure*}

The analytical fits provided in the previous section can be used to generate `advanced warning' contour plots across the parameter space. In practice, one can fix a desired precision of a given parameter and ask how long prior to merger such precision is achieved across the mass-redshift parameter space. This is particularly useful, for example, to assess under which circumstances and for which sources a specific instrument is appropriate to search for a counterpart, or to pre-select {\it on the fly} candidates from existing galaxy catalogs in the expected mass and redshift ranges to concentrate follow-up efforts on the most promising candidates. We now show the contour maps of each parameter and discuss specific examples of their use in the next section.  
 
In the two panels of Fig.~\ref{fig:sky_loc_of_time} we plot the remaining time prior to coalescence for progressively smaller values of the sky position uncertainty. For example, a MBHB with $\mtot = 10^6 \msun$ at $z \simeq 0.6 $  ($z \simeq 1.5 $) can be localized within $\Delta \Omega= 10 \degsq$ ($100 \degsq$) 2 days before merger. However the source has to be at $z < 0.5$ ($z \simeq 1.1$) for a 1 week earlier alert. Note that in this and in the following figures, contours are produced using the fit to the median values of the parameter estimation. As shown in Fig.~\ref{fig:triple-sky-loc}, the sky location estimate is subjected to large uncertainties, which affects also the contour plots and in turn the redshift at which a source can be localized with a solid angle at a given time prior to coalescence. An example of this is given in the upper panel of Fig.~\ref{fig:sky_loc_of_time} and refers to localization of $10\degsq$, 2 days before merger. Depending on the specific parameters at the source, the redshift at which a MBHB with $\mtot = 10^6 \msun$ can be localized with a  $10 \degsq$ accuracy 2 days before merger ranges from $\simeq 0.3$ up to 1.2. 
The lower panel of Fig.~\ref{fig:sky_loc_of_time} shows that median localizations of $1 \degsq$ or better ($0.4 \degsq$) can only be achieved 5 hours (1 hour) before merger for MBHBs at $z\lesssim 0.5$.  Again depending on
the source parameters, the redshift at which a MBHB with $\mtot = 10^6 \msun$ can be localized within  $0.4\degsq$ 1 hour before merging ranges similarly from less than 0.3 up to about 1.3.

However measurements improve right after merger, when the full signal is considered in the parameter estimation.
This is shown by the dashed-dotted lines, which highlight how sources in the interval between a few $10^5\msun$ up to a few $10^6\msun$ can be localized to $\rm{sub-}\degsq$ precision beyond $z=3$. 

The results of the same type of analysis for the luminosity distance are presented in Fig.~\ref{fig:dl_of_time}. For a MBHB with $\mtot = 10^6 \msun,$ 10\% $d_L$ precision can be generally achieved one week (10 hours) before merger at redshift $z\approx 1$ ($z \approx 2$). Getting to the 1\% precision level, however, is much more challenging during the inspiral, and is generally possible only few hours before merger for systems at $z \lesssim 1$. Albeit to a lesser extent than sky localization, results still depend on the specific parameter of the source, with the uncertainty range marked by the green shade in Fig.~\ref{fig:dl_of_time}. For example, the luminosity distance of a particularly favorable MBHB with $\mtot = 10^6 \msun$ can be measured with 10\% precision 1 week before merger even if it is at $z\approx 1.5$. For an unfavorable system, the same performance is achieved only if it is located at $z\approx 0.8$.

Following Eq. 3.7 in \cite{PhysRevD.71.084025} and assuming a fixed cosmology ($\Delta H_0 = \Delta \Omega_{\Lambda} = 0$), we convert the error on luminosity distance in a redshift error and give the result in the lower panel of Fig.~\ref{fig:dl_of_time}.
 For a system with $\mtot = 10^6 \msun$ the redshift can be determined to $10\%$ precision 1 week before merger if the source lies
in $z \in [0.9, \, 1.6]$.  For the same binary $1\%$ accuracy is attained 1 hour (10 hours) prior to merging for $z\sim 1$ ($z\sim 0.4$). Redshift uncertainties clearly mirror those on the luminosity distance.

In Fig. \ref{fig:dz_of_time_merger} we report the $\Delta z/z$ \uncer when the full signal is considered. Systems with $\mtot \simeq  10^6 \msun$ are localized with a precision of $\Delta z/z \simeq 0.1 \, (0.3) \%$ up to $z \lesssim 1 \, (2.3)$.
Except for the most massive systems ($\mtot > 3\times 10^6 \msun$), the redshift of all MBHB mergers can be determined by LISA to a precision of $1\%$ up to $z \simeq 4$.

We stress again that the reported \uncer in the luminosity distance does not take into account the weak lensing error. The lensing limit our error $\Delta d_L/d_L \simeq 1 \%$ already at $z=1$. At $z = 0.5, \, 1, \,  2$ this translates to a $\Delta z/z \simeq 0.83\%, \, 0.81\%, \,0.82\% $ respectively

Finally, in Fig.~\ref{fig:mchirp_of_time} we plot the same time contour levels in progression, i.e. those times at which a parameter is determined at $1\%$ and $10\%$ precision for the source's intrinsic parameters, namely, chirp mass and mass ratio.
As expected, when increasing the total mass of the system or the source redshift, the amount of time left, when the required precision is reached, is reduced  for both parameters (note that this is not always the case for extrinsic parameters discussed before). Starting from the left panel, LISA could constrain the chirp mass at $1 \%$ ($10 \%$)  1 week from coalescence for a system with $\mtot = 10^6 \msun$ at $z \simeq 0.4$ ($z  \simeq 2$).

Overall, the chirp mass is basically determined with $10\%$ precision 1 day before merger for the whole parameter space considered in this study and even 1 month before merger for sources with $\mtot\lesssim 10^6\msun$ out to $z\lesssim 1$. 
However we stress that, even for the chirp mass determination, there are large uncertainties in the progression of times to coalescence. The chirp mass of systems with $\mtot = 10^7 \msun$ can be determined with $10\%$ precision 1 week prior merger if the redshift of the source is $z \in [0.6, 1]$, depending on the exact parameters. For a systems with $\mtot = 10^6 \msun$ the same is true but out to much larger redshifts, $z \in [1.5, 2.5]$.

In the right panel of Fig.~\ref{fig:mchirp_of_time} we see that LISA can constrain the mass ratio for a MBHB with $\mtot = 10^6 \msun$ with a precision of $1\%$ ($10\%$) at 10 hours before merger if the source is at $z \simeq 0.5$ ($z \simeq 2.5$). With the exception of more massive systems ($\mtot > 3 \times 10^6 \msun$) at relatively high redshift ($z>2.5$), the mass ratio should be measured with an accuracy of $10\%$ when there are still 10 hours left before the merger. Also in this case, we highlight that the reported contours are affected by large uncertainties: the mass ratio of a MBHB with $\mtot = 10^5 \msun$ ($\mtot = 10^6 \msun$) can be determined to $10\%$ precision 1 week before the merger if the source redshift is $\simeq  [1.5, 2.7]$ ($\simeq [0.7, 1.4] $).

\section{\label{sec:multimessenger_view}Multi-messenger view}
\begin{figure*}
    \centering
    \includegraphics[width = \textwidth]{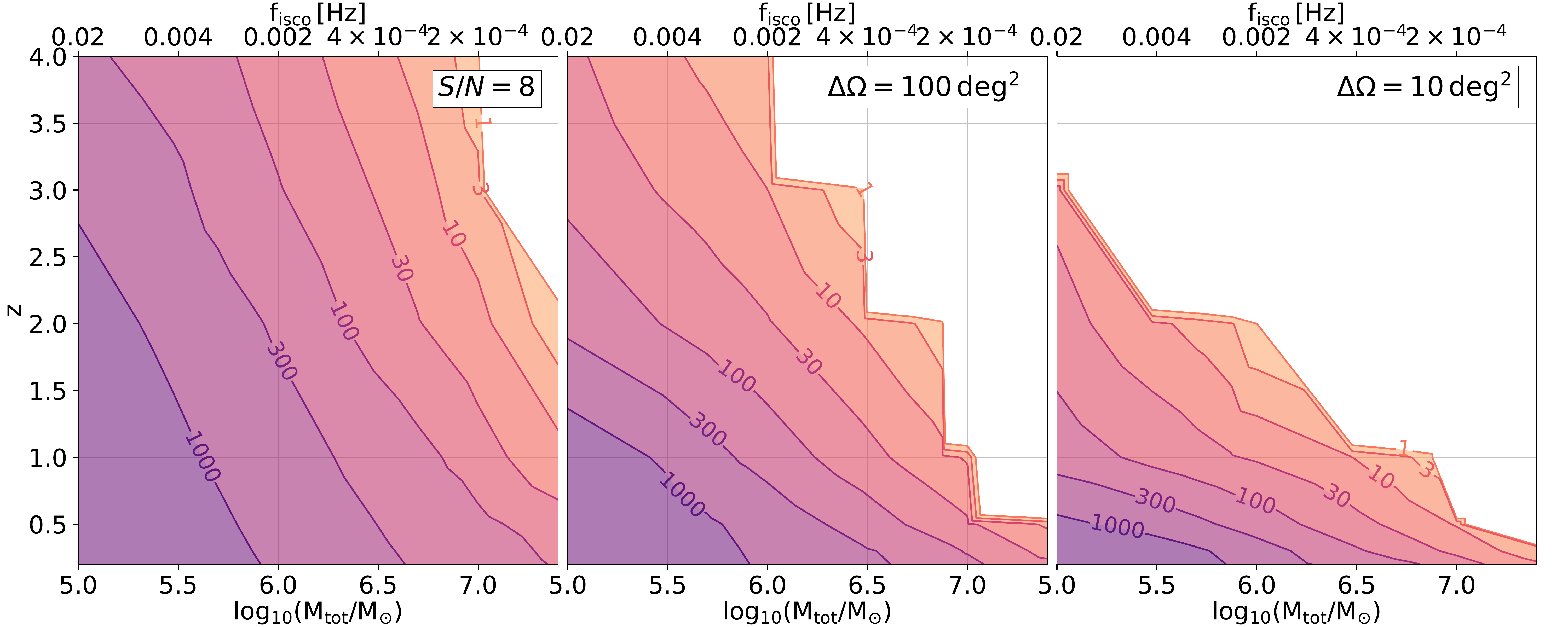}
    \caption{Contours of number of cycles spent by sources in the time interval ranging from the moment at which $S/N = 8$ (left panel), $\Delta \Omega = 100 \degsq$ (central panel) and $\Delta \Omega = 10 \degsq$ (right panel) up to ISCO as function of total mass and redshift. The values of $f_{\rm ISCO}$ are computed in the source-frame. }
    \label{fig:triple_contour_ncyc_of_time}
\end{figure*}

The analytic fits and results presented in the previous section are useful tools to explore potential synergies between LISA and EM telescopes, to select the best instrumentation to perform early warning searches and follow-ups, and to devise optimal survey strategies. Especially {\it on the fly}, the necessity of covering relatively large sky localization areas (generally of several $\degsq$) calls for wide field of view (FOV) instruments. Spanning the EM spectrum, primary candidates for fruitful synergies that are expected to be operational at the time LISA fly are the Square Kilometre Array (SKA, \cite{2009IEEEP..97.1482D}) in the radio, the Large Synoptic Survey Telescope (LSST, \cite{2009arXiv0912.0201L}) in optical and \emph{Athena} in X-ray \cite{2013arXiv1306.2307N}. We briefly discuss the potential of each of them in the following, deferring a more systematic investigation to future work.

Starting from the high frequencies, \emph{Athena} is an X-ray satellite selected as ESA L2 mission and due to fly at the same time as LISA.
\footnote{We defer the reader to 
https://www.cosmos.esa.int/documents/\\678316/1700384/Athena$\_$LISA$\_$Whitepaper$\_$Iss1.0.pdf for a preview on this theme.}
The WFI instrument on board has a FOV of $0.4 \degsq$, reaching a soft X-ray flux limit of $\approx 3\times 10^{-16}$erg cm$^{-2}$ s$^{-1}$ in 100 ks (about 1 day) of exposure \cite{2013arXiv1308.6785R}. This roughly corresponds to the received flux from a $10^6\msun$ MBH accreting at the Eddington limit at $z\approx 2$. As clear from Fig.~\ref{fig:sky_loc_of_time}, only systems at $z\lesssim0.5$ can be localized within $0.4 \degsq$ during the inspiral, and only a few hours before merger, which provides just enough time to repoint Athena in the source direction before the merger occurs. However, post-merger localization is generally better than the FOV of \emph{Athena} out to $z\gtrsim3$. Considering also the long exposure times required, \emph{Athena} is therefore optimal to search for post-merger signatures \cite{2020NatAs...4...26M} associated, e.g. to the emergence of a relativistic jet \cite{Palenzuela2010}. Particularly favorable are low redshift sources, which will allow a single field pointing of \emph{Athena} few hours before merger, enabling the detection of a putative X-ray flash at merger (e.g. \cite{2002ApJ...567L...9A}).

At the low frequency end, SKA will be surveying the radio sky. In its first operational stage, SKA1-MID is expected to have a FOV of $1 \degsq$, reaching a detection flux limit of $2\,\mu$Jy in 1 hour integration time\footnote{https://www.skatelescope.org/technical/info-sheets/}. In this case, considerations similar to those made for \emph{Athena} apply. SKA1-MID will be optimally suited to identify the launch of a post-merger radio jet. The subsequent stage, SKA2-MID, is more uncertain, but the goal is to improve both the sensitivity and FOV by roughly factor of ten, getting to $0.1\,\mu$Jy over $10 \degsq$, allowing effective pre-merger searches, at least out to $z\approx 1$.

Moving to the optical, LSST can also play primary partnership role with LISA. In its $9.6 \degsq$ FOV, this telescope can reach a limiting magnitude of about 24.5 in mere 30 seconds of pointing \cite{2008SerAJ.176....1I}, sufficient to detect a $10^6\msun$ MBH accreting at Eddington out to $z\approx 1.5$. Note that the survey speed easily allows to cover in just five minutes the $\Delta\Omega=100 \degsq$ with which LISA sources out to $z\approx 1.5$ are expected to be localized two days before merger. A viable strategy would then be to survey the whole area every few minutes with LSST for the last two days of inspiral, which would allow to construct an $\approx 1000$ point light curve of each object within the (evolving) LISA error box.
Note that in this time, the MBHB will complete several orbits, resulting in a large number of GW cycles. For a given time to coalescence, the corresponding GW frequency $f_t$ can be obtained using Eq. 2.7  of \cite{PhysRevD.74.122001}. Then, ignoring higher post-Newtonian corrections, an estimate of the number of GW cycles experienced by a binary when it sweeps from $f_t$ to ISCO is given by
\begin{equation}
    N_{\mathrm{cyc}} = \frac{1}{32\pi^{8/3}} \left( \frac{c^3}{G \mathcal{M}_z} \right)^{5/3} (f_t^{-5/3} - f_{\rm ISCO}^{-5/3}  )
\end{equation}
where $f_{\rm ISCO} = c^3/[6\sqrt{6} \pi G {\rm{M_{tot}}}(1+z)]$.
 
In Fig.~\ref{fig:triple_contour_ncyc_of_time} we show in the mass-redshift plane, the number of cycles left to merger when a source first accumulates $S/N=8$ in the LISA band, and when it is localized within $100 \degsq$ and $10 \degsq,$ respectively.
The right panel shows that a $3\times 10^5\msun$ MBHB at $z=1$ still needs to complete 100 orbits when it is localized within $10 \degsq$ about ten hours before merger. If any EM periodicity rises during the GW chirp, LSST will have the potential to effectively uncover it. We note that similar arguments apply to SKA2-MID, which might potentially detect a periodic signal due to e.g. a precessing radio jet.  

\begin{figure}
    \centering
    \includegraphics[width = 0.5\textwidth]{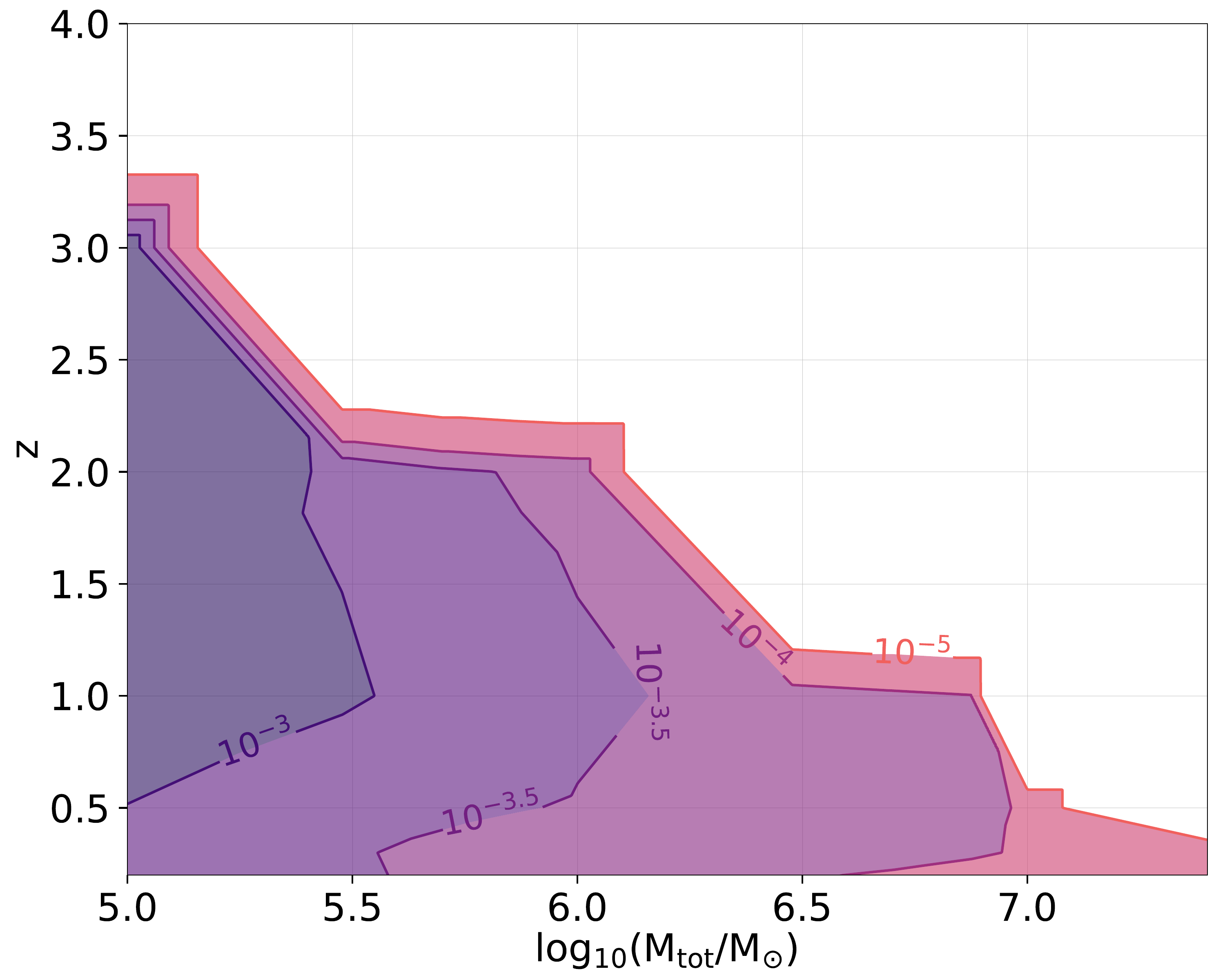}
    \caption{Contours of observed frequency $f_f$ when sources are localized with $\Delta \Omega = 10 \degsq$ accuracy as function of  total mass in the source frame and redshift.}
    \label{fig:freq_modulation}
\end{figure}

We can further correlate the frequency $f_t$ at which the median of $\Delta \Omega$ equals  $10\degsq$ with the binary separation in units of the binary Schwarzschild radius $r_{\rm schw}=2GM_{\rm tot}/c^2$

 
 \begin{equation}
    \frac{a}{r_{\rm schw}}\simeq 40 \left({10^{-3} {\rm Hz}\over f_t}\right )^{2/3}\left ({10^5\msun\over M_{\rm tot}}\right )^{2/3}
 \end{equation}
to show that light binaries are sufficiently separated that mini-disks that may form around each black hole might leave a periodic imprint on the EM light cure in the approach to merger. In Fig.~\ref{fig:freq_modulation} we show contour lines of constant $f_t$ when $\Delta\Omega=10\degsq$  as a function of redshift and total mass in the source frame.
 
We note that few days before merger, the distance and mass of the GW source is typically known within 10\% precision, as shown in Fig.~\ref{fig:dl_of_time} and Fig.~\ref{fig:mchirp_of_time}. Assuming the concordance cosmological model at the time of discovery, the luminosity distance can be converted into a redshift. This will help in reducing the number of candidates already {\it on the fly} and to cut sources outside the relevant redshift and host galaxy mass range to potentially uncover a precursor signal.  

Sources at $z\approx 1$ are localized within $\simeq 0.1 \degsq$ when the full signal is considered, as illustrated in  Fig.~\ref{fig:triple-sky-loc}.  Likewise, the luminosity distance and redshift, assuming a cosmology, are also known with a precision of a few percents. Considering that there are about $10^6$ galaxies per $\degsq$, projected on the sky, the redshift information from the GW signal can greatly help in weeding out galaxies in the sky error area using galaxy catalogs.
Ideally, once the EM counterpart is identified, optical/near-IR imaging and spectroscopy of the host galaxy would let us study in a new and unique way the relationship between the MBH mass, inferred from the GW signal, and the galaxy mass, beside having an independent measure of the redshift to carry on estimates on the cosmological parameters \citep{Schutz1986}.  Furthermore, any discrepancy between the two values of the redshifts may help detecting false positive associations, i.e. turn-on AGN not having any secure connection with the GW event, given that after merger the onset of accretion might have a delay.

A multi-objects spectrograph like MOONS on VLT, with $\sim$ 1000 fibers over a field of view of $\sim$ 500 square arcmin would be an excellent instrument for spectroscopic follow-ups. Spectroscopic surveys or follow-ups can help select a putative EM counterpart also through AGN narrow emission lines and the presence of broad lines, although the dynamical environment close to a MBH merger can alter the ``standard'' picture \citep{2019ApJ...879..110K}. A comparison between the MBH mass from the GW signal with that inferred using standard AGN techniques \citep[][and references therein]{2009ApJ...692..246D} would be invaluable as consistency between the two mass measurements might reveal how the emission region changes under the highly dynamical conditions present in the post  merger gaseous environment.

One important caveat is that LISA MBHs have relatively low masses, therefore they will not be bright sources in absolute terms, even if they accrete at the Eddington rate. A $10^6\msun$ MBH at
redshift $z\sim 1$ accreting at the Eddington limit would emit a [2-10] keV X-ray flux of about $8\times 10^{-16}$ in cgs units, and could be detected with a 5 ks exposure with {\it Athena} with an apparent magnitude of $\sim 25.4$ in the B band. A host galaxy 1000 times heavier than the MBH, and on the star formation ``main sequence'' \citep{2014ApJS..214...15S}, would have magnitude $\sim 24-25.4$ depending on its dust content.This means that if the MBH is accreting at sub-Eddington rates the galaxy will be generally brighter than the AGN at optical wavelengths. As a consequence, in terms of identifying the GW source, optical selection cannot rely on AGN features, such color selection or broad or narrow AGN emission lines, since these will be weak except for MBHs accreting close to the Eddington rate. Study of the host galaxy, however, would be facilitated because the spectral energy distribution is in this case dominated by the galaxy properties. Optical spectroscopy near the MBH where broad lines are produced would require instruments with very high sensitivity, e.g., a long-lived James Webb Space Telescope (JWST) or Wide Field Infrared Survey Telescope (WFIRST), which can look for spectroscopic signatures of the MBH-powered AGN once the host galaxy has been identified. 

\section{\label{sec:conclusion} Summary and conclusions}

 MBHBs of $10^5- 3 \times 10^7\msun$ coalescing in low redshift galaxies ($z<4$) have been the focus of this paper. In the inspiral phase the GW signal is sufficiently long-lived to enable a pre-merger astrophysical characterization of these sources. To this aim, we carried on a parameter estimation on the fly, i.e. as a function of the time to coalescence. When the GW inspiral event is evolving in time, the signal-to-noise ratio continues to rise and uncertainties in the parameter estimates reduce. In our study we selected a sequence of times, from one month down to one hour prior to coalescence and considered the source sky localization, luminosity distance, chirp mass and mass ratio as key parameters, providing ready-to-use analytical fits of their associated uncertainties as a function of time. 

Here we enumerate our key findings and concluding remarks:
\begin{itemize}
\item Between $3\times 10^5\msun$ and  $3\times 10^6\msun,$ the $S/N$ of MBHBs at $z=1$ rises above 8 - assumed as threshold for detection - 1 month to 1 week (for the heaviest system in this range) before coalescence. 5 hours prior merging the $S/N$ is in the hundreds and at coalescence it reaches values in the thousands.
\item Between $3\times 10^5\msun$ and  $3\times 10^6\msun,$ the binary chirp mass and mass ratio  are determined with a fractional error $\lesssim  10\%$  1 month to 1 week (for the heaviest system in this range) before merging. 5 hours before coalescence the accuracy narrows down at $1\%$ level or even less (for the lightest MBHBs), and continues to improve significantly down to the end of inspiral.  The luminosity distance and redshift 1 week before merging are inferred with an accuracy of about 10\%. 
\item The median of the sky localization error $\Delta \Omega$ during the inspiral phase  decreases by more than two orders of magnitude, due to the increase of the $S/N$ as time progresses. Moreover close to merger, spin-precession effects, higher harmonics and doppler modulations help breaking degeneracies and further reduces \uncer. However at any given time the uncertainty of $\Delta \Omega$ around the median value widens as the binary approaches coalescence. This allows the lightest sources (with masses of a few $10^5\msun$) in the best (worst) configurations to be localized within  $\simeq 1 \degsq$ ($\simeq 100 \degsq$) one day before coalescence. For systems of a few $10^6\msun$ the uncertainty in the sky position 
is larger, between  $\simeq 10\degsq$ and  $\simeq 10^3 \degsq,$ few days prior merging.  Only at 'merger'  the median of  $\Delta \Omega$ plummets down to  $\simeq 10^{-1}\degsq.$ Some sources can be localized with square arc-minute precision at the time of coalescence. 
\item Low-redshift ($z\lesssim 1$), low mass MBHBs ($\mtot \lesssim 5\times 10^6\msun$) can be detected first by large field-of-view telescopes as LSST and SKA from 2 days to few hours in advance  if precursor emission exists, and later as time progresses by X-ray telescopes such as {\it Athena} at merger. These sources cover about 100  to 30  cycles before coalescence, within a sky localization uncertainty $\Delta \Omega\simeq 10 \degsq$, opening the possibility of detecting any periodicity, if present, in the EM signal possibly correlated with the periodicity in the inspiral signal.
\item MBHBs with total mass around $\simeq 10^7\msun$ at $z\simeq 1$ appear in the GW sky few days before merging and display a rapid increase in the $S/N$ just in the last few hours. For these binaries the chirp mass and mass ratio is known to a precision of 10\%  about 3 days before merger.  The sky-position remains highly uncertain across the entire inspiral phase. For these systems, the sky localization can be reconstructed at `merger' by exploiting the amplitude and phase of the harmonics of the ringdown as shown in  \citep{2020arXiv200110011B}.
\item Moving to higher redshift ($1<z \le 3$) and to binaries with masses between $3\times 10^5\msun$ and  $3\times 10^6\msun,$ the information on the chirp mass (mass ratio) accumulates in the last few days  (few hours) reaching 10\% precision.  A similar trend is observed for the uncertainty in the luminosity distance. During the inspiral phase, a median sky localization of $10\degsq$ is reached about 1 hour before coalescence. The localization improves when the full signal is recovered.  MBHBs with total mass $\simeq 10^7\msun$ are localized within $10\degsq$ in their inspiral phase 10 to 1 hour prior merger up to $ z \lesssim 1.3$. 
\item The analysis post-merger of the full GW signal allows sky localization of $\simeq 0.4\degsq$ out to $z\simeq  3$, for those  sources clustering around a mass interval between $\simeq 3\times 10^5\msun$ and $\simeq  10^6\msun$ that could be detected by  \emph{Athena} at mergers and in the post-merger phase.
\end{itemize}

MBHB mergers are expected to be rare events in the Universe and there are large \uncer in the predicted number of events (see \cite{10.1093/mnras/stz3102, Volonteri:2020wkx} for a recent discussion). In this paper we focus on LISA's ability to constrain source parameters without accounting for the expected number of events. To assess LISA's possibilities to detect EM counterparts, our simulations have to be convolved with a realistic population of merging MBHBs. We defer this point to later studies. 

In this analysis we did not include neither the scheduled gaps in the data due to LISA communication with Earth, antennas re-pointing or laser locking \cite{2017arXiv170200786A}, nor the gaps due to unexpected failures requiring system's reboot. 
Gaps would degrade our results especially in the early inspiral. However, when an event is detected, a protected period might be established around the time of the merger
reducing the effect of the scheduled gaps since most of the $S/N$ is built close to merger.

In a recent work Marsat \emph{et al.} \cite{Marsat:2020rtl} reported the appearance of eight degenerate points in the posterior distribution for the sky position of the source when performing Bayesian parameter estimation for inspiral-merger-ringdown signals. These degeneracies can be broken close to merger with the inclusion of higher harmonics and the frequency-dependence in the LISA response function (that is what we observed with \texttt{ptmcmc}). 
The Fisher matrix approach is unable to track these degenerate points so our results have to be considered conservative in the early inspiral.

In this study we considered only LISA.
If other space-born GW observatories \cite{PhysRevD.100.043003, Ruan:2019tje} sensitive to the same frequencies joined LISA in the sky, they might help further reducing the \uncer, especially for the sky position of the source.

In summary, MBHBs with masses  between $\simeq 3\times 10^5\msun$ and  $\simeq 3\times 10^6\msun$ at $z\simeq 1$ carry exquisite information on their astrophysical parameters during the inspiral phase that can be inferred on the fly. The contribution of higher harmonics, included in this investigation, make these unequal and relatively-long-lived binaries the best sources for coordinated searches of EM counterparts. GRMHD/radiative transfer simulations should focus on these systems to enhance our knowledge  on their emerging spectra and variability.

\acknowledgments
We thank John Baker for useful comments and and contributions to the \texttt{PTMCMCSampler} code. A.~M. and M.~C. acknowledge partial financial support from the INFN TEONGRAV specific initiative. A.~M. acknowledges networking support by the COST Action CA16104. M.~V. and S.~B. acknowledges support by the CNES for the space mission LISA. M.~C. acknowledges funding from MIUR under the grant PRIN 2017-MB8AEZ. A.~S. is supported by the European Research Council (ERC) under the European Union’s Horizon 2020 research and innovation program ERC-2018-COG under grant agreement No 818691 (B Massive). M.~L.~K. acknowledges support from the National Science Foundation under grant DGE-0948017 and the Chateaubriand Fellowship from the Office for Science \& Technology of the Embassy of France in the United States. M.~L.~K's research was supported in part through the computational resources and staff contributions provided for the Quest/Grail high performance computing facility at Northwestern University.

\bibliography{bibliography.bib}

\clearpage
\appendix
\section{\label{sec:app_to_discuss_random_params} Parameters effect on the uncertainties distribution}
In this appendix we discuss the  role of each binary parameter in shaping the sky position, luminosity distance, chirp mass and mass ratio \uncer distributions. 
For each of these parameters we compared different cases:
\begin{enumerate}[(I)]    
\item $q = 0.1, \, 0.5, \, 1$ and randomly distributed in $[0.1, 1]$ 
\item $t_c = 0.5,  \, 2,  \, 4 \yr$ and $t_c \in [0, \, 4 \yr]$ 
\item $\chi_1 = \chi_2 = 0.1, \, 0.5, \, 0.9$ and $\chi_1, \, \chi_2 \in [0, \, 1]$ 
\item inclination $i = 0, \,  \pi/4, \, \pi/2$ and uniformly distributed 
\item different sky positions.
\end{enumerate}

For each case, we assume a MBHB with $\mtot = 10^6 \msun$ at $z = 1$ and randomize over $\rm N = 10^3$ realizations for the other parameters.

\begin{figure}
    \includegraphics[width=0.5\textwidth]{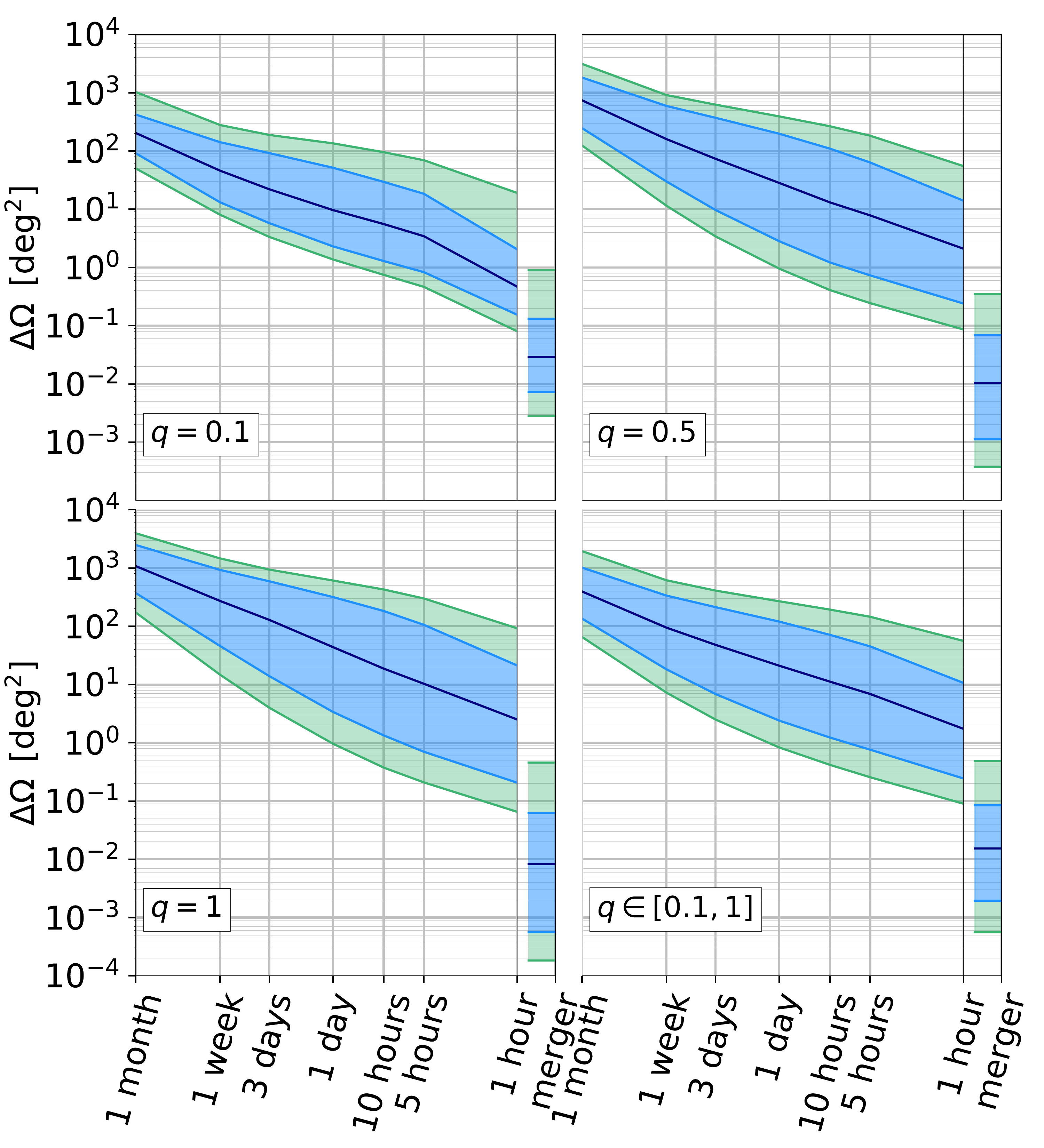} 
    \caption{Sky localization \uncer as function of time to coalescence for different mass ratios as given in each panel. Colors as in Fig.~\ref{fig:triple-sky-loc}.}
    \label{fig:sky-loc-random-q}
\end{figure}

In Fig.~\ref{fig:sky-loc-random-q} we show sky position \uncer for different values of the mass ratio and for the case of random extraction in the interval $[0.1,1]$. 
The sky position of systems with $q = 0.1$ is recovered better than the one for equal-mass system at all times before merger. At 1 month from coalescence, systems with $q = 0.1$ and $q = 1$ are localized with a median accuracy of $\simeq 200 \degsq$ and $\simeq 10^3 \degsq$, respectively. At 1 hour from merger equal-mass systems are localized with a median of $\simeq 3 \degsq$, while the uncertainties for $q = 0.1$ systems are smaller by a factor of $\simeq 5$. This is expected due to the fact that higher harmonics turns out to be important for small mass ratios and can lead to an improvement in the parameter estimation.

However at merger nearly equal-mass systems have higher $S/N$ and the recovered area is usually smaller than the one for unequal-mass systems.

The distributions for unequal-mass systems are usually narrow but they still cover several orders of magnitude. In both $q = 0.1$ and $q = 1$ cases the recovered areas cover more than an order of magnitude at 1 month from coalescence and over two orders of magnitude at 1 hour from merger. 

The cases for $q = 0.5$ and $q \in [0.1, 1]$ show intermediate behavior. From these results, we choose to randomize over the mass ratio. 

We find similar trend also for the luminosity distance, chirp mass and mass ratio uncertainties.

\begin{figure}
    \includegraphics[width=0.5\textwidth]{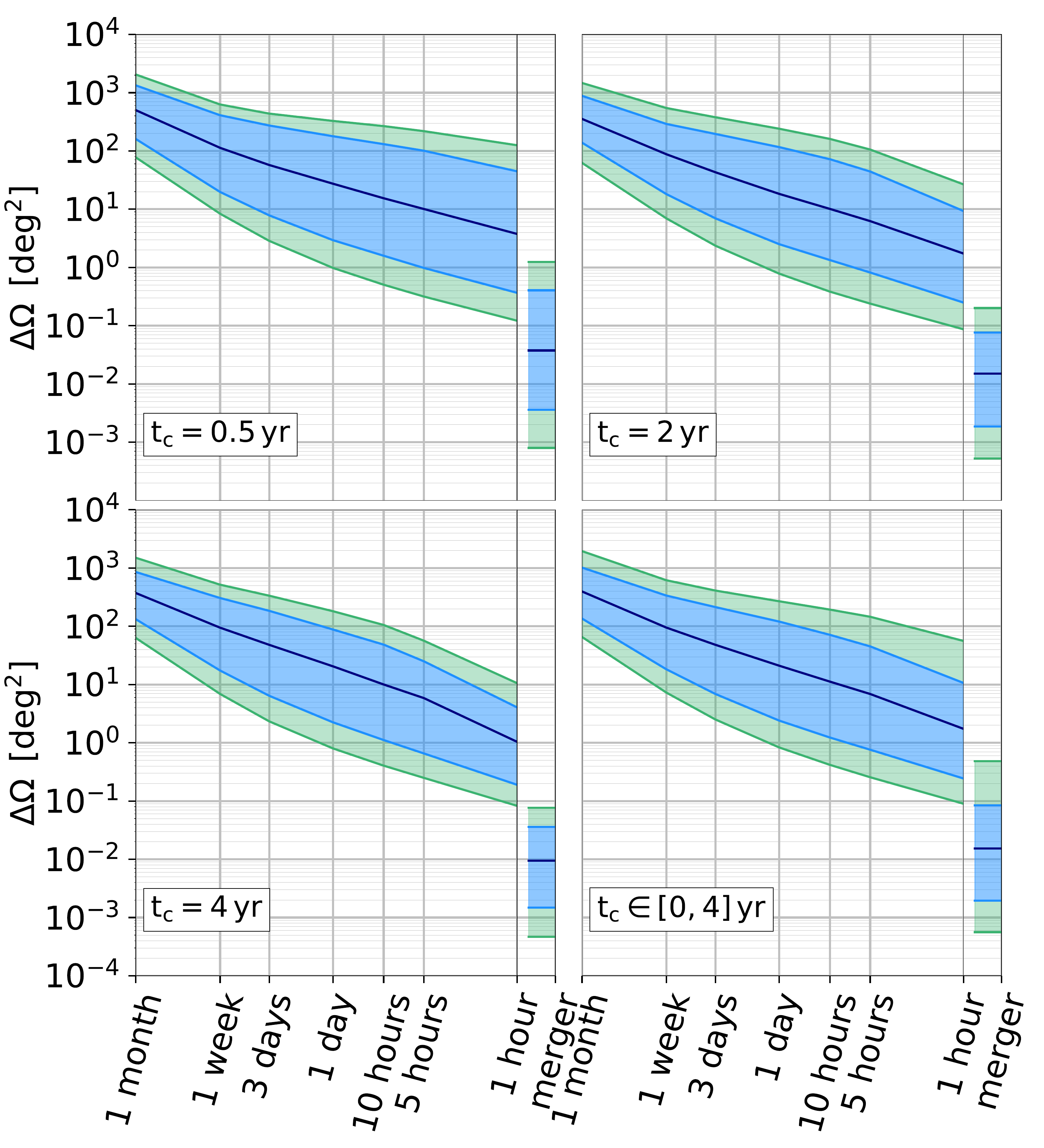}
    \caption{Same as Fig.~\ref{fig:sky-loc-random-q} for case (II), i.e. for different coalescence times as reported in each panel.}
    \label{fig:sky-loc-random-time}
\end{figure}

In Fig.~\ref{fig:sky-loc-random-time} we show the recovered binary position errors for systems at different times to coalescence. In particular, we consider systems with coalescence time fixed at $t_c = 0.5, \, 2, \, 4 \yr$ and randomly distributed in $[0,\, 4] \yr$.

MBHBs merging in $4 \yr$ are localized 4 times better than systems at only $0.5 \yr$ from coalescence at 1 hour from merger. The intermediate case with $t_c = 2 \yr$ and the case for random distributed coalescence time show intermediate behavior. Since there is small difference between the two extreme cases, we chose to randomize over coalescence time when providing analytical formulas.

\begin{figure}
    \includegraphics[width=0.5\textwidth]{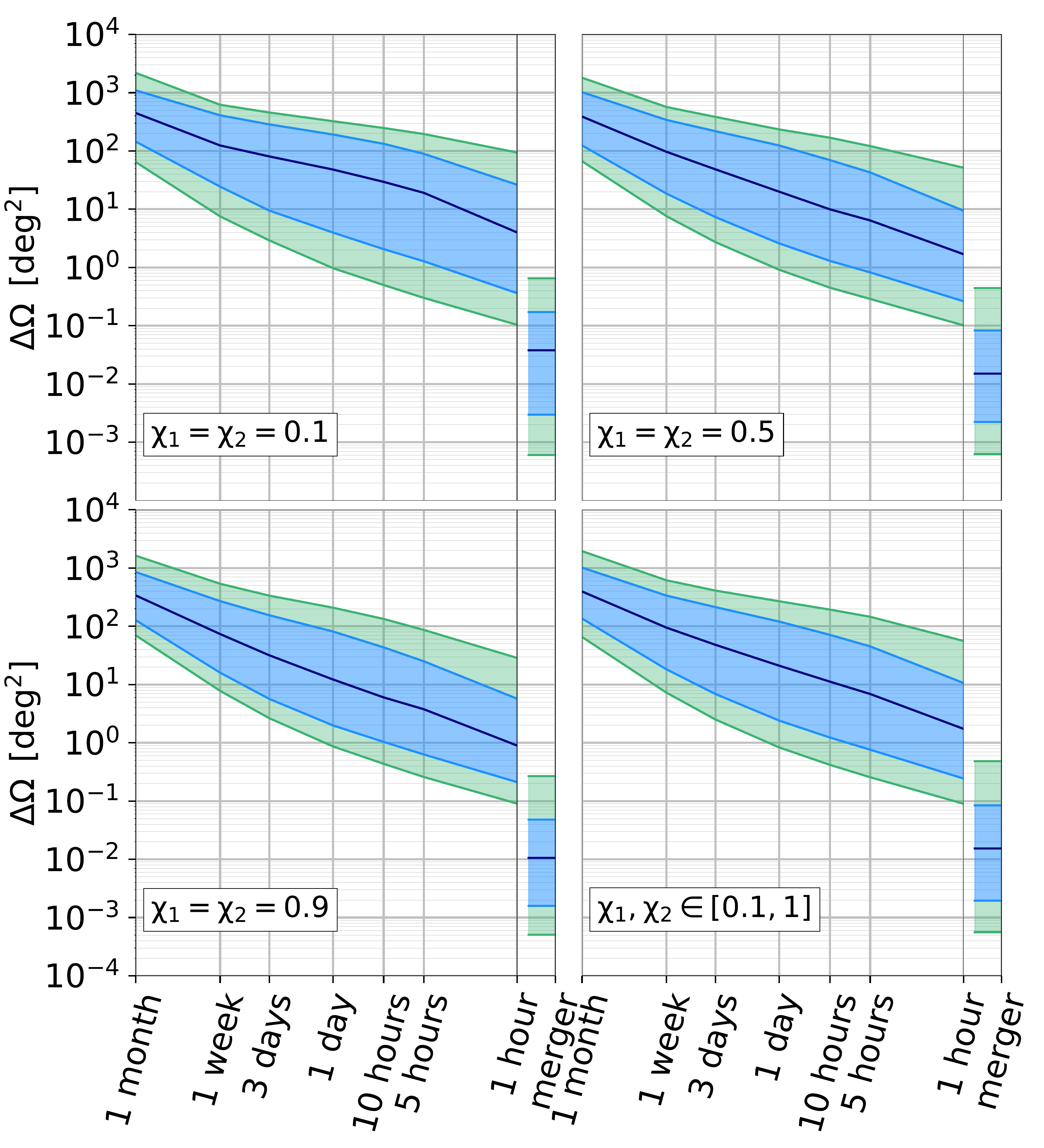}
    \caption{Same as Fig.~\ref{fig:sky-loc-random-q} for case (III), i.e. for different spin magnitudes as reported in each panel.}
    \label{fig:sky-loc-random-spin}
\end{figure}

In Fig.~\ref{fig:sky-loc-random-spin} we give the sky uncertainty distributions for three different values of spin magnitude, namely $\chi_1 = \chi_2 = 0.1, 0.5, 0.9$, and the case where spin magnitudes are both randomly extracted in $[0,1]$. In all cases, we leave spin directions uniformly distributed over a sphere.
Higher spins help in breaking degeneracies and usually allow to recover better sky positions. At the end of the inspiral, spin-precession effects become important and high-spinning (low-spinning) systems can be localized with a median value of $\simeq 0.9 \degsq$ ($\simeq 4 \degsq$). However the distributions show similar range. Since spin values do not seem to affect significantly the recovered area, we choose to randomize over the allowed range.

\begin{figure}
    \includegraphics[width=0.5\textwidth]{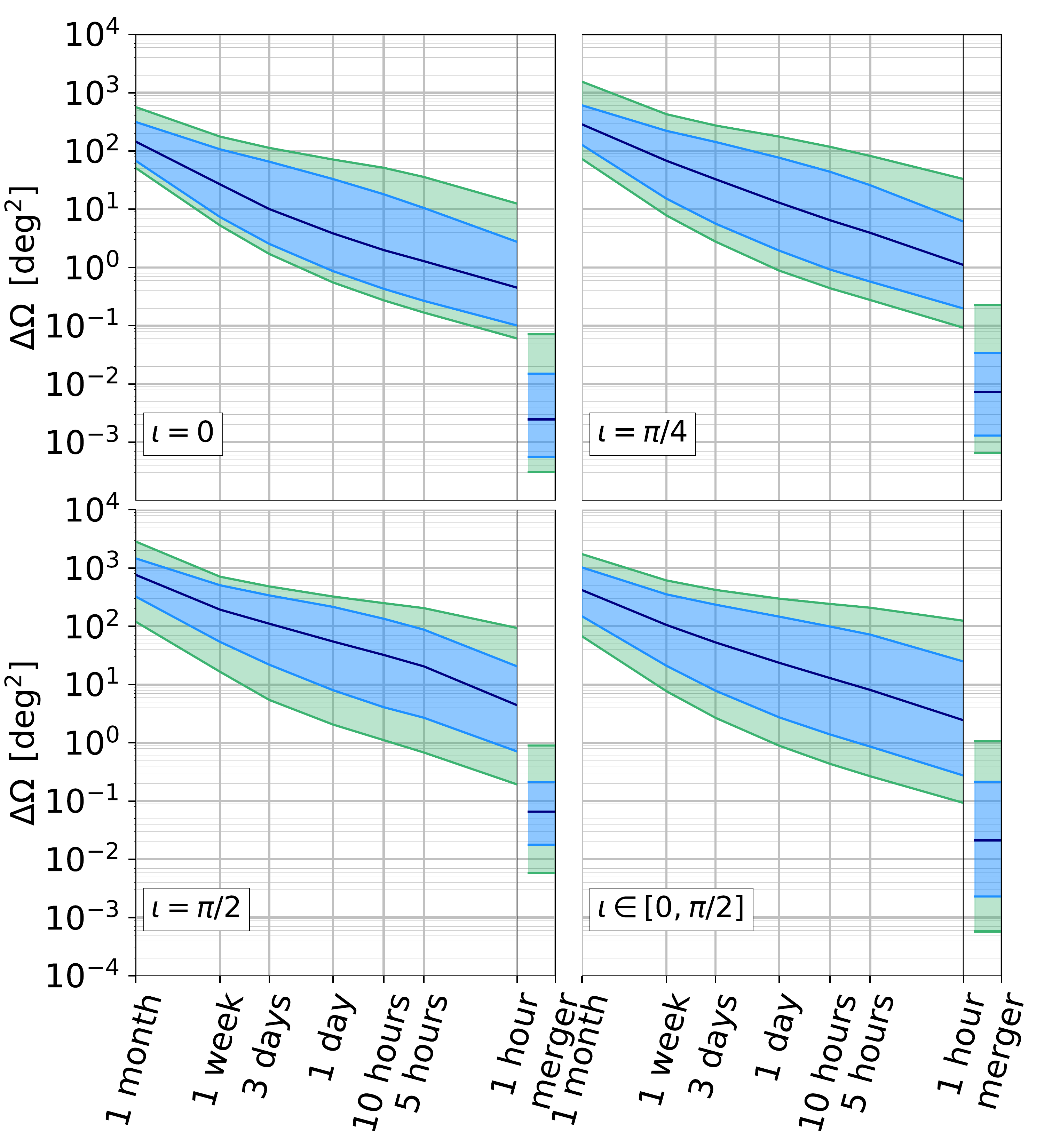}
    \caption{Same as Fig.~\ref{fig:sky-loc-random-q} for case (IV), i.e. for different inclination values as reported in each panel.}
    \label{fig:sky-loc-random-incl}
\end{figure}

In Fig.~\ref{fig:sky-loc-random-incl} we show the sky uncertainty distribution for three different inclination values and for the case $\iota \in [0, \, \pi/2]$. Face-on systems are localized with a better precision during all the inspiral with \uncer smaller than one order of magnitude at 1 hour before merger. The difference is larger at merger since face-on systems have larger $S/N$. Different inclination values affect only the median value, while the distributions present similar ranges. Even if the inclination affects the recovered error on the sky position, we choose to randomize over this parameter.

\begin{figure}
    \includegraphics[width=0.5\textwidth]{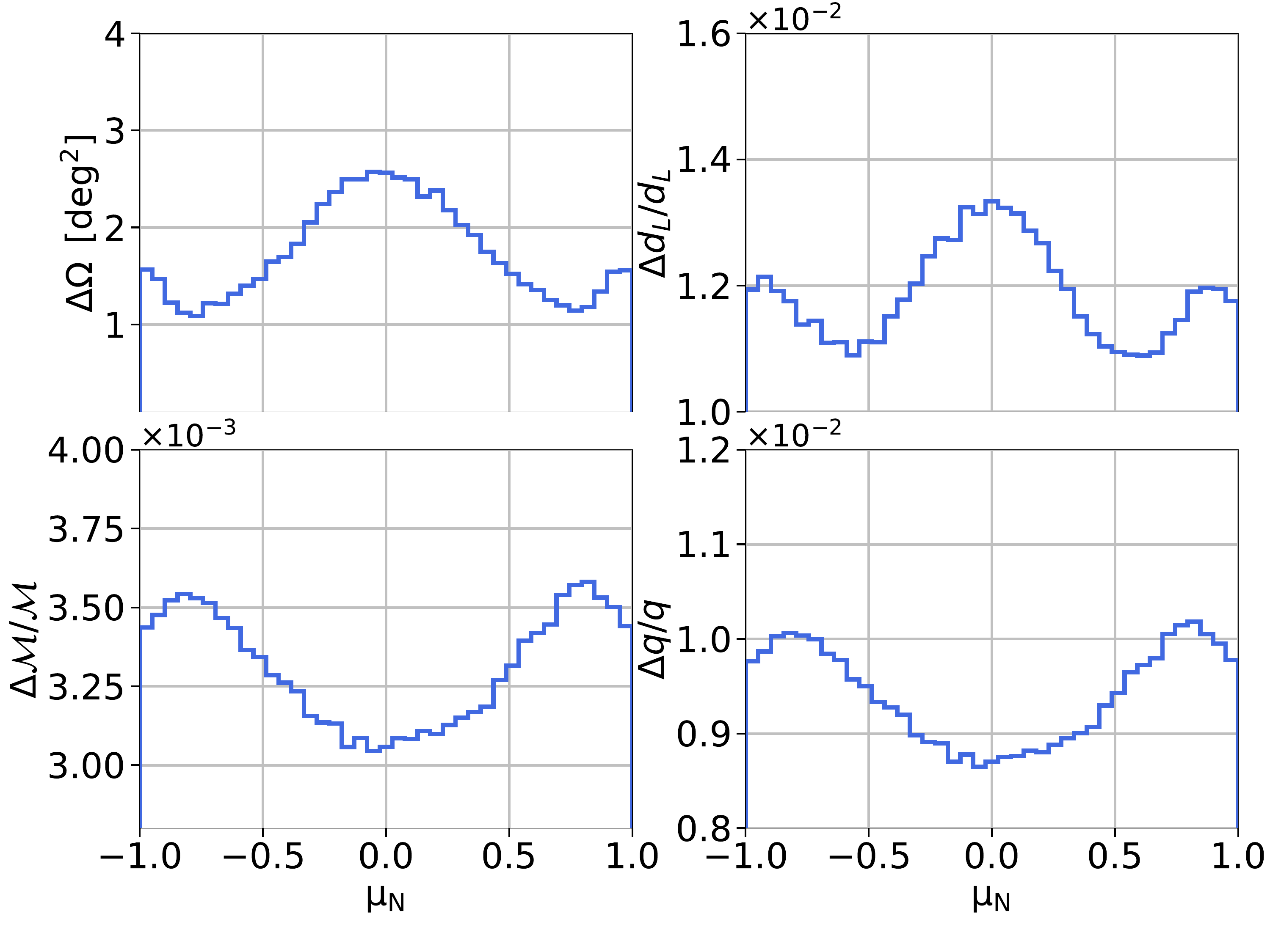}
    \caption{Distribution of sky position, luminosity distance, chirp mass and mass ratio uncertainties as function of $\mu_N$ for a system with $\mtot = 10^6$, $z = 1$.}
    \label{fig:all_distro_panels}
\end{figure}

In Fig.~\ref{fig:all_distro_panels} we show the \uncer on the sky position, luminosity distance, chirp mass and mass ratio at different sky positions. Similarly to \cite{Lang_2008}, we define $\mu_N = \cos(\theta_N)$ and divided the interval $\mu_N \in [-1, \,1]$ in 40 bins. In each bin, we perform N $=10^4$ realizations, varying $\mu_N$ in the bin range. We keep $\mtot = 10^6 \msun$ and $z = 1$. 

All recovered distribution are symmetric respect to $\mu_N = 0$ as expected and the small differences are due to statistical fluctuations.
LISA is able to constrain the sky position of sources lying outside the galactic plane better by a factor of 2 than sources lying in the plane.
Luminosity distance \uncer show three peaks at central value and approaching the poles. The central peak is slightly higher than the outlying ones. 
Chirp mass and mass ratio show opposite trends and they are recovered better for sources lying in the galactic plane with an improvement of  $\simeq 10\%$ with respect to sources lying outside.

Finally, we find no strong dependence for the recovered \uncer on the polarization or the initial phase, so we average over them.

\section{\label{sec:comparison_old_curve} Comparison with previous studies}
\begin{figure}
    \includegraphics[width=0.5\textwidth]{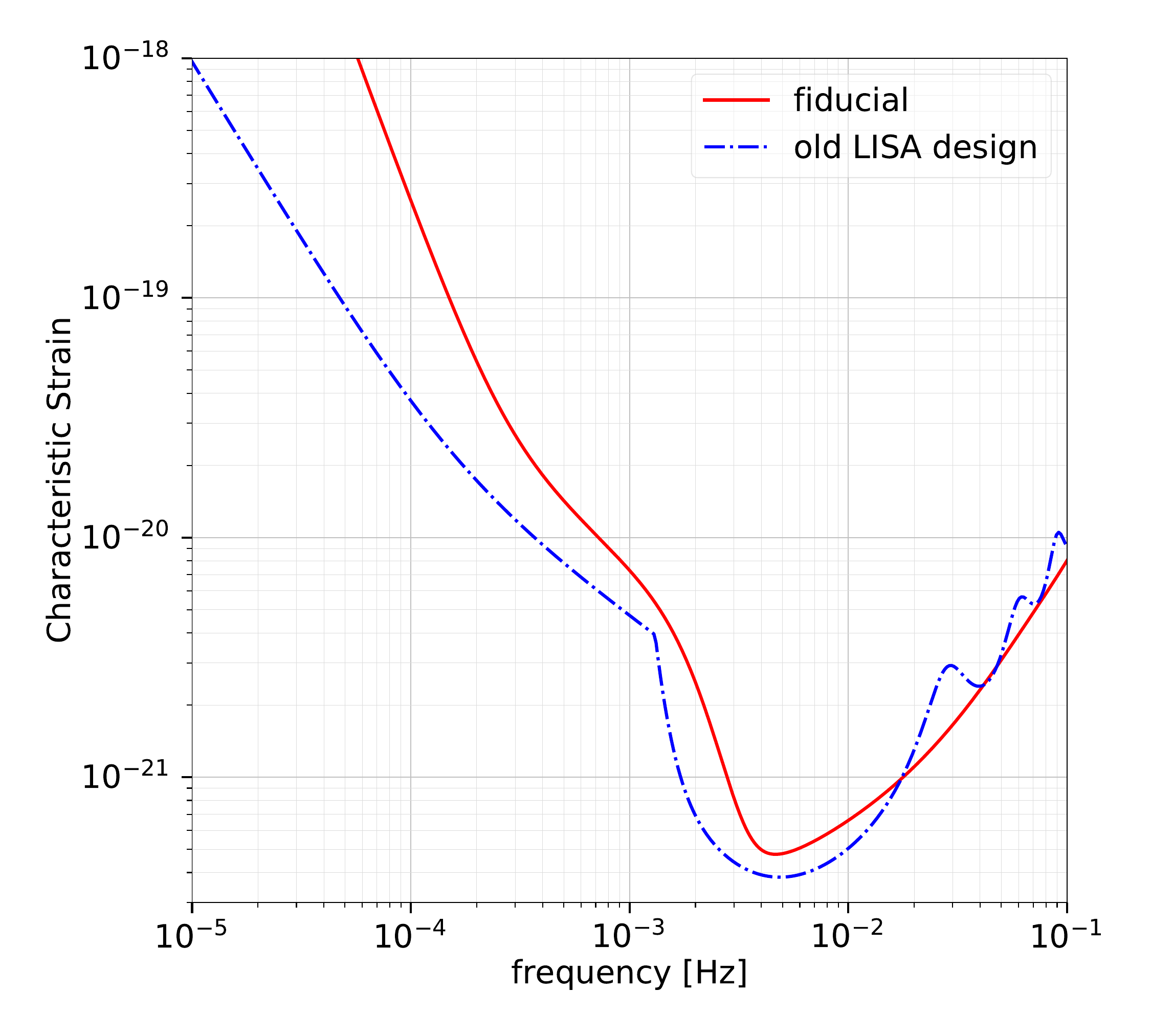}
    \caption{Old LISA design (dotted-dashed blue line) and current LISA design (continuous red line) sensitivity.}
    \label{fig:lisa_noise_comparison}
\end{figure}

In this appendix we compare our results against previous results in the literature.
In Fig.~\ref{fig:lisa_noise_comparison}, we plot the fiducial LISA design sensitivity adopted in this study and the sensitivity adopted in past studies focusing on LISA ability to constrain source parameters during the inspiral \cite{Lang_2008,Kocsis_2008}. 
At $f = 10^{-4} \, \rm {Hz}$, the old LISA design sensitivity is roughly an order of magnitude higher, i.e. has a lower characteristic strain  than the current one. Also in the bucket of the curve, $f \simeq 5 \times 10^{-3} \, \rm {Hz}$, old LISA design had a higher sensitivity.

\begin{figure}
    \includegraphics[width=0.5\textwidth]{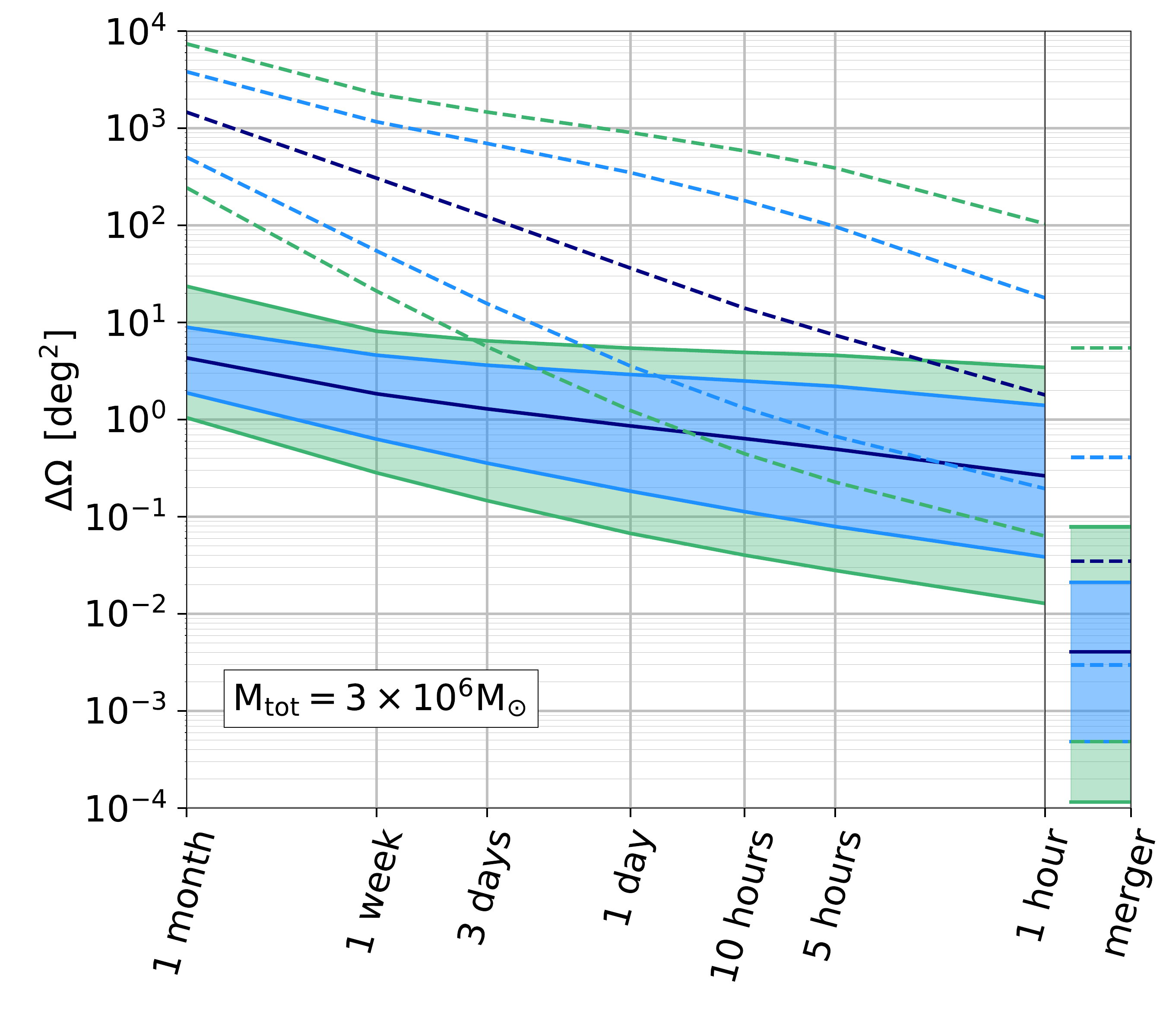}
    \caption{Sky position uncertainty as function of time to coalescence for a MBHB as labeled in the plot at $z=1$. Continued (dashed) lines are obtained with the old (current) LISA sensitivity. Colors as in Fig.~\ref{fig:triple-sky-loc}.}
    \label{fig:larson_comparison}
\end{figure}

As a consequence, in past studies, LISA sources where better localized in the sky already at 1 month from coalescence. In Fig.~\ref{fig:larson_comparison} we plot the sky position \uncer as function of time to coalescence for the two sensitivity curves reported in Fig.~\ref{fig:lisa_noise_comparison} for a MBHB with $\mtot = 3 \times 10^6$ at $z=1$. We change only the sensitivity curve, randomizing all the other parameters in the same range over $N = 10^3$ realizations.
The improved low-frequency sensitivity of the old LISA design leads to a median sky position error of $\simeq 4 \degsq$ at 1 month from coalescence, almost three orders of magnitude better compared to the current LISA design. When the system approaches merger the difference between the two configurations narrows down, leading to a median sky localization uncertainty of $\simeq 0.3 \degsq$ at 1 hour from coalescence, to be compared with a fiducial value of $\simeq 2 \degsq$. Similar considerations hold also when the full signal is considered.

\section{\label{sec:binary_angular_momentum} Estimation of binary angular momentum}
\begin{figure}
    \includegraphics[width=0.5\textwidth]{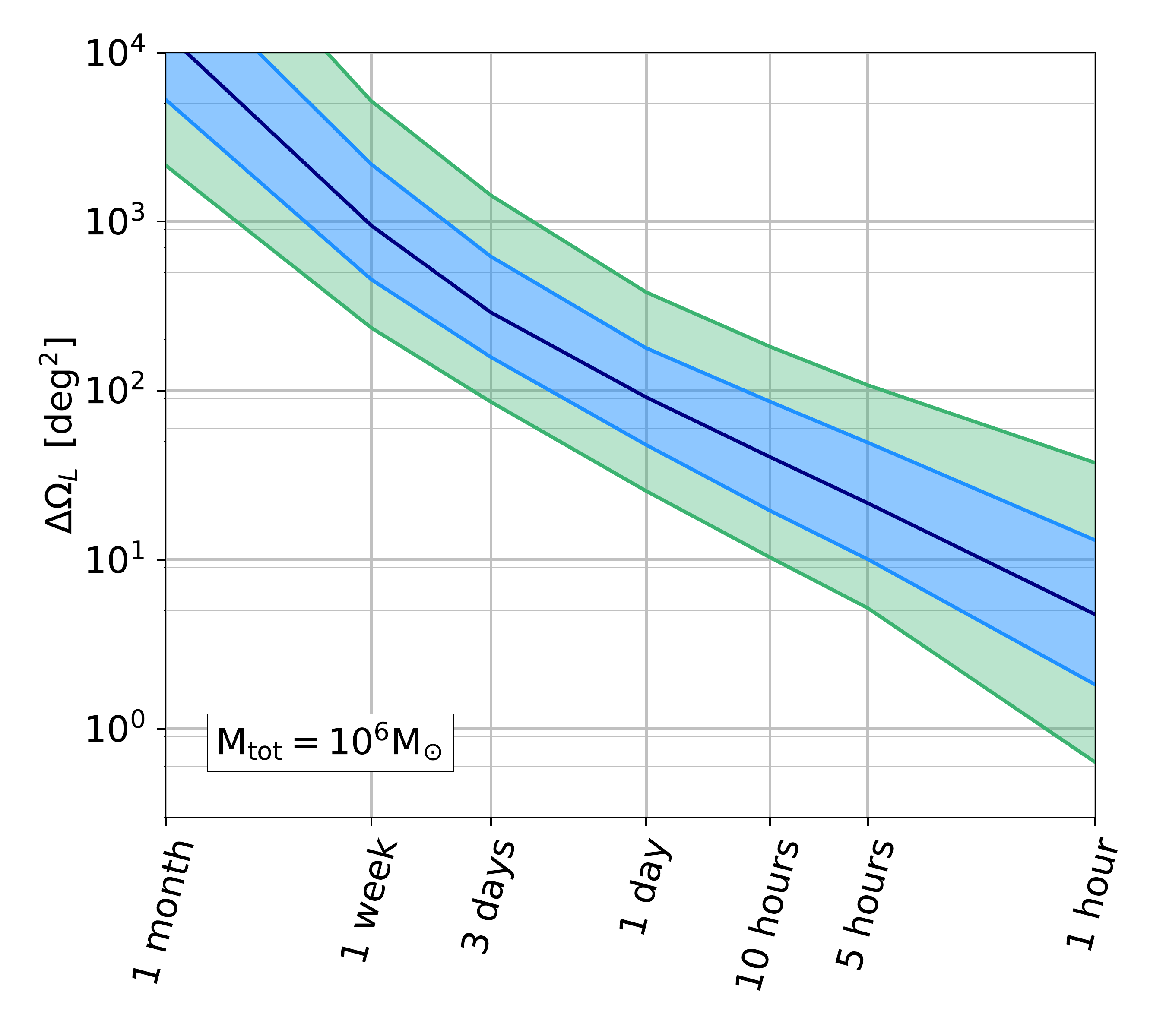}
    \caption{Binary angular momentum uncertainty as function of time to coalescence for a MBHB as labeled in the plot at $z=1$. Colors as in Fig.~\ref{fig:triple-sky-loc}.}
    \label{fig:momentum_L}
\end{figure}

In this appendix we estimate the uncertainties in the direction of the binary angular momentum $\bf{L}$, relative to the line of sight from the source to the observer. To compute $\Delta \Omega_{L}$, we adopted the same formula for the sky-position $\Delta \Omega$, but replacing the uncertainties on $[\cos(\theta_N), \phi_N]$ with the one on $[\cos(\theta_L), \phi_L]$ (neglecting the cross-correlation term).
In Fig.~\ref{fig:momentum_L} we show LISA ability to constrain the binary angular momentum as function of time left before merger. At 1 month from merger, the binary angular momentum is not constrained, but, at 1 hour from merger, it is determined with an accuracy of $\simeq 10 \degsq$, corresponding to an uncertainty of $\sim$ 3 deg. This is a key information, as the EM emission, i.e. its level of variability and spectral properties, depend on the inclination of the orbital plane relative to the observer and knowing the direction of $\bf{L}$ could help identifying 
 the EM counterpart through its expected peculiar emission \citep{d_Ascoli_2018,10.1093/mnras/sty423}.

The orbital angular momentum contributes to the total angular momentum, $\bf{J}$, defined as
\begin{equation}
    \bf{J} = \bf{L} + \bf{S_1} + \bf{S_2}
\end{equation}
where $\bf{S_1}$ and $\bf{S_2}$ are the BH spin vectors. Therefore it would also be interesting to investigate LISA ability to constrain on the fly  the directions of the individual spins  to trigger EM alerts informative of the potential directions of incipient jets. We suspect that this will be possible only when the full signal is considered. Thus, a complete and exhaustive analysis is beyond the scope of this paper.

Here, we would like to highlight few key points:
\begin{itemize}
    \item During the inspiral, if each BH produces a jet, the jet is likely to be aligned with the spin direction \cite{Palenzuela2010}. LISA ability to constrain spin orientations might play a key role in discriminating sources for which jets are pointing towards us.
    \item After the merger, a jet might be launched aligned with the spin of the remnant BH and, therefore, to the total angular momentum $J$. However we observed that the error on the individual spin orientation is larger than that on the binary angular momentum, leading to an overall degraded estimate for $J$ by more than an order of magnitude (on average) at the end of the inspiral.
    \item If spin magnitudes are small, the total angular momentum is determined basically by the binary angular momentum. This could suppress the large uncertainties connected with spin orientation.
    \item Gas accretion is expected to align BH spins to the binary angular momentum \cite{Palenzuela2010,Dotti2010}. In this case randomizing over spin orientations, as we have done in this paper, might not be ideal. Moreover in this situation the recoil kick is in the orbital plane and it might trigger EM emission \cite{Kocsis_2008}.
\end{itemize}
For these reasons, here we report the \uncer on the binary angular momentum only (these data can be be found at the aforementioned GitHub page). We commit to explore these aspects in detail in later studies.

\section{\label{sec:app_for_tables} Tables with Coefficients}
\begin{turnpage}
\begin{table*}[ht!]
\begin{ruledtabular}
\begin{tabular}{c| c c c c c | c c c c c } 
        & \multicolumn{10}{c}{Parameters}\\
 Coefficients  & \multicolumn{5}{c}{$\Delta \Omega$}  & \multicolumn{5}{c}{$\Delta \rm d_L / d_L $} \\ 
 
 & median  & $68\% \, \rm lower$ & $68\%  \,\rm upper$ &  $95\% \, \rm lower $ & $95\%  \,\rm upper $ 
 & median  & $68\% \, \rm lower$ & $68\%  \,\rm upper$ &  $95\% \, \rm lower $ & $95\%  \,\rm upper $   \\

\Xhline{0.1pt} \\
$c_1$ &	$2568.27 $ & $3512.72 $ & $1577.58$ &  $3874.1084$  & $297.1314$	&	  $1005.79$  & $1559.583 $ & $756.97$ &  $1988.84$ &  $-267.832$  \\ 
$c_2$ &	$-1208.887$	& $-1626.038$ & $-731.275$ &  $-1797.039$	&$-168.1976$	&	  $-458.186 $ & $-711.0698$ & $-348.628$ &  $-921.42$ &  $141.362$  \\ 
$c_3$ &	$184.4128$	& $245.4319$ & $109.31$ &  $272.2114$	& $28.8558$	&	  $67.2397  $ & $105.1866$ & $51.56386$ &  $139.1144$ &  $[z, \,  \log_{10} \mtot]$  \\ 
$c_4$ &	$-9.13586$	& $-12.08178$ & $-5.29714$ &  $-13.47515$	& $-1.56994$	&	  $-3.1845 $  & $-5.0494 $ & $-2.4594$ &  $-6.84734$  &  $1.4809 $  \\ 
$c_5$ &	$-1796.673$	& $-2344.484$ & $-1108.298$ &  $-2602.995$	& $-289.0075$	&	  $-718.3414$  & $ -1024.527$ & $-537.0054$  &  $-1285.656$ &  $129.117$  \\ 
$c_6$ &	$846.8472$	& $1085.9196$ & $513.703$ &  $1208.7319$	& $151.8264$	&	  $328.787 $ & $467.3178$ & $247.8523$   &  $595.444$  &  $-70.771$  \\ 
$c_7$ &	$[z, \,  \log_{10} \mtot]$ & $[z, \,  \log_{10} \mtot]$ & $[z, \,  \log_{10} \mtot]$	&  $[z, \,  \log_{10} \mtot]$	&	$[z, \,  \log_{10} \mtot]$	&  $[z, \,  \log_{10} \mtot] $ & $[z, \,  \log_{10} \mtot]$ & $[z, \,  \log_{10} \mtot]$ &  $[z, \,  \log_{10} \mtot]$ &  $12.92974$  \\ 
$c_8$ &	$6.49952$	& $8.148226$ & $3.799775$ &  $9.158112$	& $1.36678$	&	  $2.355736  $  & $3.355828$ & $1.79838$ &  $4.45595$ &  $-0.77395$  \\ 
$c_9$ &	$389.6703$	& $491.046$ & $245.2222$ &  $548.7184$	&  $79.19753$	&	  $158.2055$ & $212.1132$ & $118.1205 $ &  $263.1475$ &  $-19.987$  \\ 
$c_{10}$ &	$-183.79561$	& $-227.3143$ & $-113.8672 $ &  $-254.8002$	& $-40.3808 $	&	  $-72.62682$ & $ -96.74195$ & $-54.6334$ &  $-121.82422$ &  $11.4809$  \\ 
$c_{11}$ &	$28.282011$	& $34.4617$ & $17.2381$ &  $38.797776$	& $6.6287905$	&	  $10.85154$  & $14.38965$ & $8.211404$ &  $18.45193$ &  $-2.17817$  \\ 
$c_{12}$ &	$-1.419813$	& $-1.708847$ & $-0.8532665$ &  $-1.9359222$	& $-0.355383$	&	   $-0.527822$  & $-0.6969475 $ & $-0.4018752$ &  $-0.913203$ &  $0.133935$  \\ 
$c_{13}$ &	$-26.699734$	& $-32.74404$ & $-17.2392$ & $-36.7993$	&  $-6.44256 $	&	  $-10.96238 $  & $-14.01813$ & $-8.1937$ &  $-17.2456$ &  $1.01535$  \\ 
$c_{14}$ &	$12.5929$	& $15.13522$ & $8.01997$ &  $17.07276$	& $3.2388846  $	&	 $5.03955$ & $ 6.388211$ & $3.79471$ &  $7.97705$ &  $-0.617936$  \\ 
$c_{15}$ &	$-1.939627$	& $-2.2921364$ & $-1.218924$ &  $-2.59833$	& $-0.527868 $	&	 $-0.7551506 $ & $-0.9498658 $ & $-0.5720776 $ &  $-1.207692$ &  $0.122496$  \\ 
$c_{16}$ &	$0.097577$	& $0.1136226$ & $0.0607029$ &  $0.1296806$	& $0.02822687$	&	   $0.0369069$ & $0.0460217  $ & $0.0281455$ &  $0.0597755$ &  $-0.0077483$  \\ 
\Xhline{0.1pt} \\
$d_1$ &	$-129.9857$3 & $-164.5524$ & $-77.3729$ 	&  $-183.877$	& $-25.3058$	&	   $-48.84506 $ & $-69.43392$ & $-37.05507$ &  $-90.144$ &  $-25.2$  \\ 
$d_2$ &	$-2.54\times10^{-4}$ & $-3.18\times 10^{-4}$ & $-5.04\times 10^{-4}$	&  $-4.34\times10^{-4}$	& $-6.6\times10^{-4}$	&		 $-1.1\times 10^{-4}$  & $-4.9\times10^{-5}$ & $-1.8\times 10^{-4}$ &  $-5\times 10^{-5}$ &  $-9.4\times 10^{-4}$  \\ 
$d_3$ &	$-2.26\times 10^{-3}$ & $-2.47\times 10^{-3}$ & $-1.63\times 10^{-3}$	&  $-3.06\times 10^{-3}$	& $-2.02\times 10^{-3}$	&		  $-8.1\times 10^{-4}$ & $-5.1\times10^{-4}$ & $-6\times 10^{-4}$ &  $-4.8\times 10^{-4}$ &  $-4.1\times 10^{-3}$  \\ 
$d_4$ &	$4.82\times 10^{-4}$ & $4.98\times 10^{-4}$ & $4.3\times 10^{-4}$	&  $5.95\times 10^{-4}$	& $4.68\times 10^{-4}$	&	     $1.84\times 10^{-4}$ & $1.33\times10^{-4}$ & $1.6\times 10^{-4}$ &  $1.3\times 10^{-4}$ &  $1\times 10^{-3}$  \\ 
\Xhline{0.1pt} \\
$z_c$ &	$0.753$ & $0.744$ & $0.76$	&  $0.744$	& $0.78$	&	  $0.38$ & $0.38$ & $0.38$ &  $0.38$ &  $0.375$  \\ 

\end{tabular}
\end{ruledtabular}
\caption{\label{tab:coeff_tab_skyloc_dl} Coefficients for the fit reported in Section \ref{sec:analytical_pe} for the sky position and luminosity distance uncertainties. }
\end{table*}

\end{turnpage}

\begin{turnpage}
\begin{table*}[ht!]
\begin{ruledtabular}
\begin{tabular}{c| c c c c c | c c c c c } 
        & \multicolumn{10}{c}{Parameters}\\
 Coefficients  & \multicolumn{5}{c}{$\Delta \mathcal{M}/\mathcal{M}$}  & \multicolumn{5}{c}{$\Delta \rm q/q $} \\ 
 
 & median  & $68\% \, \rm lower$ & $68\%  \,\rm upper$ &  $95\% \, \rm lower$ & $95\%  \,\rm upper$ 
 & median  & $68\% \, \rm lower$ & $68\%  \,\rm upper$ &  $95\% \, \rm lower$ & $95\%  \,\rm upper$   \\

\Xhline{0.1pt} \\
$c_1$ &	$265.154 $ & $204.7335 $ & $-34.7932 $ &  $3031.93$	&	$-312.206 $  &  $111.22 $  & $403.8919$ & $-132.91$ &  $2670.45$ &  $-392.089 $ \\ 
$c_2$ &	$-74.012 $ & $-34.3552 $ & $63.0254 $ &  $-1446.096$	&	$184.6885$  &  $-13.238 $  & $-152.5055 $ & $100.255$ &  $-1260.664 $ &  $225.7991  $ \\
$c_3$ &	$1.539 $ & $-6.60087 $ & $-19.0848 $ &  $224.992$	&	$-36.6357$  &  $-6.2392 $  & $15.513  $ & $-23.5952$ &  $193.5704$ &  $-43.6128 $ \\
$c_4$ &	$0.52438  $ & $1.054653  $ & $1.54848 $ &  $-11.4193$	&	$2.38862$  &  $0.8533  $  & $-0.2618  $ & $1.72765$ &  $-9.65920$ &  $2.78034 $ \\
$c_5$ &	$-250.187 $ & $-212.7418   $ & $-61.196 $ &  $-2025.58$	&	$65.1703  $  &  $-192.371   $  & $-371.442 $ & $-36.698$ &  $-1793.416$ &  $119.722 $ \\ 
$c_6$ &	$[z, \,  \log_{10} \mtot]$	& $[z, \,  \log_{10} \mtot]$ & $-0.022941 $ &  $[z, \,  \log_{10} \mtot]$	&	$[z, \,  \log_{10} \mtot]$	&  $[z, \,  \log_{10} \mtot] $ & $[z, \,  \log_{10} \mtot]$ & $-4.5964$ &  $[z, \,  \log_{10} \mtot]$ &  $[z, \,  \log_{10} \mtot]$ \\ 
$c_7$ &	$-7.49936 $ & $-2.57786$ & $[z, \,  \log_{10} \mtot] $	&  $-151.76925$	&	$12.45644 $	&  $-5.81345  $ & $-19.1958 $ & $[z, \,  \log_{10} \mtot]$ &  $-131.5355$ &  $17.20657$ \\ 
$c_8$ &	$0.024866 $ & $-0.293273$ & $-0.6267496$ &  $7.763953$	&	$-0.92562$  &  $-0.0076827 $  & $0.680987  $ & $-0.567725 $ &  $6.62843 $ &  $-1.192362  $ \\
$c_9$ &	$59.854   $ & $51.8721$ & $21.8577$ &  $426.415$	&	$4.8196 $  &  $54.7792$  & $89.8498 $ & $23.238 $ &  $379.1713 $ &  $-6.76903  $ \\
$c_{10}$ &	$-21.99511 $ & $-16.97759  $ & $-4.55444 $ &  $-204.7197 $	&	$1.6225 $  &  $-21.48994$  & $-38.2546$ & $-6.8044 $ &  $-180.43504$ &  $7.5663$ \\
$c_{11}$ &	$2.249767 $ & $1.25226 $ & $-0.3938$ &  $32.18694$	&	$-1.071028$  &  $2.44623$  & $5.081692$ & $0.1906$ &  $28.059366$ &  $-2.079704 $ \\ 
$c_{12}$ &	$-0.0446262  $ & $0.018931 $ & $0.0877907$ &  $-1.655344 $	&	$0.1089012 $  &  $-0.067153$  & $-0.2032869$ & $0.047122 $ &  $-1.4233821$ &  $0.165538$ \\ 
$c_{13}$&	$-4.22054  $ & $-3.61683 $ & $-1.77072 $ &  $-28.6609 $	&	$-1.15488 $  &  $-4.2843 $  & $-6.4682$ & $-2.23887$ &  $-25.5663 $ &  $-0.36784$ \\ 
$c_{14}$ &	$1.58222  $ & $1.21883 $ & $0.455984 $ &  $13.78835 $	&	$0.30454 $  &  $1.74461 $  & $2.790075 $ & $0.791875$ &  $12.19611 $ &  $-0.09896$ \\
$c_{15}$ &	$-0.169695 $ & $-0.099666 $ & $0.0013989$ &  $-2.173974 $	&	$0.0035938 $  &  $-0.214409  $  & $-0.37913 $ & $-0.067924  $ &  $-1.903058$ &  $0.072043$ \\
$c_{16}$ &	$0.0041172 $ & $-2.4234\times 10^{-4} $ & $-0.0044783$ &  $0.1122246 $	&	$-0.00354464$  &  $0.0073079 $  & $0.015841$ & $-1.2534\times 10^{-4}$ &  $0.0969823 $ &  $-0.0073868$ \\
\Xhline{0.1pt} \\
$d_1$ &	$86.4775 $ & $62.2216$ & $5.5642 $ &  $969.4018 $	&	$-51.8627$  &  $67.7606$  & $153.149 $ & $5.2722$ &  $850.214 $ &  $-79.8362 $ \\
$d_2$ &	$2.49\times 10^{-4}  $ & $5.2\times 10^{-4}$ & $1.26\times 10^{-4}$ &  $2.05\times 10^{-3} $	&	$6.78\times 10^{-4}$  &  $8.0E-05 $  & $1.0\times 10^{-3} $ & $1.82\times 10^{-4} $ &  $1.3\times 10^{-3} $ &  $-3.22\times 10^{-4}$ \\
$d_3$ &	$8.82\times 10^{-3} $ & $1.05\times 10^{-2}$ & $2.15\times 10^{-3} $ &  $1.07\times 10^{-2}  $	&	$2.26\times 10^{-3}$  &  $8.78\times 10^{-3}$  & $9.21\times 10^{-3}  $ & $1.71\times 10^{-3}$ &  $9.84\times 10^{-3} $ &  $1.79\times 10^{-3}$ \\
$d_4$ &	$-9.16\times 10^{-4} $ & $-1.07\times 10^{-3}$ & $-2.28\times 10^{-4} $ &  $-1.13\times 10^{-3}  $	&	$1.61\times 10^{-4} $  &  $-8.3\times 10^{-4} $  & $-9.39\times 10^{-4} $ & $-1.91\times 10^{-4} $ &  $-1.08\times 10^{-3} $ &  $2.31\times 10^{-4}$ \\
\Xhline{0.1pt} \\
$z_c$ &	$0.364 $ & $0.349$ & $0.327$ &  $0.349$	&	$0.35$  &  $0.35$  & $0.36$ & $0.36$ &  $0.36$ &  $0.349$ \\

\end{tabular}
\end{ruledtabular}
\caption{\label{tab:coeff_tab_mchirp_q} Same as Tab. \ref{tab:coeff_tab_skyloc_dl} for chirp mass and mass ratio uncertainties.}
\end{table*}

\end{turnpage}

\begin{turnpage}
\begin{table*}[ht!]
\begin{ruledtabular}
\begin{tabular}{c| c c c c c | c c c c c } 
        & \multicolumn{10}{c}{Parameters}\\
 Coefficients  & \multicolumn{5}{c}{$\Delta \Omega$}  & \multicolumn{5}{c}{$\Delta \rm d_L / d_L $} \\ 
 
 & median  & $68\% \, \rm lower$ & $68\%  \,\rm upper$ &  $95\% \, \rm lower $ & $95\%  \,\rm upper $ 
 & median  & $68\% \, \rm lower$ & $68\%  \,\rm upper$ &  $95\% \, \rm lower $ & $95\%  \,\rm upper $   \\

\Xhline{0.1pt} \\
$m_1$ & $[z]$ & $[z]$ & $[z]$ & $[z]$ & $[z]$  & $[z]$ & $[z]$ & $[z]$ & $[z]$ & $[z]$   \\ 
$m_2$ & $13.296$ & $14.39$ & $11.92$ & $16.513 $ & $8.024$  & $6.088 $ & $6.567 $ & $5.934 $ & $7.134  $ & $5.511 $   \\ 
$m_3$ & $62.097$ & $86.262 $ & $-3.704 $ & $93.583$ & $-26.74$  & $26.99 $ & $32.9  $ & $24.703   $ & $33.24 $ & $36.263$   \\ 
$m_4$ & $-0.858 $ & $ -0.836$ & $ -0.7946$ & $-0.8473$ & $-0.7515$  & $-0.385$ & $-0.398 $ & $-0.3858$ & $-0.4075 $ & $-0.3928$   \\ 
$m_5$ & $-11.4027 $ & $-15.5978$ & $-0.0168$ & $-16.76$ & $3.664  $  & $-5.0583  $ & $-6.046  $ & $-4.6995   $ & $-6.0585  $ & $-6.844 $   \\ 
$m_6$ & $0.1608$ & $0.1576  $ & $0.1613$ & $0.1522$ & $0.1668$  & $0.0794  $ & $0.0771  $ & $0.08033  $ & $0.07462 $ & $0.08179$  \\ 
$m_7$ & $0.6807$ & $0.91963$ & $0.03179 $ & $0.9788$ & $-0.1593$  & $0.307$ & $0.3611   $ & $0.28916 $ & $0.3589$ & $0.42055$   \\ 
$m_8$ & $-4.1 $ & $-4.511  $ & $ -3.671 $ & $-5.2655 $ & $-2.34 $  & $-1.889  $ & $-2.044   $ & $-1.8422  $ & $-2.235 $ & $ -1.6968 $  \\ 
$m_9$ & $-0.0725$ & $-0.0726$ & $-0.0853 $ & $-0.0641 $ & $-0.0986$  & $-0.0429 $ & $-0.038 $ & $ -0.04372 $ & $-0.03373$ & $-0.0442$   \\ 
$m_{10}$ & $0.4119$ & $0.4477 $ & $0.382$ & $0.5116$ & $0.2718$  & $0.1948 $ & $0.2061$ & $0.192 $ & $0.2213$ & $0.18024$   \\ 
\Xhline{0.1pt} \\
$n_1$ & $7.222\times 10^{-6}$ & $3.852\times 10^{-6}$ & $1.31\times 10^{-3}$ & $2.998\times 10^{-6}$ & $3.743\times 10^{-5}$  & $1.91\times 10^{-5}$ & $1.315\times 10^{-5} $ & $2.696\times 10^{-5}  $ & $1.313\times 10^{-5}$ & $1.3\times 10^{-5}$   \\ 
$n_2$ & $-8.6325\times 10^{-3}$ & $-6.1675\times 10^{-3}$ & $9.8153\times 10^{-2}$ & $-5.63045\times 10^{-3}$ & $1.726\times 10^{-2}$  & $-2.0083\times 10^{-2} $ & $-1.6227\times 10^{-2}$ & $-2.234\times 10^{-2}$ & $-1.5893\times 10^{-2}$ & $-1.5404\times 10^{-2}$  \\ 
$n_3$ & $1.082\times 10^{-4}$ & $6.092\times 10^{-5}$ & $1.746\times 10^{-2}$ & $4.6124\times 10^{-5}$ & $5.2977\times 10^{-4}$  & $2.846\times 10^{-4} $ & $2.02\times 10^{-4} $ & $3.89\times 10^{-4} $ & $1.98\times 10^{-4} $ & $2.057\times 10^{-4}$   \\ 
$n_4$ & $-5.34\times 10^{-5}$ & $-2.943\times 10^{-5} $ & $-8.143\times 10^{-3}$ & $-2.236\times 10^{-5}$ & $-2.538\times 10^{-4}$  & $-1.382\times 10^{-4} $ & $-9.85\times 10^{-5} $ & $-1.83\times 10^{-4}$ & $-9.37\times 10^{-5} $ & $-9.81\times 10^{-5}$   \\
$n_5$  & $7.53\times 10^{-6}$ & $4.114\times 10^{-6} $ & $1.107\times 10^{-3}$ & $3.138\times 10^{-6}$ & $ 3.57\times 10^{-5}$  & $1.947\times 10^{-5} $ & $1.388\times 10^{-5}  $ & $2.52\times 10^{-5} $ & $1.286\times 10^{-5}  $ & $1.37\times 10^{-5}$ \\

\end{tabular}
\end{ruledtabular}
\caption{\label{tab:coeff_tab_skyloc_dl_merger_below} Coefficients for the fit reported in Section \ref{sec:analytical_pe} for the sky position and luminosity distance uncertainties at merger for systems with $\mtot \le 3 \times 10^6 \msun$. }
\end{table*}
\end{turnpage}

\begin{turnpage}
\begin{table*}[ht!]
\begin{ruledtabular}
\begin{tabular}{c| c c c c c | c c c c c } 
        & \multicolumn{10}{c}{Parameters}\\
 Coefficients  & \multicolumn{5}{c}{$\Delta \Omega$}  & \multicolumn{5}{c}{$\Delta \rm d_L / d_L $} \\ 
 
 & median  & $68\% \, \rm lower$ & $68\%  \,\rm upper$ &  $95\% \, \rm lower $ & $95\%  \,\rm upper $ 
 & median  & $68\% \, \rm lower$ & $68\%  \,\rm upper$ &  $95\% \, \rm lower $ & $95\%  \,\rm upper $   \\

\Xhline{0.1pt} \\
$m_1$ & $[z]$ & $[z]$ & $[z]$ & $[z]$ & $[z]$  & $[z]$ & $[z]$ & $[z]$ & $[z]$ & $[z]$   \\ 
$m_2$ & $54.14$ & $52.38 $ & $65.175$ & $50.41$ & $73.024$  & $30.014$ & $31.89 $ & $31.65$ & $31.075 $ & $33.47$   \\ 
$m_3$ & $202.74$ & $175.795 $ & $51.54  $ & $71.944$ & $254.464$  & $77.09$ & $74.578$ & $74.52 $ & $36.196 $ & $112.27 $   \\ 
$m_4$ & $-2.303 $ & $-2.247 $ & $-2.587  $ & $-2.195 $ & $-2.75 $  & $-1.17 $ & $-1.127  $ & $-1.236  $ & $-1.11  $ & $-1.316  $   \\ 
$m_5$ & $-27.1115$ & $-23.352$ & $-4.2706  $ & $-8.3389 $ & $-33.468  $  & $-9.9143 $ & $-9.5723  $ & $-9.3884   $ & $-4.0335  $ & $-14.8135$   \\ 
$m_6$ & $0.1833$ & $0.184$ & $0.185 $ & $0.1785$ & $0.1905$  & $0.0915$ & $0.091  $ & $0.0930  $ & $0.0894  $ & $0.09356$  \\ 
$m_7$ & $1.20935 $ & $1.0344  $ & $0.06735$ & $0.3129$ & $1.46415 $  & $0.42342  $ & $0.4075$ & $0.3914$ & $0.1419  $ & $0.65054$   \\ 
$m_8$ & $-13.878$ & $-13.425 $ & $-16.788$ & $ -12.831$ & $-18.996$  & $-7.782  $ & $-8.362  $ & $-8.1875 $ & $-8.123  $ & $-8.668 $  \\ 
$m_9$ & $0.1188$ & $0.11 $ & $0.157$ & $0.1068$ & $0.1753$  & $0.0620 $ & $0.0565$ & $0.0699 $ & $0.0554 $ & $0.08076$   \\ 
$m_{10}$ & $0.9597$ & $0.9309$ & $1.1507 $ & $0.886 $ & $1.3064 $  & $0.5401$ & $0.5843 $ & $0.565$ & $0.5668$ & $0.5966$   \\ 
\Xhline{0.1pt} \\
$n_1$ & $4.123\times 10^{-7} $ & $5.01\times 10^{-7}$ & $3.88\times 10^{-6}$ & $2.18\times 10^{-6}$ & $2.936\times 10^{-7}$  & $1.397\times 10^{-6} $ & $1.385\times 10^{-6}$ & $1.4\times 10^{-6} $ & $4.61\times 10^{-6}$ & $7.2\times 10^{-7}$   \\ 
$n_2$ & $-1.9555\times 10^{-3}$ & $-2.23144\times 10^{-3}$ & $-5.587\times 10^{-3}$ & $-4.7614\times 10^{-3}$ & $-1.5454\times 10^{-3} $  & $-4.9447\times 10^{-3} $ & $-5.0946\times 10^{-3} $ & $-5.035\times 10^{-3} $ & $-9.245\times 10^{-3}$ & $-3.5031\times 10^{-3}$  \\ 
$n_3$ & $6.46\times 10^{-6}$ & $8.121\times 10^{-6} $ & $ 6.27\times 10^{-5}$ & $3.347\times 10^{-5}$ & $4.37\times 10^{-6}$  & $2.087\times 10^{-5} $ & $2.194\times 10^{-5}$ & $2.124\times 10^{-5}  $ & $ 6.9\times 10^{-5}  $ & $1.1\times 10^{-5}$   \\ 
$n_4$ & $-3.052\times 10^{-6}$ & $-3.892\times 10^{-6}$ & $-3.115\times 10^{-5} $ & $-1.542\times 10^{-5}$ & $-2.09\times 10^{-6} $  & $-9.85\times 10^{-6} $ & $-1.04\times 10^{-5}$ & $-1.027\times 10^{-5} $ & $-3.25\times 10^{-5}$ & $-5.12\times 10^{-6}$   \\
$n_5$  & $4.2\times 10^{-7}$ & $5.435\times 10^{-7} $ & $4.35\times 10^{-6}$ & $2.097\times 10^{-6} $ & $2.857\times 10^{-7}$  & $1.346\times 10^{-6}$ & $1.44\times 10^{-6}$ & $1.425\times 10^{-6}$ & $4.47\times 10^{-6}$ & $6.92\times 10^{-7}$ \\

\end{tabular}
\end{ruledtabular}
\caption{\label{tab:coeff_tab_skyloc_dl_merger_above} Coefficients for the fit reported in Section \ref{sec:analytical_pe} for the sky position and luminosity distance uncertainties at merger for systems with $\mtot > 3 \times 10^6 \msun$. }
\end{table*}
\end{turnpage}

\end{document}